%% file: gen-model-multichannel.tex
\documentclass[10pt,journal]{IEEEtran}

\input{include/preamble}
\input{include/acronyms}

\input{include/macros}

\usepackage{comment}

\title{Multibranch Generative Models for Multichannel Imaging with an Application to PET/CT Synergistic Reconstruction}

\author{Noel Jeffrey Pinton, Alexandre Bousse,~\IEEEmembership{Member,~IEEE},  Catherine Cheze-Le-Rest,  
	Dimitris Visvikis,~\IEEEmembership{Senior Member,~IEEE}
	
    \thanks{This work did not involve human subjects or animals in its research.}
	\thanks{This work was supported by the French National Research Agency (ANR) under grant No ANR-20-CE45-0020 and by France Life Imaging under grant No ANR-11-INBS-0006}
	\thanks{All authors are with Univ. Brest, LaTIM, INSERM, UMR 1101, 29238 Brest, France.}
    \thanks{C. Cheze-Le-Rest is also with Nuclear Medicine Department, Poitiers University Hospital, F-86022, Poitiers, France.}
	\thanks{Corresponding authors: A. Bousse, \texttt{bousse@univ-brest.fr} }
}

\newlength{\dimvert}
\newcommand{\rowname}[1]
{\rotatebox{90}{\makebox[\dimvert][c]{#1}}}
\newcommand{\rownametwo}[1]
{\rotatebox{-90}{\makebox[-\dimvert][c]{#1}}}

\begin{document}
	
\maketitle

\input{content/abstract}	
\input{content/intro}

\input{content/method}

\input{content/results}

\input{content/discussion}

\input{content/conclusion}

\appendices

\section{Training of the MVAE}\label{sec:appB}

\input{./content/training_vae}

\section{Additional Reconstructed Images}\label{sec:appA}

In this appendix we display the reconstructed images for the nine other patient from data  \ac{HL} data (Figure~\ref{fig:bigfig1}) and \ac{LH} data (Figure~\ref{fig:bigfig2}).

\input{./content/big_fig1}
\input{./content/big_fig2}

\section*{Acknowledgment}

All authors declare that they have no known conflicts of interest in terms of competing financial interests or personal relationships that could have an influence or are relevant to the work reported in this paper. 

\AtNextBibliography{\footnotesize} 
\printbibliography

\end{document}

%% file: include/preamble.tex
\usepackage{amsmath,amsfonts,amssymb,mathrsfs}
\usepackage{breqn}
\usepackage{algorithm}
\usepackage{array}
\usepackage{textcomp}
\usepackage{stfloats}
\usepackage{url}
\usepackage{verbatim}
\usepackage{graphicx}
\usepackage{stmaryrd}
\usepackage{microtype}
\usepackage{datatool}
\usepackage{siunitx}
\usepackage{pgf}

\usepackage{multirow}

\usepackage{bm} 
\usepackage{lipsum}

\usepackage{acro} 
\acsetup{single}
\usepackage{bm}
\usepackage{amsfonts}
\usepackage{soul}

\usepackage[
	style=ieee,
	dashed=false,
	maxbibnames=20,
	minbibnames=2,
	maxcitenames=1,
	mincitenames=1,
	backend=biber,
	sorting=none,
	useprefix=false
	]{biblatex}

\usepackage{tikz}
\usepackage{overpic}

\usetikzlibrary{arrows.meta,shapes,intersections,calc,angles,decorations.pathreplacing,intersections,quotes,angles,positioning,fit,petri,through,shadows,plotmarks,spy}

\usepackage{pgfplots}
\usepackage{pgfplotstable}
\usetikzlibrary{pgfplots.groupplots}

\usepackage{algpseudocode}

\usepackage{caption}
\usepackage{subcaption}
\usepackage[labelformat=simple,font=footnotesize]{subcaption}

\usepackage[font=small]{caption}


%% file: include/acronyms.tex
\DeclareAcronym{2D}{
	short=2-D,
	long=2-dimensional,
}

\DeclareAcronym{3D}{
	short=3-D,
	long=3-dimensional,
}

\DeclareAcronym{dD}{
	short=$d$-D,
	long=$d$-dimensional,
}

\DeclareAcronym{CT}{
	short=CT, 
	long=computed tomography,
}

\DeclareAcronym{PET}{
	short=PET, 
	long=positron emission tomography,
}

\DeclareAcronym{SPECT}{
	short=SPECT, 
	long=single-photon emission CT,
}

\DeclareAcronym{MRI}{
	short=MRI, 
	long=magnetic resonance imaging,
}

\DeclareAcronym{MR}{
	short=MR, 
	long=magnetic resonance,
}

\DeclareAcronym{PCCT}{
	short=PCCT, 
	long=photon-counting computed tomography,
}

\DeclareAcronym{DECT}{
	short=DECT, 
	long=dual-energy computed tomography,
}

\DeclareAcronym{MBIR}{
	short=MBIR, 
	long=model-based iterative reconstruction,
}

\DeclareAcronym{WLS}{
	short=WLS, 
	long=weighted least squares,
}
\DeclareAcronym{PWLS}{
	short=PWLS, 
	long=penalized weighted least squares,
}

\DeclareAcronym{PLS}{
	short=PLS, 
	long=parallel level set,
}

\DeclareAcronym{PML}{
	short=PML, 
	long=penalized maximum likelihood,
}

\DeclareAcronym{SPS}{
	short=SPS, 
	long=separable paraboloidal surrogates,
}

\DeclareAcronym{CS}{
	short=CS,
	long=compressive sensing,
}

\DeclareAcronym{TV}{
	short=TV,
	long=total variation,
}

\DeclareAcronym{TNV}{
	short=TNV, 
	long=total nuclear variation,
}

\DeclareAcronym{JTV}{
	short=JTV, 
	long=joint total variation,
}

\DeclareAcronym{DTV}{
	short=DTV, 
	long=directional total variation,
}

\DeclareAcronym{SQS}{
	short=SQS, 
	long=separable quadratic surrogate,
}

\DeclareAcronym{ADMM}{
	short=ADMM, 
	long=alternating direction method of multipliers,
}

\DeclareAcronym{ML}{
	short=ML,
	long= machine learning,
}

\DeclareAcronym{DL}{
	short=DL,
	long= deep learning,
}

\DeclareAcronym{DiL}{
	short=DiL,
	long= dictionary learning,
}

\DeclareAcronym{CDL}{
	short=CDL,
	long=convolutional DiL,
}

\DeclareAcronym{MCDL}{
	short=CDL,
	long=multichannel convolutional dictionary learning,
}

\DeclareAcronym{CAOL}{
	short=CAOL, 
	long=convolutional analysis operator learning,
}

\DeclareAcronym{MCAOL}{
	short=MCAOL, 
	long=multichannel convolutional analysis operator learning,
}

\DeclareAcronym{PDF}{
	short=PDF,
	long=probability distribution function,
}

\DeclareAcronym{PSNR}{
	short=PSNR, 
	long=peak signal-to-noise ratio,
}

\DeclareAcronym{SSIM}{
	short=SSIM, 
	long=structural similarity index measure,
}

\DeclareAcronym{SNR}{
	short=SNR, 
	long=signal-to-noise ratio,
}

\DeclareAcronym{CNN}{
	short=CNN, 
	long=convolutional NN,
}

\DeclareAcronym{NN}{
	short=NN, 
	long=neural network,
}

\DeclareAcronym{GAN}{
	short=GAN, 
	long=generative adversarial network,
}

\DeclareAcronym{WGAN}{
	short=W-GAN, 
	long=Wasserstein GAN,
}

\DeclareAcronym{VAE}{
	short=VAE, 
	long=variational autoencoder,
}

\DeclareAcronym{MVAE}{
	short=MVAE, 
	long=multibranch VAE,
}

\DeclareAcronym{beta-VAE}{
	short=$\beta$-VAE, 
	long=beta-variational autoencoder,
}

\DeclareAcronym{LoR}{
	short=LoR,
	long=line of response,
	long-plural-form = lines of response,
}

\DeclareAcronym{EM}{
	short=EM,
	long=expectation-maximization,
}

\DeclareAcronym{MLEM}{
	short=MLEM,
	long=maximum-likelihood expectation-maximization,
}

\DeclareAcronym{OSEM}{
	short=OSEM, 
	long=ordered subsets expectation maximization,
}

\DeclareAcronym{MAPEM}{
	short=MAPEM, 
	long=maximum \textit{a posteriori} expectation maximization,
}

\DeclareAcronym{FBP}{
	short=FBP, 
	long=filtered backprojection,
}

\DeclareAcronym{IFFT}{
	short=IFFT, 
	long=inverse fast Fourier transform,
}

\DeclareAcronym{GT}{
	short=GT,
	long=ground truth,
}

\DeclareAcronym{HU}{
	short=HU,
	long=Houndsfield Units,
}

\DeclareAcronym{LAC}{
	short=LAC,
	long=linear attenuation coefficients,
}

\DeclareAcronym{MNIST}{
	short=MNIST,
	long=Modified National Institute of Standards and Technology,
}

\DeclareAcronym{LBFGS}{
	short=L-BFGS,
	long=limited-memory  Broyden-Fletcher-Goldfarb-Shanno,
}

\DeclareAcronym{KL}{
	short=KL,
	long=Kullback-Leibler
}

\DeclareAcronym{ReLU}{
	short=RelU,
	long=rectified linear unit
}

\DeclareAcronym{PSO}{
	short=PSO,
	long=particle swarm optimization
}

\DeclareAcronym{DM}{
	short=DM,
	long=diffusion model
}

\DeclareAcronym{DPS}{
	short=DPS,
	long=diffusion posterior sampling
}

\DeclareAcronym{PCA}{
	short=PCA,
	long=principal component analysis 
}
\DeclareAcronym{MSE}{
	short=MSE,
	long=mean squared error 
}

\DeclareAcronym{LH}{
	short=LC-PET/HC-CT,
	long=low-count PET/high-count CT, 
}

\DeclareAcronym{HL}{
	short=HC-PET/LC-CT,
	long=high-count PET/low-count CT, 
}

\DeclareAcronym{VPET}{
	short=MVAE-PET,
	long=MVAE-reconstructed PET, 
}

\DeclareAcronym{VCT}{
	short=MVAE-CT,
	long=MVAE-reconstructed CT, 
}

\DeclareAcronym{PPET}{
	short=PLS-PET,
	long=PLS-reconstructed PET, 
}

\DeclareAcronym{PCT}{
	short=PLS-CT,
	long=PLS-reconstructed CT, 
}

\DeclareAcronym{EMPET}{
	short=MLEM-PET,
	long=MLEM-reconstructed PET, 
}

\DeclareAcronym{WCT}{
	short=WLS-CT,
	long=WLS-reconstructed CT, 
}

\DeclareAcronym{PETunet}{
	short=MLEM-PET+UNet,
	long=MLEM PET reconstruction  followed by U-Net denoising, 
}

\DeclareAcronym{CTunet}{
	short=WLS-CT+UNet,
	long=WLS CT reconstructtion CT followed by U-Net denoising, 
}

\DeclareAcronym{DIP}{
	short=DIP,
	long=deep image prior, 
}

\DeclareAcronym{DIPPET}{
	short=DIP-PET,
	long=DIP-reconstructed PET, 
}

\DeclareAcronym{DIPCT}{
	short=DIP-CT,
	long=DIP-reconstructed CT, 
}

\DeclareAcronym{ELBO}{
	short=ELBO,
	long=maximizing the evidence lower bound, 
}

%% file: include/macros.tex
\newcommand{\boldr}{\bm{r}}
\newcommand{\bolds}{\bm{s}}

\newcommand{\boldu}{\bm{u}}

\newcommand{\boldx}{\bm{x}}
\newcommand{\boldy}{\bm{y}}
\newcommand{\boldz}{\bm{z}}

\newcommand{\boldmu}{\bm{\mu}}

\newcommand{\boldA}{\bm{A}}

\newcommand{\boldG}{\bm{G}}

\newcommand{\boldP}{\bm{P}}

\newcommand{\boldU}{\bm{U}}

\newcommand{\boldZ}{\bm{Z}}

\newcommand{\calN}{\mathcal{N}}

\newcommand{\calU}{\mathcal{U}}

\newcommand{\calX}{\mathcal{X}}
\newcommand{\calY}{\mathcal{Y}}
\newcommand{\calZ}{\mathcal{Z}}

\newcommand{\rmd}{\mathrm{d}}

\newcommand{\rmn}{\mathrm{n}}

\newcommand{\R}{\mathbb{R}}

\newcommand{\boldybar}{\bar{\boldy}}

\newcommand{\boldxhat}{\hat{\boldx}}
\newcommand{\boldxcheck}{\check{\boldx}}
\newcommand{\boldzhat}{\hat{\boldz}}

\newcommand{\boldtheta}{\bm{\theta}}
\newcommand{\boldsigma}{\bm{\sigma}}

\newcommand{\boldpsi}{\bm{\psi}}

\newcommand{\boldeps}{\bm{\epsilon}}

\newcommand{\argmin}{\operatornamewithlimits{arg\,min}}

\newcommand{\boldzero}{\bm{0}}

\algnewcommand{\Inputs}[1]{%
	\State \textbf{Inputs:}
	\Statex \hspace*{\algorithmicindent}\parbox[t]{.8\linewidth}{\raggedright #1}
}
\algnewcommand{\Initialize}[1]{%
	\State \textbf{Initialize:}
	\Statex \hspace*{\algorithmicindent}\parbox[t]{.8\linewidth}{\raggedright #1}
}

%% file: content/abstract.tex
\begin{abstract}
	This paper presents a novel approach for learned synergistic reconstruction of medical images using multibranch generative models. Leveraging \acp{VAE}, our model learns from pairs of images simultaneously, enabling effective denoising and reconstruction. Synergistic image reconstruction is achieved by incorporating the trained models in a regularizer that evaluates the distance between the images and the model. We demonstrate the efficacy of our approach on both \ac{MNIST} and \ac{PET}/\ac{CT} datasets, showcasing improved image quality for low-dose imaging. Despite challenges such as patch decomposition and model limitations, our results underscore the potential of generative models for enhancing medical imaging reconstruction.
\end{abstract}

\begin{IEEEkeywords}
	Multibranch Generative Models, Multichannel Imaging, Synergistic Reconstruction
\end{IEEEkeywords}  

%% file: content/intro.tex
\section{Introduction}

\IEEEPARstart{M}{ultimodal} imaging refers to acquiring data from different sources or from different techniques to capture complementary information about the object or scene being observed. Multimodal imaging is used in various fields such as  remote sensing \cite{mm_rs_gomez2015, Dalla_Mura_Prasad_Pacifici_Gamba_Chanussot_Benediktsson_2015} robotics \cite{xue_mm_hri_2020, Su_Qi_Chen_Yang_Sandoval_Laribi_2023} and medical imaging  \cite{Pichler_Judenhofer_Pfannenberg_2008, Decazes_Hinault_Veresezan_Thureau_Gouel_Vera_2021}. 
Some of the modalities used in the latter field include \ac{PET}, \ac{CT}, \ac{MRI}, ultrasound, and various optical imaging techniques. \Ac{PET} is a powerful medical imaging technique that uses a small amount of radioactive material to visualize and track various processes in the body. It provides valuable insights into cancer detection and other areas of medicine such as cardiology and neurology \cite{bailey2005positron}. \Ac{PET} is  often completed with \ac{CT} and \ac{MRI} which provide anatomical information.

Images are reconstructed by solving  modality-specific inverse problems. In medical imaging, early methods were based on inversion formulas such as \ac{FBP} \cite{natterer2001mathematics} for \ac{CT} and \ac{PET} as well as \ac{IFFT} for \ac{MRI}. These methods were then followed by \ac{MBIR} algorithms which consist of iteratively minimizing a cost function comprising a data fidelity term that encompasses the physics and statistics of the measurement and  a regularizer to control the noise. Such methods include \ac{EM} for \ac{PET}  and \ac{SPECT} \cite{mlem_shepp} and its regularized versions \cite{green1990bayesian,depierro1995modified}, as well as penalized \ac{WLS} for \ac{CT} \cite{elbakri2002statistical}.

Multimodal imaging systems produce multiple images of the same underlying object. In general, each modality is reconstructed individually. However, they correspond to images of the same object. Therefore, it is natural to take advantage of the intermodality information to reconstruct the images together, or \emph{synergistically}, in order to improve the \ac{SNR}. In medical imaging, this approach can give room for dose reduction and/or scan time reduction. 

Early synergistic techniques use handcrafted multichannel regularizers embedded within a \ac{MBIR} framework, such as \ac{JTV} for color images \cite{blomgren1998color} and  \ac{PET}/\ac{MRI} \cite{mehranian2017synergistic}, the \ac{PLS} prior for \ac{PET}/\ac{MRI}  \cite{ehrhardt2014joint} and  \ac{TNV} for multienergy  (or spectral) \ac{CT} \cite{rigie2015joint} (see \cite{arridge2021overview} for a review). These handcrafted regularizers promote structural similarities between the images, and therefore may not be suitable for modalities with different intrinsic resolution. For example, in \ac{PET}/\ac{CT} or \ac{PET}/\ac{MRI}, the resolution of the \ac{PET} can be artificially enhanced, and while this enhancement may improve aesthetics and quantification in applications such as brain \ac{PET}/\ac{MRI} \cite{thomas2016petpvc}, it may not accurately represent the actual distribution of radiotracers  the rest of the body where higher sensitivity is required at the expense of spatial resolution (e.g., metastases detection).

Alternatively, the intermodality information can be learned with \ac{ML} techniques. \Ac{DiL} techniques have been used in image reconstruction for single-energy \ac{CT} \cite{xu2012low} but also for spectral \ac{CT}  through tensor \ac{DiL} to sparsely represent the images in a joint multidimensional dictionary \cite{zhang2016tensor,zhang2016spectral,wu2018low,li2022tensor} (see \cite{bousse2023systematic} for a review). A similar multichannel \ac{DiL} approach was proposed for \ac{PET}/\ac{MRI} \cite{sudarshan2020joint}. All these works reported better performances using multichannel \ac{DiL} as compared with single-channel \ac{DiL}. \Ac{DiL} relies on patch decomposition, which is not efficient for joint sparse representation across channels. To remedy this, multichannel \ac{CDL} was proposed (see \cite{perelli2022multi} for dual-energy  \ac{CT}). 

Multichannel \ac{DiL} is limited to encoding structural information only. In that sense, synergistic multichannel image reconstruction could benefit from the deeper architectures used in \ac{DL}. However, few researchers addressed synergistic reconstruction using \ac{DL}. For example in a recent work, \citeauthor{corda2023single}~\cite{corda2023single} proposed an unrolling framework for synergistic \ac{PET}/\ac{MRI} reconstruction. The training of unrolling models is supervised and computationally expensive as it integrates the imaging system forward model at each layer.

In this work, which is an extension of our previous work \cite{pinton2023synergistic,pinton2023joint}, we investigate the feasibility of \ac{DL} for synergistic multichannel image reconstruction through the utilization of a deep generative model (i.e., a \ac{VAE}) incorporated within a \ac{MBIR} framework through a regularizer in a similar fashion as proposed by \citeauthor{duff2021regularising}~\cite{duff2021regularising} (and to some extend \citeauthor{kelkar2021prior}~\cite{kelkar2021prior}). For our approach, we used multiple-branch generators to map a single latent variable to multiple images. The training is unsupervised and performed on an image pair basis and does not involve the forward model, thus enabling the possibility to use the same model to any imaging system using the same modalities.

In Section~\ref{sec:method} we present our \ac{MVAE} architecture and the corresponding reconstruction algorithm.  Section~\ref{sec:results} demonstrates the capability of our architectures to generate multiple images and to convey information across channels, and shows the results of their utilization in a denoising framework with data generated from the \ac{MNIST} database \cite{deng2012mnist} and in for synergistic  \ac{PET}/\ac{CT} reconstruction with a comparison with the \ac{PLS} technique \cite{ehrhardt2014joint}. Section~\ref{sec:discussion} discusses the limitations of our method and experiments. Finally, Section~\ref{sec:conclusion} concludes this paper.

%% file: content/method.tex
\section{Method}\label{sec:method}

\subsection{Background on Medical Image Reconstruction}\label{sec:mult_imaging}

Image reconstruction corresponds to the task of estimating an image $\boldx \in \calX\triangleq \R^m $ from a random measurement $\boldy \in \calY\triangleq \R^n$, where $m$ and $n$ are respectively the dimension of the image (number of pixels or voxels) and the dimension of the measurement (number of detectors). The image $\boldx$ is a visual representation of the interior of an object (e.g., the patient). 

The measurement is modeled with a forward model which takes the form of a mapping $\boldybar \colon \calX \to \calY$ that incorporates the physics of the measurement such that given a \ac{GT} image $\boldx^\star$ the expected measurement matches the forward model, i.e., $\mathbb{E}[\boldy] = \boldybar(\boldx^\star)$. When the measurement consists of photon counting  (e.g., \ac{PET}, \ac{SPECT} and \ac{CT}),
$\boldy$ is a random vector that follows a Poisson distribution with independent entries, i.e.,
\begin{equation}\label{eq:poisson}
	\boldy \sim \mathrm{Poisson}(\boldybar(\boldx^\star)) \, .
\end{equation}

The forward model $\boldybar$ depends on the imaging system. In \ac{PET}, it is traditionally written as 
\begin{equation}\label{eq:pet_model}
	\boldybar (\boldx)  =  \tau (\boldP \boldx + \boldr)
\end{equation}
where $\boldx$ is the radiotracer distribution image, $\boldP \in \R^{n\times m}$ is the \ac{PET} system matrix such that each entry $[\boldP]_{i,j}$ is the probability that a positron-electron annihilation in voxel $j$ is detected by the $i$th detector pair (with incorporation of the attenuation factors, resolution and sensitivity), $\tau$ is the acquisition time and $\boldr \in \R^n$ is a ``background term'' representing the expected scatter and randoms per unit of time. In \ac{CT}, the standard (monochromatic) model is
\begin{equation}\label{eq:ct_model}
	\boldybar(\boldx)  =   I\cdot \exp(-\boldA \boldx)     
\end{equation}
where $\boldx$ is the attenuation image, $\boldA \in \R^{n\times m}$ is the \ac{CT} system matrix, $I$ is the X-ray intensity and the $\exp$ function applied to a vector should be understood as operating on each element individually.

Image reconstruction is achieved by matching $\boldybar(\boldx)$ to $\boldy$, i.e.,
\begin{equation}\label{eq:inv_prob}
	\text{find $\boldx$ s.t.} \quad \boldybar(\boldx) \approx \boldy   \, ,
\end{equation}	
which is an (ill-posed) inverse problem. As solving \eqref{eq:inv_prob} cannot be achieved with an inversion formula without amplifying the noise,  \ac{MBIR} techniques have been prevalent over the last decades. \Ac{MBIR} consists in solving an optimization problem of the form 
\begin{equation}\label{eq:PML}
	\min_{\boldx \in \calX} \, L(\boldy , \boldybar(\boldx)) + \beta R(\boldx)  
\end{equation}
where $L$ is a loss function that evaluates the goodness of fit between the measurement  $\boldy$ and the expectation $\boldybar$, $R\colon \calX \to \R$ is a regularizer, and $\beta>0$ is a weight, with an iterative algorithm. The loss function $L$ is usually defined a negative Poisson log-likelihood, although it can be approximated by a \ac{WLS} loss (see for example \cite{elbakri2002statistical} in \ac{CT}). The regularizer $R$ promotes images that have desired properties, such as piecewise smoothness or sparsity of the gradient.

The choice of the algorithm to solve \eqref{eq:PML} largely depends on $\boldybar$, $L$,  and $R$. Examples from the literature include \ac{SPS} for \ac{CT} \cite{elbakri2002statistical}, \ac{MLEM}, \ac{OSEM}  \cite{mlem_shepp,Hudson1994} and modified \ac{MLEM} \cite{depierro1995modified} for \ac{PET} with smooth regularizers. Non-smooth regularizers can be addressed for example with a primal-dual algorithm \cite{sidky2012convex}.

\subsection{Synergistic Reconstruction in Multimodal Imaging}

Multimodal hybrid imaging systems such as \ac{PET}/\ac{CT}, \ac{PET}/\ac{MRI},  \ac{SPECT}/\ac{CT}  (and to some extent, spectral \ac{CT})  can acquire multiple measurement $\{\boldy_k\}=\{\boldy_1,\dots,\boldy_K\}$, $\boldy_k  \in \R^{n_k}\triangleq \calY_k$, to reconstruct several images $\{\boldx_k\}=\{\boldx_1,\dots,\boldx_K\}$. For simplicity, we assume that the images $\boldx_k$ are all $m$-dimensional. In general, each channel $k$ is individually reconstructed  by solving  \eqref{eq:PML} using its corresponding forward model $\boldybar_k \colon \calX \to \calY_k$, loss function $L_k\colon \calY_k\times \calY_k \to \R$ and regularizer $R_k$. Another approach consists of reconstructing the images simultaneously by solving  
\begin{equation}\label{eq:PMLsyn}
	\min_{\{\boldx_k\} \in \calX^K} \, \sum_{k=1}^K  \eta_k L_k(\boldy_k , \boldybar_k(\boldx_k))  + \beta R_{\mathrm{syn}}(\boldx_1,\dots,\boldx_K)  
\end{equation}
where  $R_{\mathrm{syn}} \colon  \calX^K \to \R$ is a \emph{synergistic} regularizer that promotes structural and/or functional dependencies between the multiple images and the $\eta_k$s are positive normalized weights ($\eta_k>0$, $\sum_k \eta_k = 1$) that tune the strength of the regularizer for each channel $k$ independently\footnote{In \eqref{eq:PMLsyn} the weight $\beta$ may also be incorporated in the $\eta_k$s. However, in this work, it is more convenient to keep them separated, cf. footnote in Section~\ref{sec:algo}.}. 

A classical regularizer is  \ac{JTV} \cite{blomgren1998color} which encourages joint sparsity of the image gradients. Similarly, \ac{TNV}, which encourages common edge locations and a shared gradient direction among image channels, was used in spectral \ac{CT} \cite{rigie2015joint}. Another example is the \ac{PLS} prior which was used in \ac{PET}/\ac{MRI} \cite{ehrhardt2014joint}. By promoting common features between the images, synergistic regularizers can convey information across channels in a way that each image $\boldx_k$ leverages the entire raw data $\boldy_1,\dots,\boldy_K$, thus improving the \ac{SNR}. However, enforcing structural similarities may not be appropriate for modalities that do not have the same intrinsic resolutions, such as in \ac{PET}/\ac{CT} and \ac{PET}/\ac{MRI}.

\subsection{Learned Regularizers with Generative Models}\label{sec:learned_reg}

\Ac{ML} and \ac{DL} techniques have been used in inverse problem-solving and image reconstruction \cite{arridge2019solving}. These approaches have changed the paradigm of image reconstruction in the sense that they are trained to deliver the reconstructed images. For example, unrolling methods extend conventional iterative algorithms into a deep architecture for end-to-end reconstructions \cite{monga2021algorithm}, while other techniques directly map the raw data into the image space \cite{Kandarpa_Bousse_Benoit_Visvikis_2021,Ma2022_direct, Cao_Xu_2022}. Another category of technique aims at training a penalty $R_{\boldtheta}\colon  (\R^m)^K \to \R$ with respect to some parameter $\boldtheta$ such that it promotes plausible multichannel image $\{\boldx_k\}$, that is to say, images that are plausible not only individually but also together. 

\subsubsection{Proposed Regularizer}

This section focuses on generative model-based regularizers, i.e., based on a multichannel image patch generator $\boldG_{\boldtheta}^{\mathrm{mult}}$ with trainable parameter $\boldtheta$, of the form
\begin{equation}\label{eq:genmodel}
	\boldG_{\boldtheta}^{\mathrm{mult}} \triangleq \left\{ \boldG_{\boldtheta}^1 , \dots , \boldG_{\boldtheta}^K \right\} \colon \calZ \to \calU^K 
\end{equation}
where for each $k=1,\dots,K$, $\boldG_{\boldtheta}^k \colon \calZ \to \calU$ is a generative model that maps a latent variable $\boldz$ in the latent space $\calZ\triangleq \R^s$ to a \ac{dD} image, $d<m$, corresponding to a patch (a portion of an image) in channel $k$, and $\calU = \R^d$ is the patch space. The $s$-dimensional latent variable $\boldz$, $s<d$, encodes the information of the image patch. Note that $\boldG^{\mathrm{mult}}_{\boldtheta}$ takes a single $\boldz$ as input such that the $K$ images correspond to the same $\boldz$.

We used a patch decomposition with overlaps to reduce training complexity and minimize hallucinations. Hallucinations arise when the generative model produces outputs that are not constrained by the range-space of the training data, leading to artifacts or erroneous features in the generated images. By using overlapping patches, the effect of local hallucinations in a patch are averaged out by neighboring patches.

In the following,  $\boldP_p \colon \calX \to \calU  $ is the $p$th patch extractor, $p=1,\dots,P$, such that for each channel $k$, $\boldu_k = \boldP_p \boldx_k \in \calU $ is a ``portion'' of $\boldx_k$. The patches cover the entire image, with possible overlaps.  The corresponding synergistic regularizer  $R_{\boldtheta}$ is then defined as
\begin{equation}\label{eq:regulariser}
	\begin{split}
		& R_{\boldtheta}( \{\boldx_k\} ) \triangleq	\\
		& \min_{\{\boldz_p\} \in \calZ^P}   \, \sum_{p=1}^P \sum_{k=1}^K  \frac{\eta_k}{2} \left\|  \boldG^k_{\boldtheta}(\boldz_p) - \boldP_p\boldx_k\right\|^2  + \alpha H(\boldz_p)
	\end{split}
\end{equation}
where $H$ is a regularizer on the latent variable with weight $\alpha>0$ and the $\eta_k$ are the same as in \eqref{eq:PMLsyn}. The regularizer  $R_{\boldtheta}$ is minimized when for all patch $p$, each $\boldP_p\boldx_k$, $k=1,\dots,L$, is approximately generated from a same $\boldz_p$ that is `regularized' in the sense of $H$. Solving the \ac{PML} problem \eqref{eq:PMLsyn} with $R_{\mathrm{syn}}= R_{\boldtheta}$ requires to alternate minimization in $\{\boldz_p\}$ and $\{\boldx_k\}$. The penalty $R_{\boldtheta}$ allows some flexibility for the solution in the sense that the reconstructed image patches (i.e., obtained by solving \eqref{eq:PMLsyn}) are not necessarily in the range of $\boldG^{\mathrm{mult}}_{\boldtheta}$. Another approach discussed in \citeauthor{duff2021regularising}~\cite{duff2021regularising} consists in imposing the images to be in the range of $\boldG^{\mathrm{mult}}_{\boldtheta}$. However, minimizing this regularizer is unpractical when using overlapping patches.

\subsubsection{Proposed Generative Model}

In this work we propose to utilize a multichannel image generator $\boldG_{\boldtheta} \colon \calZ \to \calU^K$ trained as a \ac{MVAE} with a multibranch architecture inspired from \citeauthor{duff2021regularising}~\cite{duff2021regularising} and \citeauthor{wang2024uconnect}~\cite{wang2024uconnect}. While standard \acp{VAE} are trained with a single encoder and a single generator (or decoder), our \ac{MVAE} uses a multibranch encoder mapping the $K$ channels to a single latent variable $\boldz$ which is then mapped to $K$ images with $\boldG_{\boldtheta} = \left\{ \boldG_{\boldtheta}^k, \, k=1,\dots,K \right\}$ with a multibranch decoder. Details on the training are given in Appendix~\ref{sec:appA}. In contrast with multichannel single-branch models, multibranch generative models introduce parallel pathways, each specializing in generating specific components of the data. By implementing this segregation, the original attributes unique to each modality can be better preserved in both the encoding and decoding components with minimal cross-talk, with interactions only taking place within the latent space.

\begin{figure*}
	\centering
	\includegraphics[width=.85\textwidth]{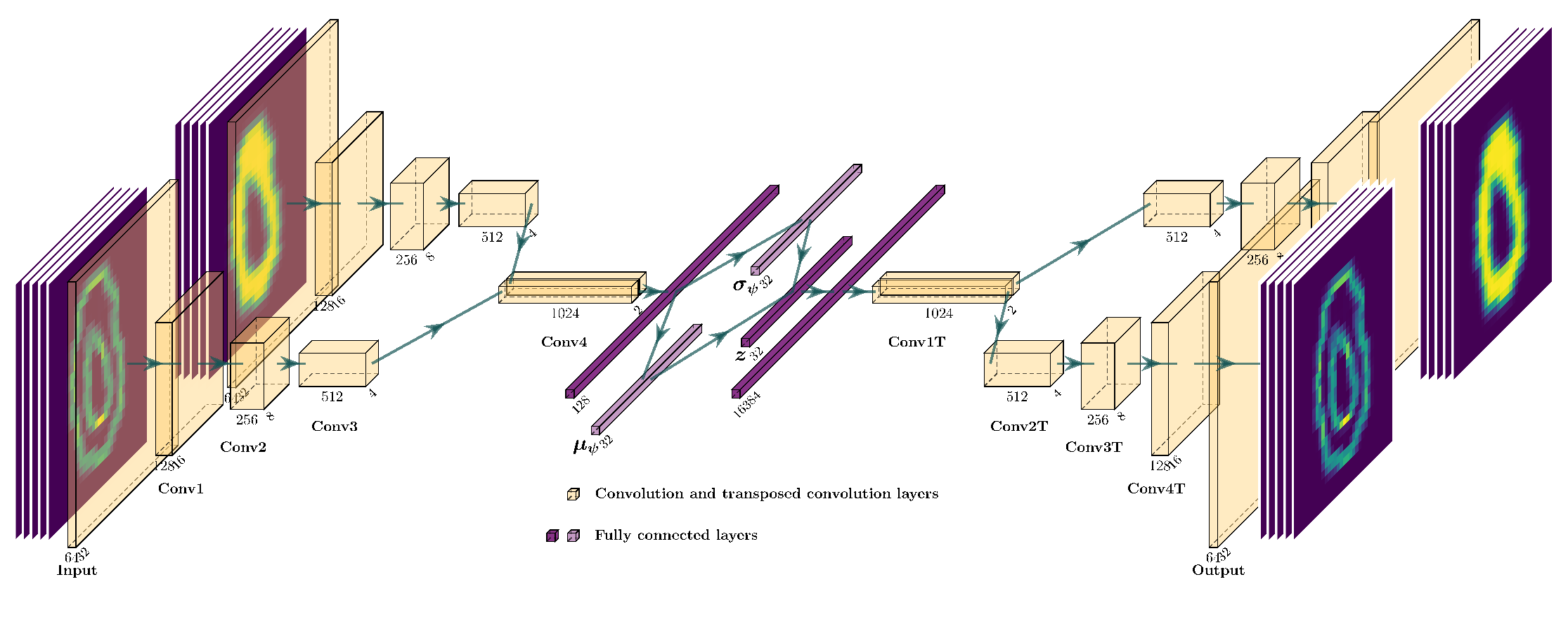}
	\caption{Architecture of our proposed  \ac{MVAE}. }
	\label{fig:mvae-archi}
\end{figure*}

The \ac{MVAE} architecture for $K=2$, 32$\times$32 patches, and $s=32$  is represented in Figure~\ref{fig:mvae-archi}. In this work, we employed a four-layer \ac{CNN} network for the encoder and decoder branches with \ac{ReLU} \cite{agarap2018deep} activation. All encoder branches have identical architecture and the same principle is also applied to all decoder branches. The \ac{VAE} network was trained on bimodal ($K=2$) image pairs with little to no processing on the original image. The model is designed to be trained on matching images, such that it can learn to convey information from one image to another. In the \ac{PET}/\ac{CT} case, the image pairs need to be co-registered as patient motion and respiratory motion may cause misalignment between the \ac{PET} and the \ac{CT} which may affect the training.

\subsection{Reconstruction Algorithm}\label{sec:algo}

Solving \eqref{eq:PMLsyn} using the synergistic regularizer $R_{\mathrm{syn}} = R_{\boldtheta}$ defined in \eqref{eq:regulariser} is achieved by alternating minimization in $\{\boldz_p\}$ and $\{\boldx_k\}$. Given a current estimate $\boldx_k^{(q)}$ at iteration $q$, the new estimate at iteration $q+1$ is given by
\begin{align}
	\boldz_p^{(q)} = {} & \argmin_{\boldz \in \calZ}  \, \sum_{k=1}^K  \frac{\eta_k}{2} \left\|  \boldG^k_{\boldtheta}(\boldz) - \boldP_p\boldx_k^{(q)} \right\|_2^2  + \alpha H(\boldz)  \nonumber \\
	&   \quad \forall p=1,\dots,P \label{eq:z-update} \\
	\boldx_k^{(q+1)} = {} & \argmin_{\boldx\in \calX} \,  L_k(\boldy_k , \boldybar_k(\boldx))   \nonumber \\
	&  + \frac{\beta}{2} \sum_{p=1}^P\left\| \boldG^k_{\boldtheta}\left(\boldz_p^{(q)}\right) - \boldP_p\boldx \right\|^2_2 \quad \forall k=1,\dots,K \label{eq:x-update} 
\end{align}
Both sub-minimizations can be achieved with iterative algorithms initialized from the previous estimates  $\boldz_p^{(q)}$ and $\boldx_k^{(q)}$. The $\boldx$-update \eqref{eq:x-update} depends on the loss $L_k$ and the forward model $\boldybar_k$\footnote{The minimization w.r.t. $\boldx_k$ \eqref{eq:x-update} does not depend on $\eta_k$ as each loss $L_k$ is multiplied by $\eta_k$ in \eqref{eq:PMLsyn}}. For \ac{PET}/\ac{CT} reconstruction, we used   a modified \ac{MLEM} algorithm \cite{depierro1995modified} for the \ac{PET} update (10 sub-iterations) while used using a \ac{SPS} algorithm \cite{elbakri2002statistical} for the \ac{CT} update (10 sub-iterations). Note that it is possible to use different $\beta$-values for each $k$ in \eqref{eq:x-update} to adjust the strength of $R_{\boldtheta}$ for each channel. Finally, we implemented  the $\boldz$-update step \eqref{eq:z-update}  with a \ac{LBFGS} algorithm \cite{zhu1997algorithm} (50 sub-iterations). A total of 20 outer-iterations was used.

%% file: content/results.tex
\section{Results}\label{sec:results}

In this section, we show some results of our methodology for $K=2$. Section~\ref{sec:mnist} presents an evaluation based on the \ac{MNIST} image dataset \cite{deng2012mnist} (resized to 32$\times$32), while Section~\ref{eq:recon} presents results  for \ac{PET}/\ac{CT} joint reconstruction from synthetic projection data generated from real patient images. The two parameters $\eta_1$ and $\eta_2$ in the optimization problem \eqref{eq:PMLsyn}   and the regularizer \eqref{eq:regulariser} are rewritten as  
\begin{equation}
	\eta_1 = \eta \quad \text{and} \quad \eta_2 = 1-\eta, \quad \eta\in[0,1] \, .
\end{equation}
For simplicity, and to solely focus on the interchannel dependencies, we used $H\equiv 0$, i.e., no regularization on $\boldz$. Our generative model-based regularizer simplifies to 
\begin{dmath}\label{eq:regulariser2mod}
	R_{\boldtheta}(\boldx_1,\boldx_2) = \min_{\{\boldz_p\} \in \calZ^P}   \, \Bigg( \sum_{p=1}^P  \frac{\eta}{2} \left\|  \boldG^1_{\boldtheta}(\boldz_p) - \boldP_p\boldx_1\right\|^2  + \frac{1-\eta}{2} \left\|  \boldG^2_{\boldtheta}(\boldz_p) - \boldP_p\boldx_1\right\|^2 \Bigg)  \, .
\end{dmath}	
While we used a range of $\eta$-values for \ac{MNIST}, we used $\eta=1/2$ for \ac{PET}/\ac{CT}. 

The quality of the reconstructed images was assessed using the \ac{PSNR} and \ac{SSIM} with respect to the \ac{GT} images $(\boldx_1^\star,\boldx_2^\star)$. We used the functions \verb|peak_signal_noise_ratio}| and \verb|structural_similarity| from the Python package \texttt{skimage.metrics} to compute the \ac{PSNR} and \ac{SSIM}. The training of the architectures was implemented on a single Nvidia RTX A6000 GPU.

\input{./content/mnist}
\input{./content/petct}

%% file: content/mnist.tex
\subsection{MNIST Data}\label{sec:mnist}

\subsubsection{Data and Training}

\begin{figure}
	\centering
	
	\subfloat[$\boldx_1$  images \label{subfig:x1_mnist}]{
		\begin{tikzpicture}
			\begin{scope}[spy using outlines={rectangle,magnification=3,size=12mm,connect spies}]
				\node {
        \includegraphics[width=0.45\linewidth]{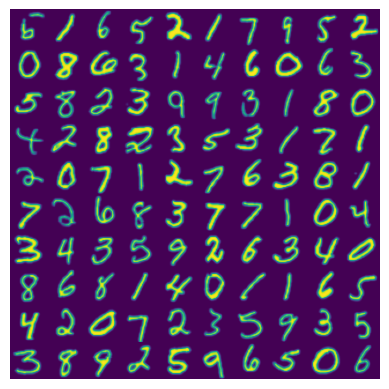}
				};
				\spy [red] on (-0.16,1.35) in node [left,red] at (1.85,0.7);
				\spy [green] on (-0.16,-0.95) in node [left,green] at (1.85,-0.7);
			\end{scope}
		\end{tikzpicture}
	}
	\subfloat[$\boldx_2$  images\label{subfig:x2_mnist}]{
		\begin{tikzpicture}
			\begin{scope}[spy using outlines={rectangle,red,magnification=3,size=12mm,connect spies}]
				\node {
    \includegraphics[width=0.45\linewidth]{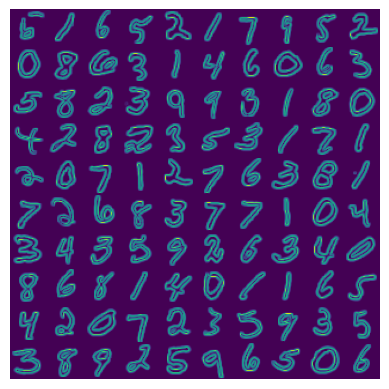}
				};
				\spy [red] on (-0.16,1.35) in node [left,red] at (-0.65,0.7);
				\spy [green] on (-0.16,-0.95) in node [left,green] at (-0.65,-0.7);
			\end{scope}
		\end{tikzpicture}
	}

	\caption{Example of $\boldx_1,\boldx_2$ image pairs derived from the \ac{MNIST} dataset used to train the \ac{MVAE} model. }\label{fig:mniest}
\end{figure}

The database consists of a collection of 70,000 32$\times$32 image pairs representing digits from 0 to 9 with various shapes (see Figure~\ref{subfig:x1_mnist}). These images play the role of the first channel, i.e., $\boldx_1$. The second channel images $\boldx_2$ are derived from $\boldx_1$ using a Roberts edge detection filter from \verb"scikit-image" \cite{van2014scikit} followed by a Gaussian filter. Figure~\ref{fig:mniest} shows an example of training image pairs. We used $P=1$, $d=m=32^2$, and $\boldP_1 = \mathrm{id}_{\calX}$ (identity operator on $\calX$).

The \ac{MVAE} models were trained in an unsupervised manner using 60,000 image pairs for training and 10,000 image pairs for testing. All models were trained using the Adam optimizer with a learning rate of $10^{-4}$. The batch size was chosen experimentally to balance between memory and time constraints and we used batches of 10,240 for 10,000 epochs.

\subsubsection{Results}\label{sec:results_mnist}

\paragraph{Image Generation}

\begin{figure}
	\centering
	\subfloat[$\boldG^1_{\boldtheta}(\boldz)$ \label{subfig:vae_1}]{
		\begin{tikzpicture}
			\begin{scope}[spy using outlines={rectangle,magnification=3,size=12mm,connect spies}]
				\node {
					\includegraphics[width=0.45\linewidth]{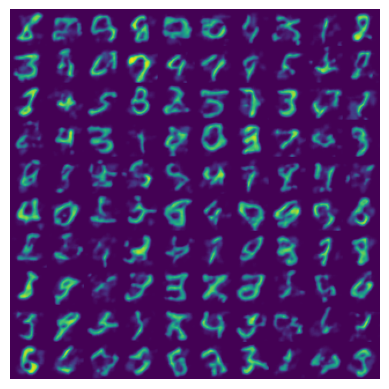}
				};
				\spy [red] on (-0.15,1.35) in node [left,red] at (1.85,0.7);
				\spy [green] on (-0.15,-0.95) in node [left,green] at (1.85,-0.7);
			\end{scope}
		\end{tikzpicture}
	}
	\subfloat[$\boldG^1_{\boldtheta}(\boldz)$ \label{subfig:vae_2}]{
		\begin{tikzpicture}
			\begin{scope}[spy using outlines={rectangle,magnification=3,size=12mm,connect spies}]
				\node {
					\includegraphics[width=0.45\linewidth]{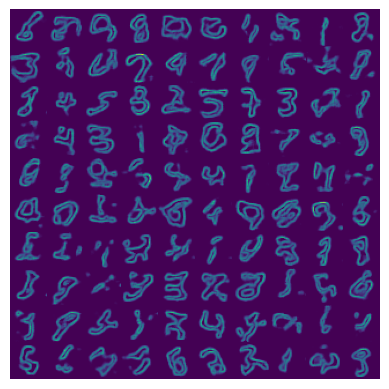}
				};
				\spy [red] on (-0.15,1.35) in node [left,red] at (-0.6,0.7);
				\spy [green] on (-0.15,-0.95) in node [left,green] at (-0.6,-0.7);
				
			\end{scope}
		\end{tikzpicture}
	}
	
	\caption{\Ac{MVAE}-generated image pairs $(\boldG^1_{\boldtheta}(\boldz),\boldG^2_{\boldtheta}(\boldz))$,  using the \ac{MNIST}-trained models, with random $\boldz\in \calZ$. The sub-images on \subref{subfig:vae_1} and \subref{subfig:vae_2} at same position were generated from the same $\boldz$.}\label{fig:rdm_z_vae}

\end{figure}

Figures~\ref{fig:rdm_z_vae}  shows generated images using  the \ac{MVAE} model using a random $\boldz$ generated by uniformly sampling each of its coordinates on $[-3,3]$. The images are distorted digits, similar to the training dataset. We observe that images generated from the same $\boldz$ correspond to each other, which suggests that both generators were able to learn from the pairs as opposed to each image individually.

\paragraph{Image Prediction: $\boldx_1$ to $\boldx_2$ and $\boldx_2$ to $\boldx_1$}

We define the ``model-fitting'' function $f_\eta$, which evaluates the goodness of the fit between a pair of target images  $(\boldx^\star_1,\boldx^\star_2)$ and the generated pair from a trained two-channel model $(\boldG^1_{\boldtheta}(\boldz),\boldG^2_{\boldtheta}(\boldz))$ as
\begin{dmath}
	f_\eta(\boldz,\boldx_1^\star,\boldx_2^\star)  \triangleq  \eta  \left\|  \boldG^1_{\boldtheta}(\boldz) - \boldx^\star_1\right\|_2^2  + (1-\eta) \left\|  \boldG^2_{\boldtheta}(\boldz) - \boldx^\star_2\right\|_2^2	
\end{dmath}	
The optimal latent variable, denoted $\check{\boldz}_\eta$, is defined as
\begin{equation}\label{eq:z-fitting}
	\check{\boldz}_\eta  \triangleq \argmin_{\boldz \in \calZ} f_\eta(\boldz,\boldx_1,\boldx_2) \, ,
\end{equation}	
where we dropped the $(\boldx_1,\boldx_2)$-dependency on the left-hand side to lighten the notation. Finally, the ``predicted images'' are given by 
\begin{equation}\label{eq:gen_images}
	\boldxcheck_k^\eta = \boldG^k_{\boldtheta}( \check{\boldz}_\eta  ) \, ,  \quad  k=1,2  \,.
\end{equation}
Thus, $( \boldxcheck_1^\eta, \boldxcheck_2^\eta )$ represents the ``best copy'' of $(\boldx^\star_1,\boldx^\star_2)$ where the weight $\eta$  dictates which target image the generative model should prioritize. In particular for $\eta=0$  the model-fitting process \eqref{eq:z-fitting} is oblivious to $\boldx_1$ so that $\boldxhat_1^0$ is a ``prediction'' of $\boldx_1^\star$ from $\boldx_2^\star$ using the model (and conversely with $\eta=1$).

\begin{figure}
	\centering
	\subfloat[Target  $\boldx^\star_1$\label{subfig:target_vae_1}]{
		\includegraphics[width=0.22\linewidth]{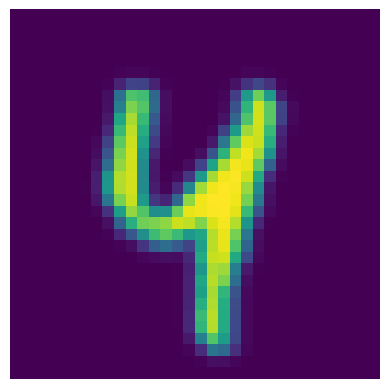}
	}
	\subfloat[$\boldxcheck^{0}_1$]{
		\includegraphics[width=0.22\linewidth]{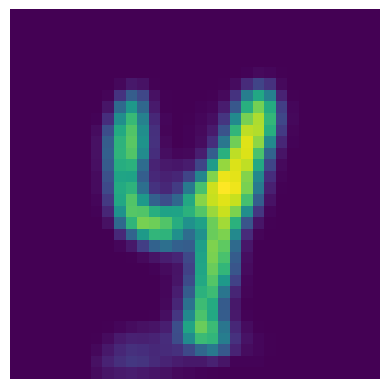}
	}
	\subfloat[$\boldxcheck^{0.5}_1$]{
		\includegraphics[width=0.22\linewidth]{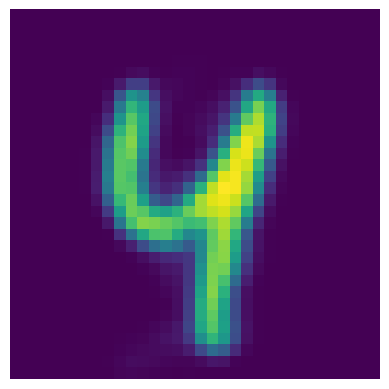}
	}
	\subfloat[$\boldxcheck^{1}_1$]{
		\includegraphics[width=0.22\linewidth]{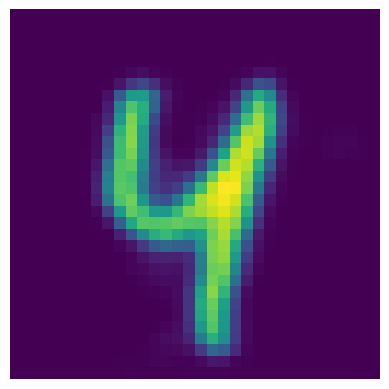}
	}
	
	
	\subfloat[Target  $\boldx_2^\star$\label{subfig:target_vae_2}]{
		\includegraphics[width=0.22\linewidth]{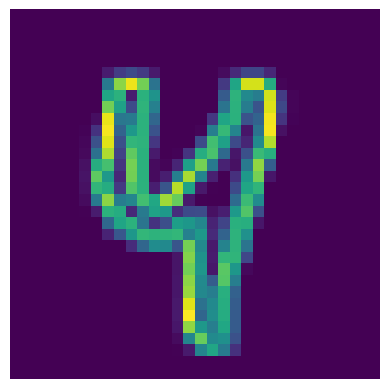}
	}
	\subfloat[$\boldxcheck^{0}_2$]{
		\includegraphics[width=0.22\linewidth]{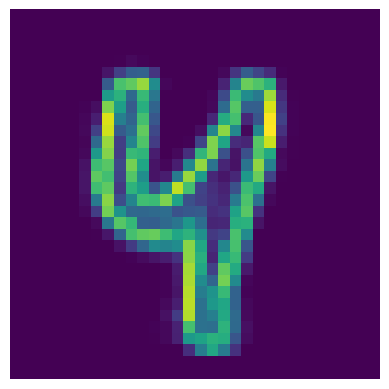}
	}
	\subfloat[$\boldxcheck^{0.5}_2$]{
		\includegraphics[width=0.22\linewidth]{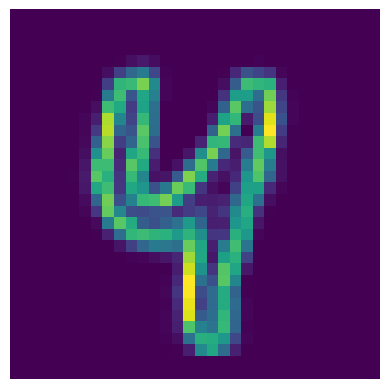}
	}
	\subfloat[$\boldxcheck^{1}_2$]{
		\includegraphics[width=0.22\linewidth]{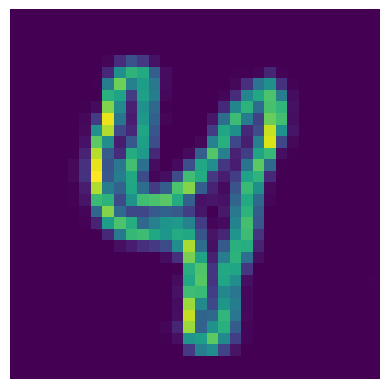}
	}

	\caption{
		\Ac{MVAE}-generated images $(\boldxcheck_1^\eta,\boldxcheck_2^\eta)$ obtained by model-fitting to a target \ac{MNIST} digit pair $\boldx_1^\star,\boldx_2^\star$ for different values of the parameter $\eta\in[0,1]$ which balances the contribution of each image to the fitting. When $\eta=0$ (resp. $\eta=1$), the model-fitting is oblivious of $\boldx_1^\star$ (resp. $\boldx_2^\star$) and focuses $\boldx_2^\star$ (resp. $\boldx_1^\star$), such that $\boldxcheck^{0}_1$ (resp. $\boldxcheck^{1}_2$) is a prediction of $\boldx_1^\star$ (resp. $\boldx_2^\star$) from $\boldx_2^\star$ (resp. $\boldx_1^\star$). When $\eta=0.5$, the model uses $\boldx_1^\star$ and $\boldx_2^\star$ equally.
	}\label{fig:prediction_vae}
\end{figure}

Figure~\ref{fig:prediction_vae} shows \ac{MVAE}-generated images $(\boldxcheck_1^\eta,\boldxcheck_2^\eta)$ obtained using model-fitting to a target \ac{MNIST} digit pair $(\boldx_1^\star,\boldx_2^\star)$ (from the testing dataset) for different values of $\eta$.   When $\eta=0.5$, both generated images $(\boldxcheck_1^{0.5},\boldxcheck_2^{0.5})$ correspond to the targets with almost no visible mismatch. When $\eta=1$, $\boldxcheck_1^{1}$ is  similar to $\boldx_1^\star$  as expected, while $\boldxcheck_2^1$ is somehow similar to $\boldx_2^\star$ (with some distortions), which shows that the model managed to predict an image fairly similar to $\boldx_2^\star$ using $\boldx_1^\star$ only. The converse result is observed with $\eta=0$. 

\paragraph{Image Denoising}\label{sec:denoising}

In this section, we focus on denoising two noisy images $\boldx_1^{\mathrm{n}}$ and $\boldx_2^{\mathrm{n}}$,
\begin{equation}\label{eq:noise_mnist}
	\boldx^{\mathrm{n}}_k = \boldx^\star_k  + \boldeps_k \, , \quad k=1,2 
\end{equation}	
where $(\boldx_1^\star,\boldx_1^\star)$ is a \ac{GT} image pair (from the testing dataset) and $\epsilon_k\sim\calN(\boldzero_m, \sigma_k^2 \mathrm{id}_{\calX})$, using a penalized least squares approach, i.e.,  solving \eqref{eq:PMLsyn}   with $\boldy_k =\boldx^{\mathrm{n}}_k$, $\boldybar_k = \mathrm{id}_{\calX}$, $L_k(\boldy_k,\boldybar_k) = \frac{1}{2}\|\boldy_k - \boldybar_k\|_2^2$ and using the trained regularizer $R_{\mathrm{syn}} = R_{\boldtheta}$ defined in \eqref{eq:regulariser2mod}. As $P=1$, the image update \eqref{eq:x-update} simplifies to $\boldx_k^{(q+1)} = (1+\beta)^{-1}\left(\boldx^{\mathrm{n}}_k + \beta \boldG^k_{\boldtheta}\left(\boldz^{(q+1)}\right)  \right)$, $k=1,2$ . We used $\beta=1$ and  $\sigma_2>\sigma_1$ in order to assess if the second image benefits from the first. 

The denoised images are denoted $\boldxhat_1^\eta$ and $\boldxhat_2^\eta$ their corresponding latent encoding is denoted $\boldzhat^\eta$. Figure~\ref{fig:vae_mnist_denoise_images_eta}  shows the input noisy images $\boldx^{\mathrm{n}}_k$ and the denoised images $\boldxhat_k^\eta$. For $\eta=0$, the model ignores the first channel and focuses on the second channel only, which is the noisiest. Therefore the model fails to generate the second image, which results in a poor prediction for the first image. The quality of $\boldxhat_1^\eta$ seems to increase as $\eta$ approaches 0.9 (with a slight decrease at $\eta=1$). In contrast, the quality of $\boldxhat_2^\eta$ seems to increase between $\eta=0$ and $0.5$ then slowly decrease from $0.5$ and $1$. These observations are confirmed  by the \ac{PSNR}-$\eta$  curves (Figure~\ref{fig:vae_PSNRvsEta}). This experiment suggests that the noisier channel benefits from the less noisy one (best results obtained with $\eta=0.5$). Conversely, the less noisy channel does not benefit much from the noisier one.

\newlength{\tempdimd}
\setlength{\tempdimd}{0.17\linewidth}	
\setlength{\dimvert}{\tempdimd}	
\begin{figure*}	
	\begin{center}
		\begin{tikzpicture}
			\node(table) at (0,0) [] { 
				\begin{tabular}{p{0.01\tempdimd}p{\tempdimd}p{0.8\tempdimd}p{0.8\tempdimd}p{0.8\tempdimd}p{0.8\tempdimd}p{0.8\tempdimd}p{0.8\tempdimd}p{0.8\tempdimd}p{0.8\tempdimd}p{0.8\tempdimd}p{0.01\tempdimd}}
					\rowname{$\boldx_1^{\rmn}$}&
					\includegraphics[width=\tempdimd]{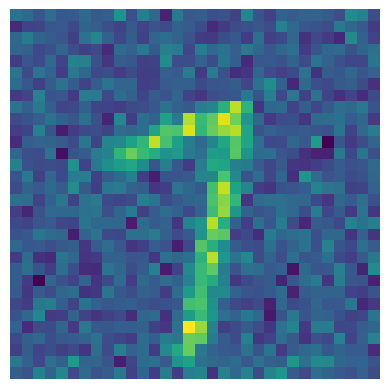}&
					\includegraphics[width=\tempdimd]{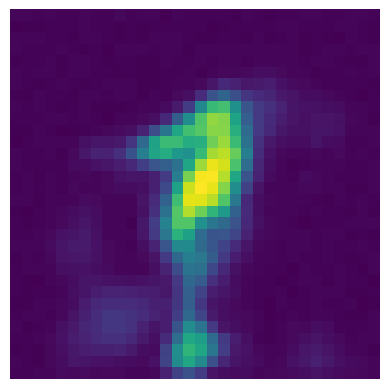} & 
					\includegraphics[width=\tempdimd]{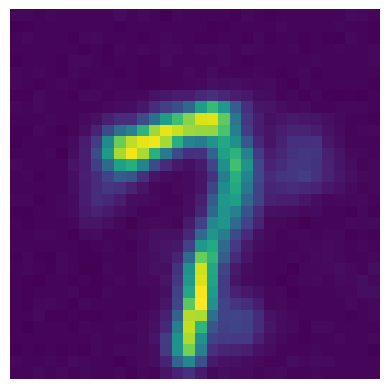}&
					\includegraphics[width=\tempdimd]{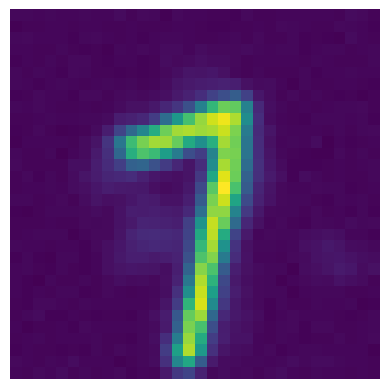}&
					\includegraphics[width=\tempdimd]{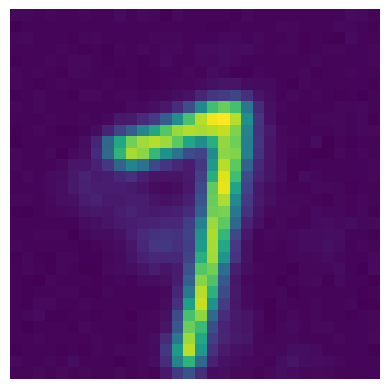}&
					\includegraphics[width=\tempdimd]{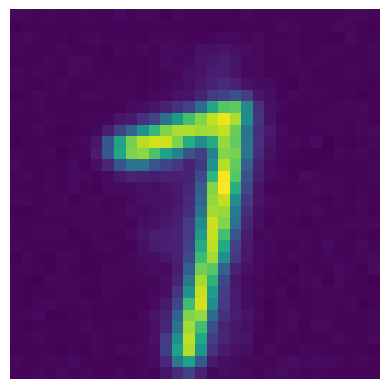}&
					\includegraphics[width=\tempdimd]{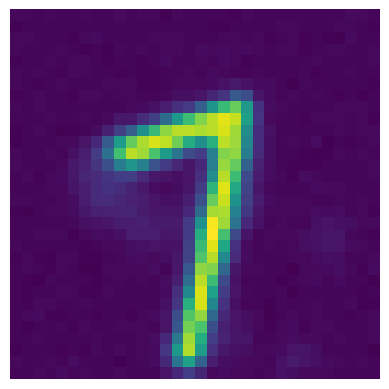}&
					\includegraphics[width=\tempdimd]{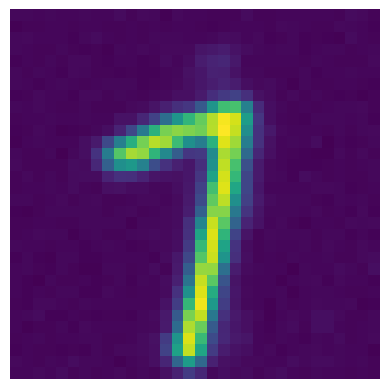}&
					\includegraphics[width=\tempdimd]{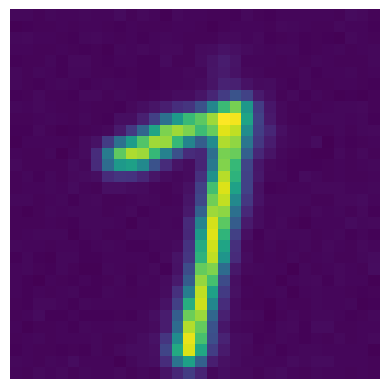}&
					\includegraphics[width=\tempdimd]{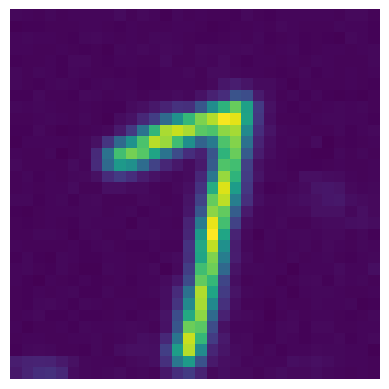}& \rownametwo{$\boldxcheck_1^\eta$} \\
					
					\rowname{$\boldx_2^{\rmn}$}&
					\includegraphics[width=\tempdimd]{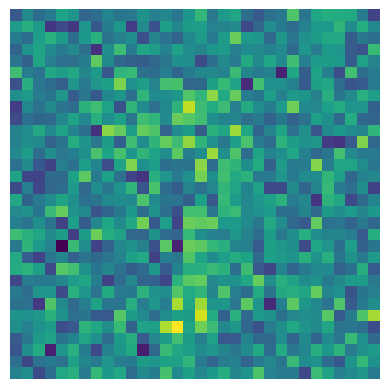}&
					\includegraphics[width=\tempdimd]{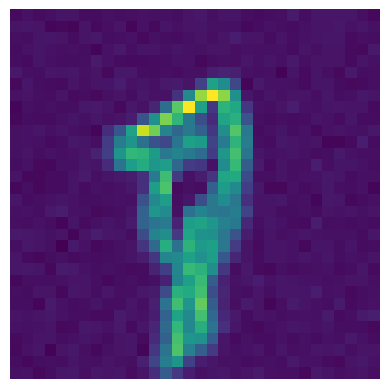} & 
					\includegraphics[width=\tempdimd]{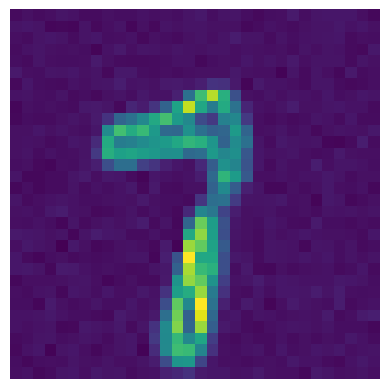}&
					\includegraphics[width=\tempdimd]{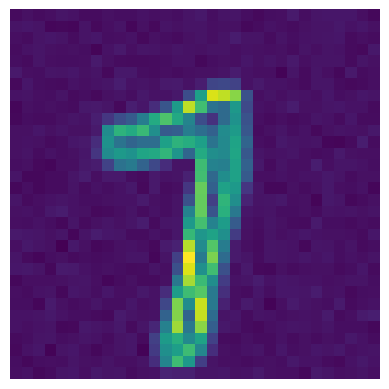}&
					\includegraphics[width=\tempdimd]{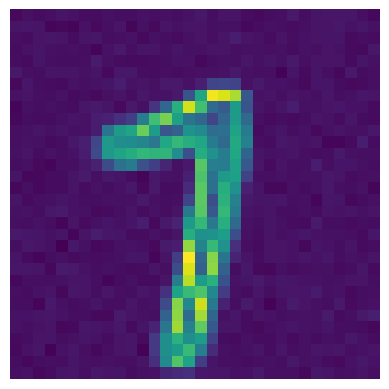}&
					\includegraphics[width=\tempdimd]{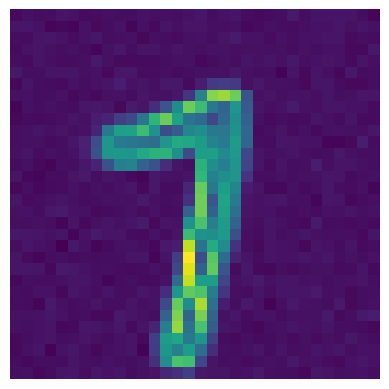}&
					\includegraphics[width=\tempdimd]{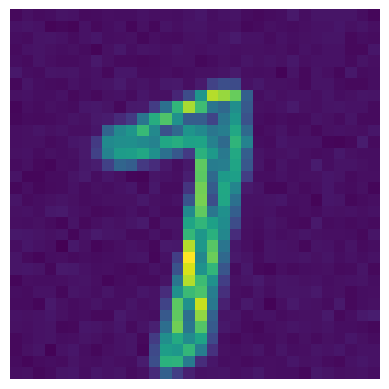}&
					\includegraphics[width=\tempdimd]{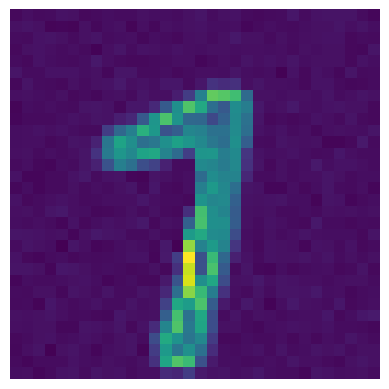}&
					\includegraphics[width=\tempdimd]{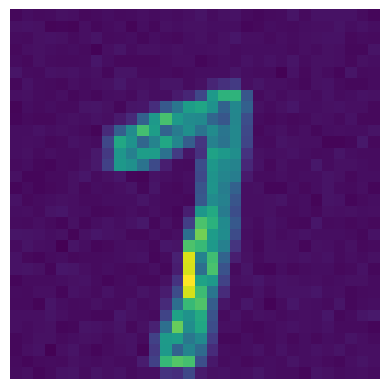}&
					\includegraphics[width=\tempdimd]{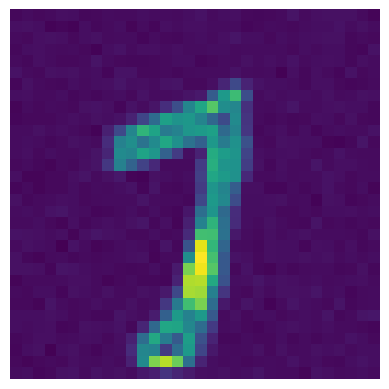}& 
					\rownametwo{$\boldxcheck_2^\eta$}\\
				\end{tabular}
			} ;
			\node(a) at (-3.8\tempdimd,1.2\tempdimd) [] {} ;
			\node(b) at (5.5\tempdimd,1.2\tempdimd) [] {} ;
			\draw[-Stealth,line width = 0.18em,red!70!black]   (a)--(b) node[pos=0.95,above] () { $\eta=1$  } ;
			\path[red!70!black]   (b)--(a) node[pos=0.95,above] () { $\eta=0$  } ;
			\node(c) at (-3.8\tempdimd,-1.11\tempdimd) [] {} ;
			\node(d) at (5.5\tempdimd,-1.11\tempdimd) [] {} ;
			\draw[-Stealth,line width = 0.18em,green!70!black]   (d)--(c) node[pos=0.95,below] () { $\eta=0$  } ;
			\path[green!70!black]   (c)--(d) node[pos=0.95,below] () { $\eta=1$  } ;
			
			\node(label1) at (-4.85\tempdimd,1.6\tempdimd) [align = center] {noisy \\ input} ;
			\node(label2) at (0.65\tempdimd,1.6\tempdimd) [align = center] {denoised images} ;
			
		\end{tikzpicture}
	\end{center}	
	\caption{Noisy images $(\boldx_1^{\mathrm{n}},\boldx_2^{\mathrm{n}})$  and
		\ac{MVAE}-denoised images $(\boldxcheck_1^\eta,\boldxcheck_2^\eta)$ for different values of $\eta$ ranging from $0.01$ to $0.99$ $\beta=1$.}\label{fig:vae_mnist_denoise_images_eta}
\end{figure*}

\begin{figure}
	\centering
	\begin{tikzpicture}[scale=0.55] 
		
		\begin{axis}[
			mark options={mark size = 3pt},
			xlabel={$\eta$},
			ylabel={PSNR},
			xmin = 0,
			xmax = 1,
			legend cell align=left,
			legend style={at={(0.92,0.83)},anchor=north},
			x post scale=2,
			]
			
			\addplot[color=red, style={thick}, mark=square*] table[x=eta, y=psnr_recon, col sep=comma] {./images/vae/mnist_denoise/s42/pet_beta1.0_sigma0.15.txt};
			\addlegendentry{$\boldxhat_1$}
			
			\addplot[color=blue, style={thick}, mark=square*] table[x=eta, y=psnr_recon, col sep=comma] {./images/vae/mnist_denoise/s42/ct_beta1.0_sigma0.35.txt};
			\addlegendentry{$\boldxhat_2$}
			
		\end{axis}

	\end{tikzpicture}
	\caption{\Ac{PSNR} values of denoised \ac{MNIST} images using \ac{MVAE} for a range of $\eta$ values.}
	\label{fig:vae_PSNRvsEta}
\end{figure}

We performed a \ac{PCA} on the latent variables $\boldz$s obtained by encoding the training dataset image pairs $(\boldx_1,\boldx_2)$ with the trained encoder. A target latent variable $\boldz^\star$ was obtained by encoding the target image pair $(\boldx_1^\star,\boldx_2^\star)$ from \eqref{eq:noise_mnist}. Figure~\ref{fig:pca} shows a projection of the latent space onto the \ac{2D} subspace spanned by the two first principal components. The blue crosses represent the $\boldzhat^\eta$ latent variables corresponding to the denoised image pairs $(\boldxhat_1^\eta,\boldxhat_1^\eta)$ for different values of $\eta$. We observe that the distance to the target increases as $\eta$ moves away from $1/2$.

\begin{figure}
	\centering
	\begin{tikzpicture}[scale=0.55] 
		\scriptsize
		\begin{axis}[
			mark options={mark size = 3pt},
			xlabel={PC 1},
			ylabel={PC 2},
			legend cell align=left,
			legend style={at={(0.8,0.3)},anchor=north},
			x post scale=2,
			]
			
			\addplot[color=green, only marks,mark options={scale=0.2,line width=2.5},] 
			table[x=ax1, y=ax2, col sep=comma, 
			] {./images/pca_latent_vectors.txt};
			\addlegendentry{\normalsize {$\boldz$ (from training dataset)}}
			
			\addplot[color=blue, 
					 only marks, 
					 scatter src=explicit symbolic,
					 mark=x,
					 mark options={scale=1.3,line width=1},
					 nodes near coords,		
			]table[x=ax1, y=ax2, meta=label, col sep=comma,] {./images/mnist_latent.txt};
			\addlegendentry{{\normalsize  $\boldzhat^\eta$ (from denoised images)}}
			
			\addplot[color=red, 
					 mark=+,
					 mark options={scale=1.5,line width=1.5},
					 only marks] 
			table[x=ax1, y=ax2, col sep=comma] {./images/gt_encoding.txt};
			\addlegendentry{{\normalsize target $\boldz^\star$}}
			
		\end{axis}

	\end{tikzpicture}
	\caption{\Ac{2D} representation of the latent space by \ac{PCA}. The \ac{PCA} was performed on the $\boldz$ latent variables of the training dataset (green dots). The red cross corresponds to the target latent variable $\boldz^\star$ obtained by encoding the target pair $(\boldx_1^\star,\boldx_2^\star)$ in \eqref{eq:noise_mnist}. The blue crosses correspond to the latent variables $\boldzhat^\eta$ from the denoised image pair $(\boldx_1^\eta,\boldx_2^\eta)$ for $\eta\in \{0.1,0.5,0.75,1\}$, the label on each blue cross corresponding to the $\eta$-value.  }\label{fig:pca}

\end{figure}

In conclusion of this experiment, we observe our \ac{MVAE} regularizer manages denoise while conveying information  between the two channels. 

%% file: content/petct.tex
\subsection{Synergistic PET/CT Reconstruction}\label{eq:recon}

\subsubsection{Data and Training}

A collection of 328 abdomen \ac{PET}/\ac{CT} image volumes were acquired by Siemens Biograph mCT \ac{PET}/\ac{CT} scanner at \emph{Centre Hospitalier Universitaire Poitiers}, Poitiers, France. Each volume comprises a set of 512$\times$512 slices (0.97-mm pixel size), for a total  of 41,000 slices per modality. A $(\boldx_1,\boldx_2)$ pair  correspond to a \ac{PET}/\ac{CT} slice pair ($\boldx_1$ for \ac{PET}, $\boldx_2$ for \ac{CT}). A total of 318 pairs were used for training while 10 pairs were used for testing---testing images and training images came from different patients.

\begin{figure}
	\centering
	
	\subfloat[\ac{PET} training patches\label{subfig:gt_patches_pet64}]{
		\begin{tikzpicture}
			\begin{scope}[spy using outlines={rectangle,magnification=3,size=11.5mm,connect spies}]
				\node {
					\includegraphics[width=0.45\linewidth]{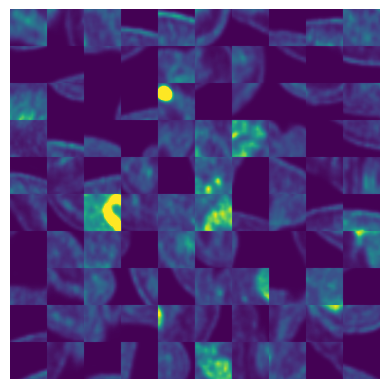}
				};		
				\spy [red] on (-1.33,-0.185) in node [left,red] at (1.85,1.2);
				\spy [green] on (0.2,-1.703) in node [left,green] at (1.85,-1.2);
			\end{scope}
		\end{tikzpicture}
	}%
	\subfloat[\ac{CT} training patches\label{subfig:gt_patchesct64}]{
		\begin{tikzpicture}
			\begin{scope}[spy using outlines={rectangle,magnification=3,size=11.5mm,connect spies}]
				\node {
					\includegraphics[width=0.45\linewidth]{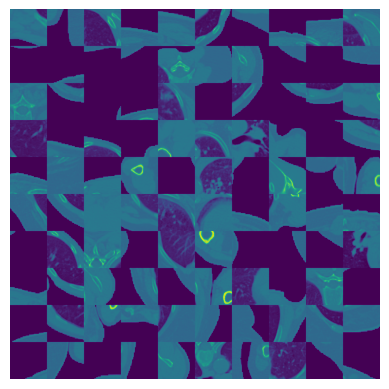}
				};
				\spy [red] on (-1.33,-0.185) in node [left,red] at (-0.69,1.2);
				\spy [green] on (0.2,-1.703) in node [left,green] at (-0.69,-1.2);
			\end{scope}
		\end{tikzpicture}
	}
	\caption{Examples for \ac{PET}/\ac{CT} patch pairs, i.e., $\boldu_1,\boldu_2$ in \eqref{eq:vae}, used to train the \ac{MVAE} model.}\label{fig:petct_gt_patches}
\end{figure}

$64\times64$ patches were randomly extracted from each image $2\cdot 10^5$ patch pairs $(\boldu_1,\boldu_2)$, see Figure~\ref{fig:petct_gt_patches} for an example of patch pairs) then were used to train the \ac{MVAE} model $\boldG_{\boldtheta}$ and to define $R_{\boldtheta}$ \eqref{eq:regulariser2mod}. The \ac{PET} and \ac{CT} were normalized before training, and the two normalizing constants were  incorporated in $R_{\boldtheta}$ \eqref{eq:regulariser2mod}. The Adam optimizer was used for training with a learning rate of 10$^{-\mathrm{4}}$ and batch size of 4,096 for 1,000 epochs. 

\begin{figure}
	\centering
	\subfloat[\ac{GT} \ac{PET} $\boldx_1^\star$\label{subfig:gt_pet64}]{
		\includegraphics[width=0.45\linewidth]{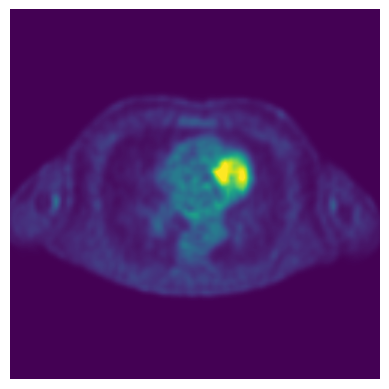}
	}%
	\subfloat[\ac{GT} \ac{CT} $\boldx_2^\star$\label{subfig:gt_ct64}]{
	\includegraphics[width=0.45\linewidth]{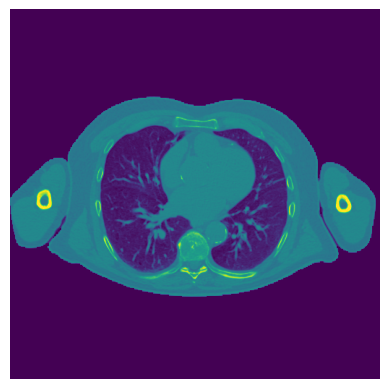}
}
	\caption{
		First of the 10 testing \ac{PET}/\ac{CT} \ac{GT} image pairs $(\boldx_1^\star,\boldx_2^\star)$ used to generate the scanner raw data following \eqref{eq:poisson}.
		}\label{fig:petct_gt}
\end{figure}

We used 10 \ac{PET}/\ac{CT} image pairs  $(\boldx_1^\star,\boldx_1^\star)$ from 10 patients  as  \ac{GT} images to test other \ac{MVAE} reconstruction. Figure~\ref{fig:petct_gt} shows the \ac{GT} images for Patient 1 which we used as main example. The \ac{PET} data $\boldy_1\in \R^{n_1}$ and \ac{CT} data $\boldy_2\in \R^{n_2}$ were generated following \eqref{eq:poisson} using $\boldybar = \boldybar_k$, $k=1,2$, where $\boldybar_1 \colon \R^m \to \R^{n_1}$ and $\boldybar_2 \colon \R^m \to \R^{n_2}$ are respectively the \ac{PET} and \ac{CT} forward models (cf.~\eqref{eq:pet_model} and \eqref{eq:ct_model}), and $n_1$ and $n_2$ are respectively the number of \ac{PET} \acp{LoR} and the number of \ac{CT} beams. The systems were implemented with ASTRA   \cite{vanAarle:16} with $n_1=120\times{}512$ for the \ac{PET}  (parallel geometry) and $n_2 = 120\times{} 750$ for the \ac{CT}  (standard fan beam geometry with 1.2-mm detector size, 600-mm origin-to-source and origin-to-detector distances).

The scanner data $\boldy_1$ and $\boldy_2$ were acquired  with 2 settings: (i) \ac{HL} with $\tau=700$ and $I=2,000$ , and (ii) \ac{LH} with $\tau=10$ and $I=1.4\cdot{}10^5$. 

The $64\times 64$ patch extractors $\boldP_p$ were defined with 75\% overlap along each axis. The attenuation correction factors were obtained from a scout reconstruction of the attenuation image $\boldx_2$ from $\boldy_2$ using an unregularized \ac{WLS} reconstruction, then converted into 511-keV images using the method proposed by \citeauthor{Oehmigen_Lindemann_Tellmann_Lanz_Quick_2020}~\cite{Oehmigen_Lindemann_Tellmann_Lanz_Quick_2020}. 

The parameter $\beta$  was finely tuned to optimize image quality in terms of \ac{PSNR} and \ac{SSIM}, and we used $\eta=1/2$. The \ac{MVAE}-reconstructed images are refered to \ac{VPET} and \ac{VCT}.

Additionally, we implemented two other reconstruction techniques for comparison:
\begin{itemize}
	\item The \ac{PLS} approach proposed by \citeauthor{ehrhardt2014joint}~\cite{ehrhardt2014joint} (the quadratic version) which consists of utilizing a synergistic penalty (i.e., $R_{\mathrm{syn}}$ in \eqref{eq:PMLsyn}) that promotes images with parallel gradient in order to enforce structural similarities between the images. We used a \ac{LBFGS} algorithm to perform the joint minimization in $\boldx_1$ and $\boldx_2$. The reconstructed images are referred to as \ac{PPET} and \ac{PCT}.
	\item \Ac{PETunet} and \ac{CTunet}, where both image-to-image U-Nets were trained to map  low-count \ac{CT} images to  the \ac{GT} images. The training of these methods is supervised and therefore they are expected to deliver better results.
\end{itemize}

Finally, we reconstructed the images using \ac{MLEM} for \ac{PET} and \ac{WLS} for \ac{CT} (using a \ac{SPS} algorithm), which corresponds to \ac{MVAE} with $\beta=0$. The reconstructed images are referred to as \ac{EMPET} and \ac{WCT}.

\subsubsection{Results}

\paragraph{Image Generation}

\begin{figure}
	\centering
	
	\subfloat[$\boldG^1_{\boldtheta}(\boldz)$ \label{subfig:vae_pet}]{
		\begin{tikzpicture}
			\begin{scope}[spy using outlines={rectangle,magnification=3,size=11.5mm,connect spies}]
				\node {
					\includegraphics[width=0.45\linewidth]{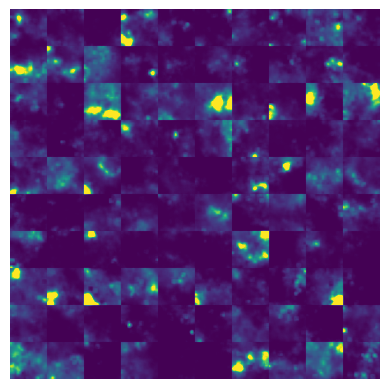}
				};
				\spy [red] on (0.95,0.195) in node [left,red] at (1.85,1.2);
				\spy [green] on (-1.33,0.195) in node [left,green] at (1.85,-1);
			\end{scope}
		\end{tikzpicture}
	}
	\subfloat[$\boldG^2_{\boldtheta}(\boldz)$ \label{subfig:vae_ct}]{
		\begin{tikzpicture}
			\begin{scope}[spy using outlines={rectangle,magnification=3,size=11.5mm,connect spies}]
				\node {
					\includegraphics[width=0.45\linewidth]{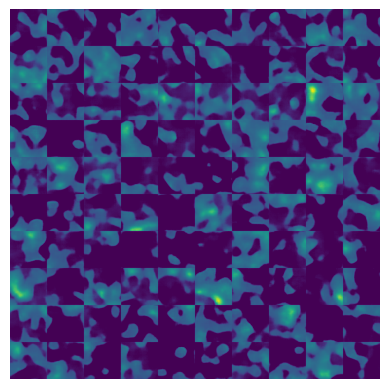}
				};
				\spy [red] on (0.95,0.195) in node [left,red] at (-0.7,1.2);
				\spy [green] on (-1.33,0.195) in node [left,green] at (-0.7,-1);
			\end{scope}
		\end{tikzpicture}
	}
	\caption{
		\Ac{MVAE}-generated image pairs $(\boldG^1_{\boldtheta}(\boldz),\boldG^2_{\boldtheta}(\boldz))$ with random $\boldz\in\calZ$,  using the \ac{PET}/\ac{CT}-trained models, with random $\boldz\in \calZ$. The sub-images on \subref{subfig:vae_pet} and \subref{subfig:vae_ct} at same position were generated from the same $\boldz$.
		}
	\label{fig:rdm_z_vae_petct}
\end{figure}

Figure~\ref{fig:rdm_z_vae_petct} shows random images of generated images patches using the trained \ac{MVAE} \ac{PET}/\ac{CT} model, using a random $\boldz$, in a similar fashion as for the \ac{MNIST}-trained model  (Section~\ref{sec:results_mnist}). The images generated from the \ac{PET} generator $\boldG^1_{\boldtheta}$ (Figure~\ref{subfig:vae_pet}) appear blurry while those  generated from the \ac{CT} generator $\boldG^2_{\boldtheta}$  (Figure~\ref{subfig:vae_ct})  are sharper.  Structural similarities can be observed (cf. the magnified areas in Figure~\ref{fig:rdm_z_vae_petct}), which suggests that information is shared between the two modalities. However, these similarities are less pronounced than  the  \ac{MNIST}-trained model.

\paragraph{Image Reconstruction: High-count \ac{PET}, Low-count \ac{CT}}

\begin{figure}
    \centering
	\subfloat[\acs{VPET}, $\eta=1/2$ \label{subfig:hl_pet_mvae}]{
		\begin{overpic}[width=0.45\linewidth]{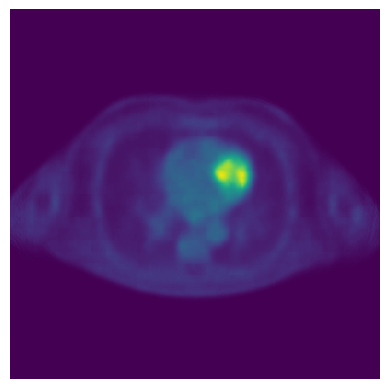}
			\put(1,4){
				\textcolor{red}{\small PSNR=\DTLfetch{table1}{patient_no}{372}{psnr_pet_vae}}
			}
			\put(1,90){
				\textcolor{red}{\small SSIM=\DTLfetch{table1}{patient_no}{372}{ssim_pet_vae}}
			}
		\end{overpic}	
		}%
	\subfloat[\acs{VCT}, $\eta=1/2$ \label{subfig:hl_ct_mvae}]{
		\begin{overpic}[width=0.45\linewidth]{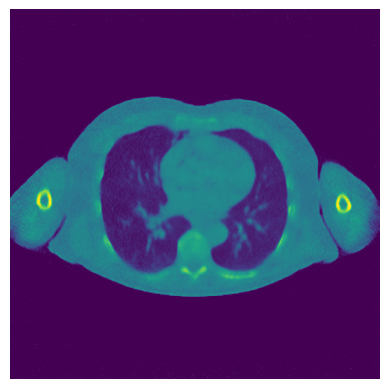}
			\put(1,4){
				\textcolor{red}{\small PSNR=\DTLfetch{table1}{patient_no}{372}{psnr_ct_vae}}
			}
			\put(1,90){
				\textcolor{red}{\small SSIM=\DTLfetch{table1}{patient_no}{372}{ssim_ct_vae}}
			}
		\end{overpic}
	}%
	
	\subfloat[\acs{PPET}\label{subfig:hl_pet_pls}]{
		\begin{overpic}[width=0.45\linewidth]{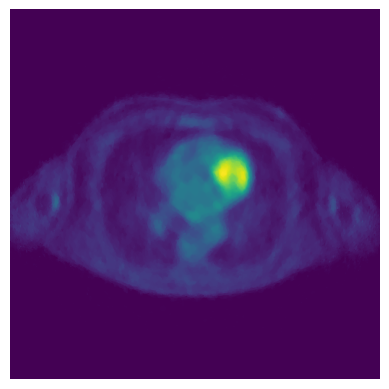}
			\put(1,4){
				\textcolor{red}{\small PSNR=\DTLfetch{table1}{patient_no}{372}{psnr_pet_mat}}
			}
			\put(1,90){
				\textcolor{red}{\small SSIM=\DTLfetch{table1}{patient_no}{372}{ssim_pet_mat}}
			}
		\end{overpic}
		}%
	\subfloat[\acs{PCT}\label{subfig:hl_ct_pls}]{
		\begin{overpic}[width=0.45\linewidth]{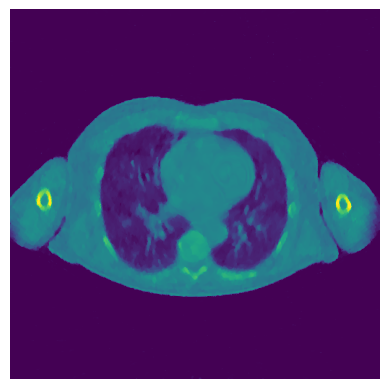}
			\put(1,4){
				\textcolor{red}{\small PSNR=\DTLfetch{table1}{patient_no}{372}{psnr_ct_mat}}
			}
			\put(1,90){
				\textcolor{red}{\small SSIM=\DTLfetch{table1}{patient_no}{372}{ssim_ct_mat}}
			}
		\end{overpic}
	}%

	\subfloat[\acs{PETunet} (individual recon.) \label{subfig:hl_dip_pet}]{
		\begin{overpic}[width=0.45\linewidth,height=0.45\linewidth]{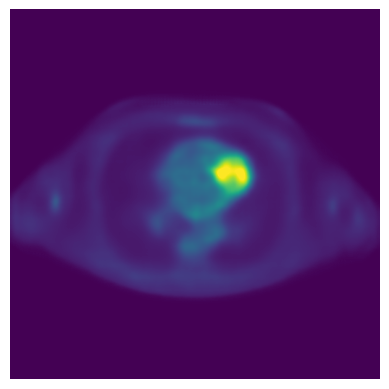}
			\put(1,4){
				\textcolor{red}{\small PSNR=\DTLfetch{table2}{patient_no}{372}{psnr_pet} }
			}
			\put(1,90){
				\textcolor{red}{\small SSIM=\DTLfetch{table2}{patient_no}{372}{ssim_pet} }
			}
		\end{overpic}
	}%
	\subfloat[\acs{CTunet} (individual recon.) \label{subfig:hl_dip_ct}]{
		\begin{overpic}[width=0.45\linewidth,height=0.45\linewidth]{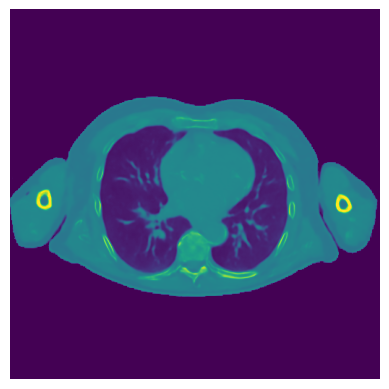}
			\put(1,4){
				\textcolor{red}{\small PSNR=\DTLfetch{table3}{patient_no}{372}{psnr_ct} }
			}
			\put(1,90){
				\textcolor{red}{\small SSIM=\DTLfetch{table3}{patient_no}{372}{ssim_ct} }
			}
		\end{overpic}
	}%

	\subfloat[\acs{EMPET} (individual recon.) \label{subfig:hl_mlem}]{
		\begin{overpic}[width=0.45\linewidth]{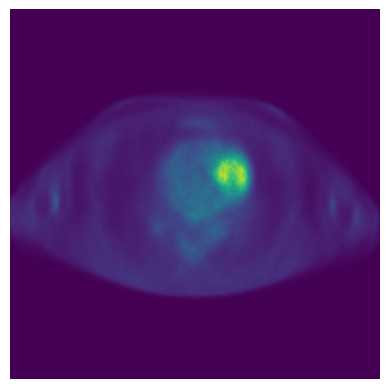}
			\put(1,4){
				\textcolor{red}{\small PSNR=\DTLfetch{table1}{patient_no}{372}{psnr_pet_solo}}
			}
			\put(1,90){
				\textcolor{red}{\small SSIM=\DTLfetch{table1}{patient_no}{372}{ssim_pet_solo}}
			}
		\end{overpic}
		}%
	\subfloat[\acs{WCT} (individual recon.) \label{subfig:hl_wls}]{
		\begin{overpic}[width=0.45\linewidth]{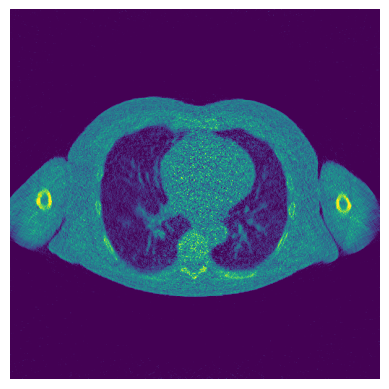}
			\put(1,4){
				\textcolor{red}{\small PSNR=\DTLfetch{table1}{patient_no}{372}{psnr_ct_solo}}
			}
			\put(1,90){
				\textcolor{red}{\small SSIM=\DTLfetch{table1}{patient_no}{372}{ssim_ct_solo}}
			}
		\end{overpic}
	}%

	\caption{
		\Ac{HL}---Reconstructed \ac{PET}/\ac{CT} images using \ac{MVAE} and \ac{PLS} (\ac{PET}/\ac{CT}), \ac{MLEM} (\ac{PET}) and \ac{WLS} (\ac{CT}). \Ac{VPET}/\ac{VCT}  (\subref{subfig:hl_pet_mvae} \& \subref{subfig:hl_ct_mvae}) and  \ac{PPET}/\ac{PCT}  (\subref{subfig:hl_pet_pls} \& \subref{subfig:hl_ct_pls}) were reconstructed synergistically, while  \acs{PETunet}   and \acs{CTunet} (\subref{subfig:hl_dip_pet} \& \subref{subfig:hl_dip_ct}) as well as  \ac{PETunet} and \ac{CTunet}  (\subref{subfig:hl_mlem} \& \subref{subfig:hl_wls})  were reconstructed individually. The \ac{GT} images $\boldx_1^\star$ and $\boldx_2^\star$ are shown in Figure~\ref{fig:petct_gt}.
		} \label{fig:hl_main_recon}
\end{figure}

Figure~\ref{fig:hl_main_recon} shows the images reconstructed from \ac{LH} simulated data (Patient 1). The \ac{EMPET} and \ac{WCT} images exhibit significant noise amplification. In contrast, \ac{VPET}/\ac{VCT}, \ac{PPET}/\ac{PCT}, and \ac{PETunet}/\ac{CTunet} appear free of noise. This example demonstrates that \ac{PLS} outperforms \ac{MVAE} on \ac{PET}, while \ac{MVAE} shows a slight advantage over \ac{PLS} on \ac{CT}. U-Net denoising surpasses all methods except \ac{PPET}, which achieves better \ac{SSIM} than \ac{PETunet}. Reconstructed images of the nine other patients are shown in Figure~\ref{fig:bigfig1} (Appendix~\ref{sec:appB}).

\newlength{\tempdima}
\setlength{\tempdima}{0.32\linewidth}

\DTLloaddb
{table1}
{images/patients/hplc.txt}
\DTLloaddb
{table4}
{images/patients/lphc.txt}
\begin{filecontents*}{images/patients/munet_hc.txt}
patient_no,psnr_pet,ssim_pet,psnr_ct,ssim_ct
340,58.09,0.9987,39.82,0.9856
341,59.37,0.9985,38.75,0.9825
345,55.61,0.9976,38.49,0.9821
348,58.23,0.9986,40.5,0.9884
351,56.43,0.9982,39.71,0.9858
370,55.69,0.9977,39.5,0.9856
371,56.91,0.9982,38.28,0.9814
372,54.14,0.9971,37.66,0.9796
374,58.17,0.9986,39.19,0.9841
375,57.04,0.998,38.78,0.9844
\end{filecontents*}
\DTLloaddb
{table2}
{images/patients/munet_hc.txt}
\begin{filecontents*}{images/patients/munet_lc.txt}
patient_no,psnr_pet,ssim_pet,psnr_ct,ssim_ct
340,58.38,0.9988,39.72,0.9864
341,59.73,0.9986,38.27,0.9823
345,55.6,0.9977,37.8,0.9812
348,58.23,0.9987,39.84,0.987
351,57.08,0.9985,39.49,0.9856
370,56.14,0.998,39.05,0.9843
371,57.26,0.9984,37.89,0.9806
372,54.45,0.9974,37.07,0.9778
374,58.61,0.9986,38.93,0.9841
375,57.62,0.9983,38.18,0.983
\end{filecontents*}
\DTLloaddb
{table3}
{images/patients/munet_lc.txt}

Figure~\ref{fig:hl_scatter} presents a scatter plot of \ac{SSIM} versus \ac{PSNR} for the 10 patients, comparing \ac{VPET}/\ac{PPET} (Figure~\ref{subfig:hl_ssim_vs_psnr_pet}) and \ac{VCT}/\ac{PCT} (Figure~\ref{subfig:hl_ssim_vs_psnr_pet}). To enhance visibility, \ac{EMPET} and \ac{WCT} were omitted, as they are significantly outperformed. Consistent with the observations from Figure~\ref{fig:hl_main_recon}, \ac{PPET} slightly outperforms \ac{VPET},  achieving \ac{PSNR}$\approx$56--64dB and \ac{SSIM}$\approx$0.999--0.9999, compared to \ac{PSNR}$\approx$51--59dB and \ac{SSIM}$\approx$0.997--0.999 for \ac{VPET}. Additionally, \ac{PETunet} shows a slight improvement over \ac{VPET} in terms of \ac{PSNR} ($\approx$54--60~dB).

On the \ac{CT} side, \ac{PCT} outperforms \ac{VCT}, with both methods achieving comparable \ac{PSNR} values ($\approx$35--39~dB), but with \ac{SSIM} values slightly favoring \ac{PCT} (\ac{SSIM}$\approx$0.96--0.985 for \ac{PCT}, compared to \ac{SSIM}$\approx$0.96--0.98 for \ac{VCT}). The underperformance of \ac{VCT} can be attributed to the tendency of \ac{VAE}-based methods to produce blurry images, which are less suited for \ac{CT} reconstruction. Finally, \ac{CTunet} surpasses all other methods, as it is specifically trained to  low-count \ac{CT} images to the \ac{GT} \ac{CT} images.

\newlength{\tempdimb}
\setlength{\tempdimb}{0.48\linewidth}

\begin{figure}
	\subfloat[\ac{PET}\label{subfig:hl_ssim_vs_psnr_pet}]{
		\centering
		\begin{tikzpicture}[scale=0.55]
			\begin{axis}[
				mark options={mark size = 3pt},
				xlabel={SSIM},
				ylabel={PSNR},
				legend cell align=left,
				legend style={at={(0.15,0.9)},anchor=north},
				x post scale=2,
				xtick distance = 0.0005,
				x tick label style={
					/pgf/number format/.cd,
					precision=4,
					/tikz/.cd,
				}
				]
				
				\addplot[red, only marks] 
				table[x=ssim_pet_mat, y=psnr_pet_mat, col sep=comma]{images/patients/hplc.txt};
				\addlegendentry{\ac{PPET}}
				
				\addplot[blue, only marks] 
				table[x=ssim_pet_vae, y=psnr_pet_vae, col sep=comma]{images/patients/hplc.txt};
				\addlegendentry{\ac{VPET}}
				
				\addplot[green, only marks]
				table[x=ssim_pet, y=psnr_pet, col sep=comma]
				{images/patients/munet_hc.txt};
				\addlegendentry{UNet-PET}

			\end{axis}

		\end{tikzpicture}
	}

	\subfloat[\ac{CT}\label{subfig:hl_ssim_vs_psnr_ct}]{
		\centering
		\begin{tikzpicture}[scale=0.55]
			\begin{axis}[
				mark options={mark size = 3pt},
				xlabel={SSIM},
				ylabel={PSNR},
				legend cell align=left,
				legend style={at={(0.15,0.9)},anchor=north},
				x post scale=2,
				xtick distance = 0.005,
				x tick label style={
					/pgf/number format/.cd,
					precision=4,
					/tikz/.cd,
				}
				]
				
				\addplot[red, only marks] 
				table[x=ssim_ct_mat, y=psnr_ct_mat, col sep=comma]{images/patients/hplc.txt};
				\addlegendentry{\ac{PCT}}
				
				\addplot[blue, only marks] 
				table[x=ssim_ct_vae, y=psnr_ct_vae, col sep=comma]{images/patients/hplc.txt};
				\addlegendentry{\ac{VCT}}
				
				\addplot[green, only marks]
				table[x=ssim_ct, y=psnr_ct, col sep=comma]
				{images/patients/munet_lc.txt};
				\addlegendentry{UNet-CT}

			\end{axis}

		\end{tikzpicture}
	}
	\caption{\Ac{HL}---\Ac{PSNR} vs \ac{SSIM} scatter plot of the 10 \ac{PET}/\ac{CT} images reconstructed using \ac{MVAE} and \ac{PLS}: \subref{subfig:hl_ssim_vs_psnr_pet} reconstructed \ac{PET} and \subref{subfig:hl_ssim_vs_psnr_ct} reconstructed \ac{CT}.	} \label{fig:hl_scatter}
\end{figure}

Table~\ref{table:hl_table} show the metrics averaged over the 10 patients.

\DTLloaddb
{table_hplc_mean}
{images/patients/hplc_mean.txt}

\DTLloaddb
{table_lphc_mean}
{images/patients/lphc_mean.txt}

\DTLloaddb
{table_hc_mean}
{images/patients/hc_mean.txt}

\DTLloaddb
{table_lc_mean}
{images/patients/lc_mean.txt}

\begin{table}[]
	\tiny
	\centering
	\begin{tabular}{|l|llll|llll|}
		\hline
		\multirow{2}{*}{} & \multicolumn{4}{c|}{HC-\acs{PET}}  & \multicolumn{4}{c|}{LC-\acs{CT}}   \\
		\hline
		& \multicolumn{1}{l}{\acs{MVAE}}   & \multicolumn{1}{l}{\acs{PLS}}  & \multicolumn{1}{l}{U-Net} & \acs{MLEM}   & \multicolumn{1}{l}{\acs{MVAE}}  & \multicolumn{1}{l}{\acs{PLS}}    & \multicolumn{1}{l}{U-Net} & \acs{WLS}    \\
		\hline
		\acs{PSNR}  & \multicolumn{1}{l}{\DTLfetch{table_hplc_mean}{row}{0}{psnr_pet_vae}}  & \multicolumn{1}{l}{\bf \DTLfetch{table_hplc_mean}{row}{0}{psnr_pet_mat}}  & \multicolumn{1}{l}{\DTLfetch{table_hc_mean}{row}{0}{psnr_pet}}    & \DTLfetch{table_hplc_mean}{row}{0}{psnr_pet_solo}  & 
		\multicolumn{1}{l}{ \DTLfetch{table_hplc_mean}{row}{0}{psnr_ct_vae}}  & \multicolumn{1}{l}{\DTLfetch{table_hplc_mean}{row}{0}{psnr_ct_mat}}  & \multicolumn{1}{l}{\DTLfetch{table_lc_mean}{row}{0}{psnr_ct}}    & 
		\DTLfetch{table_hplc_mean}{row}{0}{psnr_ct_solo}  \\
		
		\acs{SSIM}   & \multicolumn{1}{l}{\DTLfetch{table_hplc_mean}{row}{0}{ssim_pet_vae}} & 
		\multicolumn{1}{l}{ \DTLfetch{table_hplc_mean}{row}{0}{ssim_pet_mat}} & 
		\multicolumn{1}{l}{\DTLfetch{table_hc_mean}{row}{0}{ssim_pet}}    & 
		\DTLfetch{table_hplc_mean}{row}{0}{ssim_pet_solo} & 
		\multicolumn{1}{l}{ \DTLfetch{table_hplc_mean}{row}{0}{ssim_ct_vae}} & 
		\multicolumn{1}{l}{\DTLfetch{table_hplc_mean}{row}{0}{ssim_ct_mat}} & 
		\multicolumn{1}{l}{\DTLfetch{table_lc_mean}{row}{0}{ssim_ct}}    & 
		\DTLfetch{table_hplc_mean}{row}{0}{ssim_ct_solo}    \\
		\hline
	\end{tabular}
	\caption{\Ac{HL}---Metric for each method averaged over the 10 patients.} \label{table:hl_table}
\end{table}

\paragraph{Image Reconstruction: Low-count \ac{PET}, High-count \ac{CT}}

\begin{figure}
    \centering
	\subfloat[\ac{VPET}, $\eta=1/2$ \label{subfig:lh_pet_mvae}]{
		\begin{overpic}[width=0.45\linewidth]{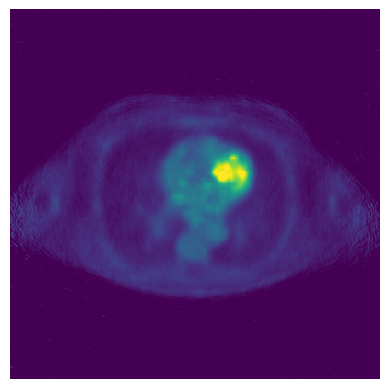}
			\put(1,4){
				\textcolor{red}{\small PSNR={\DTLfetch{table4}{patient_no}{372}{psnr_pet_vae}} }
			}
			\put(1,90){
				\textcolor{red}{\small SSIM={\DTLfetch{table4}{patient_no}{372}{ssim_pet_vae}} }
			}
		\end{overpic}	
	}%
	\subfloat[\ac{VCT}, $\eta=1/2$ \label{subfig:lh_ct_mvae}]{
		\begin{overpic}[width=0.45\linewidth]{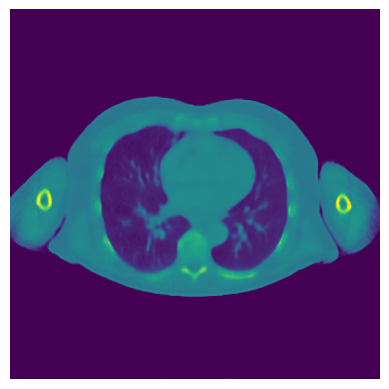}
			\put(1,4){
				\textcolor{red}{\small PSNR={\DTLfetch{table4}{patient_no}{372}{psnr_ct_vae}} }
			}
			\put(1,90){
				\textcolor{red}{\small SSIM={\DTLfetch{table4}{patient_no}{372}{ssim_ct_vae}} }
			}
		\end{overpic}
	}%

	\subfloat[\acs{PPET} \label{subfig:lh_pet_pls}]{
		\begin{overpic}[width=0.45\linewidth]{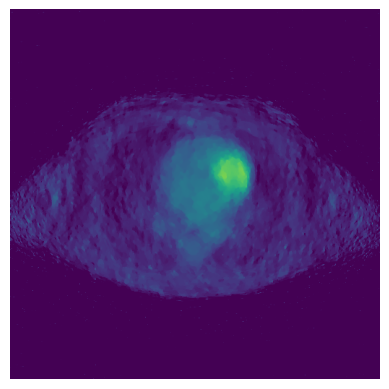}
			\put(1,4){
				\textcolor{red}{\small PSNR={\DTLfetch{table4}{patient_no}{372}{psnr_pet_mat}} }
			}
			\put(1,90){
				\textcolor{red}{\small SSIM={\DTLfetch{table4}{patient_no}{372}{ssim_pet_mat}} }
			}
		\end{overpic}
	}%
	\subfloat[\acs{PCT}\label{subfig:lh_ct_pls}]{
		\begin{overpic}[width=0.45\linewidth]{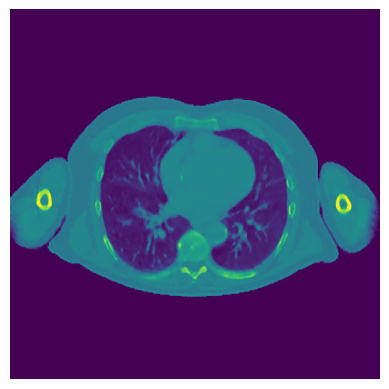}
			\put(1,4){
				\textcolor{red}{\small PSNR={\DTLfetch{table4}{patient_no}{372}{psnr_ct_mat}} }
			}
			\put(1,90){
				\textcolor{red}{\small SSIM={\DTLfetch{table4}{patient_no}{372}{ssim_ct_mat}} }
			}
		\end{overpic}
	}%

	\subfloat[\acs{PETunet} (individual recon.) \label{subfig:lh_dip_pet}]{
		\begin{overpic}[width=0.45\linewidth,height=0.45\linewidth]{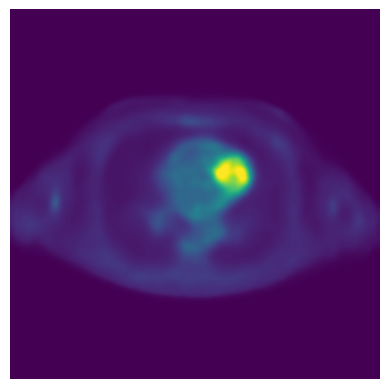}
			\put(1,4){
				\textcolor{red}{\small PSNR=\DTLfetch{table3}{patient_no}{372}{psnr_pet} }
			}
			\put(1,90){
				\textcolor{red}{\small SSIM=\DTLfetch{table3}{patient_no}{372}{ssim_pet} }
			}
		\end{overpic}
	}%
	\subfloat[\acs{CTunet} (individual recon.) \label{subfig:lh_dip_ct}]{
		\begin{overpic}[width=0.45\linewidth,height=0.45\linewidth]{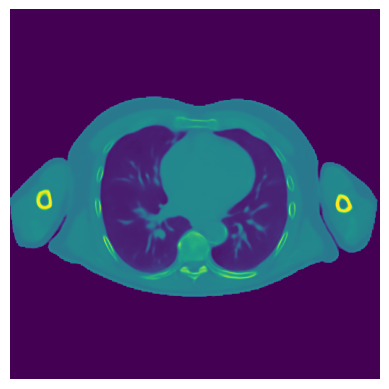}
			\put(1,4){
				\textcolor{red}{\small PSNR=\DTLfetch{table2}{patient_no}{372}{psnr_ct} }
			}
			\put(1,90){
				\textcolor{red}{\small SSIM=\DTLfetch{table2}{patient_no}{372}{ssim_ct} }
			}
		\end{overpic}
	}%

	\subfloat[\acs{EMPET} (individual recon.) \label{subfig:lh_mlem}]{
		\begin{overpic}[width=0.45\linewidth]{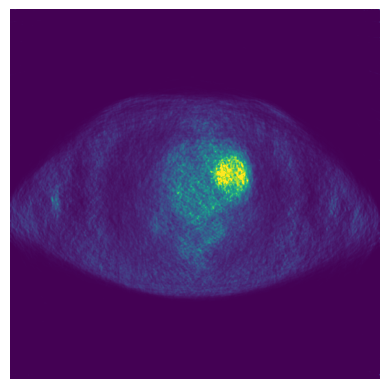}
			\put(1,4){
				\textcolor{red}{\small PSNR={\DTLfetch{table4}{patient_no}{372}{psnr_pet_solo}} }
			}
			\put(1,90){
				\textcolor{red}{\small SSIM={\DTLfetch{table4}{patient_no}{372}{ssim_pet_solo}} }
			}
		\end{overpic}
	}%
	\subfloat[\acs{WCT} (individual recon.) \label{subfig:lh_wls}]{
		\begin{overpic}[width=0.45\linewidth]{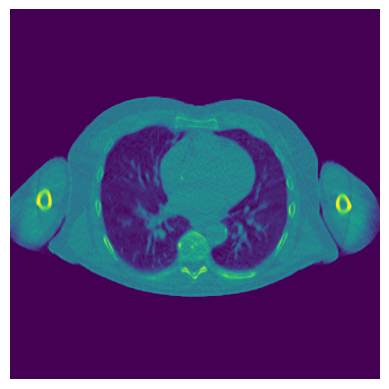}
			\put(1,4){
				\textcolor{red}{\small PSNR={\DTLfetch{table4}{patient_no}{372}{psnr_ct_solo}} }
			}
			\put(1,90){
				\textcolor{red}{\small SSIM={\DTLfetch{table4}{patient_no}{372}{ssim_ct_solo}} }
			}
		\end{overpic}
	}%

	\caption{
		\Ac{LH}---Reconstructed \ac{PET}/\ac{CT} images using \ac{MVAE} and \ac{PLS} (\ac{PET}/\ac{CT}), \ac{MLEM} (\ac{PET}) and \ac{WLS} (\ac{CT}). \Ac{VPET}/\ac{VCT}  (\subref{subfig:lh_pet_mvae} \& \subref{subfig:lh_ct_mvae}) and  \ac{PPET}/\ac{PCT}  (\subref{subfig:lh_pet_pls} \& \subref{subfig:lh_ct_pls}) were reconstructed synergistically, while  \acs{PETunet}   and \acs{CTunet} (\subref{subfig:lh_dip_pet} \& \subref{subfig:lh_dip_ct}) as well as  \ac{EMPET} and \ac{WCT}  (\subref{subfig:lh_mlem} \& \subref{subfig:lh_wls})  were reconstructed individually. The \ac{GT} images $\boldx_1^\star$ and $\boldx_2^\star$ are shown in Figure~\ref{fig:petct_gt}
	} \label{fig:lh_main_recon}

\end{figure}

Figure~\ref{fig:lh_main_recon} shows the  images obtained from \ac{LH} simulated data (Patient 1/10). Similarly to the \ac{HL} experiment, \ac{EMPET} and \ac{WCT}  suffer from noise amplification (especially \ac{EMPET}) while \ac{VPET}/\ac{VCT}, \ac{PPET}/\ac{PCT} and \ac{PETunet}/\ac{CTunet} appear noise-free. However this time \ac{MVAE} outperforms \ac{PLS} on the \ac{PET} while \ac{PLS} outperforms \ac{MVAE} on the \ac{CT}. \Ac{PETunet} denoising outperforms all other methods on the \ac{PET}  but \ac{CTunet} is outperformed by \ac{PCT}. Reconstructed images of the nine other patients are shown in Figure~\ref{fig:bigfig2} (Appendix~\ref{sec:appB}).

\begin{figure}
	\subfloat[\ac{GT} \ac{PET} $\boldx_1 + \bolds$ \label{subfig:gt_lesion}]{
		\begin{tikzpicture}
			\node (image) at (0,0) {
				\includegraphics[width=0.3\linewidth,height=0.3\linewidth]{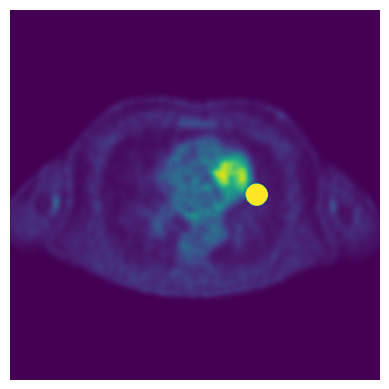}
			} ;
			\draw [-latex, ultra thick, green] (0.9,0.3) to[] (0.55,0.08);
		\end{tikzpicture}
	}
	\subfloat[\ac{VPET}, $\eta=1/2$  \label{subfig:pet_mvae_lesion}]{
		\begin{tikzpicture}
			\begin{scope}[spy using outlines={rectangle,magnification=3,size=11.5mm,connect spies}]
				\node {
					\includegraphics[width=0.3\linewidth,height=0.3\linewidth]{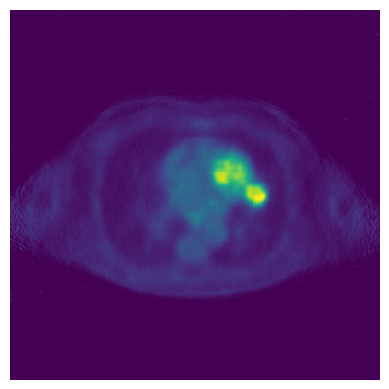}
				};				
				\spy [red] on (0.44,0.012) in node [left,red] at (-0.2,0.8);
			\end{scope}
		\end{tikzpicture}
	}
	\subfloat[\ac{VCT}, $\eta=1/2$  \label{subfig:ct_mvae_lesion}]{
		\begin{tikzpicture}
			\begin{scope}[spy using outlines={rectangle,magnification=3,size=11.5mm,connect spies}]
				\node {
					\includegraphics[width=0.3\linewidth,height=0.3\linewidth]{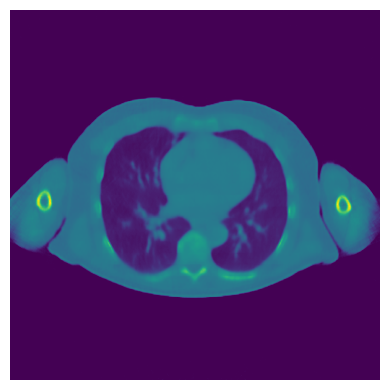}
				};				
				\spy [red] on (0.44,0.012) in node [left,red] at (-0.2,0.8);
			\end{scope}
		\end{tikzpicture}
	}	
	\caption{\Ac{LH}---\Ac{PET}/\ac{CT} mismatch experiment: \subref{subfig:gt_lesion} \ac{GT} \ac{PET} with a lesion in the lung that is absent from the \ac{CT} and the corresponding \subref{subfig:pet_mvae_lesion}  \ac{VPET} image and \subref{subfig:ct_mvae_lesion} \ac{VCT} image.  }\label{fig:lesion}
\end{figure}

In addition to this experiment, we assessed the ability of \ac{MVAE} synergistic reconstruction to deal with mismatches between the \ac{PET} and the \ac{CT} data. For this purpose, we modified the \ac{GT} \ac{PET} image $\boldx_1^\star$ from Figure~\ref{subfig:gt_pet64} by adding a 3-mm radius hot lesion $\bolds$ in the lung while keeping the \ac{GT} \ac{CT}  $\boldx_2^\star$ untouched (Figure~\ref{subfig:gt_ct64}), and we simulated projection data in  the \ac{LH} setting. The \ac{MVAE}-reconstructed images are shown in Figure~\ref{fig:lesion}. We observe that the hot lesion is present in the reconstructed \ac{PET} image but is absent from the reconstructed \ac{CT} image, with no evidence of crosstalks. This shows that  use of separate branches for each modality allows the model to address the inconsistencies between \ac{PET} and \ac{CT} images. As a result, the \ac{MVAE} model preserves unique from the \ac{PET} while avoiding  the introduction of artifacts in the \ac{CT}.

Figure~\ref{fig:lh_scatter} shows the same scatter plot as Figure~\ref{fig:hl_scatter}.  \Ac{MVAE} outperforms \ac{PLS} on the \ac{PET} (\ac{PSNR}$\approx$52--57~dB \& \ac{SSIM}$\approx$0.9975--0.999 for \ac{VPET}, \ac{PSNR}$\approx$49--53 \& \ac{SSIM}$\approx$0.994--0.997 for \ac{PPET})--- \ac{PETunet} gives the best results as it was trained to map low-dose \ac{PET} images to \ac{GT} \ac{PET} images.

Regarding \ac{CT}, \ac{PCT} gives the best results (\ac{PSNR}$\approx$38--42 \& \ac{SSIM}$\approx$0.98--0.998) while \ac{VCT} is on par with \ac{CTunet} (\ac{PSNR}$\approx$37--41 \& \ac{SSIM}$\approx$0.98--0.99), with the exception of Patient 1 for which \ac{VCT} seems to underperform.

\setlength{\tempdimb}{0.48\linewidth}

\begin{figure}
	\subfloat[\ac{PET} \label{subfig:lh_ssim_vs_psnr_pet}]{
		\centering
		\begin{tikzpicture}[scale=0.55]
			\begin{axis}[
				mark options={mark size = 3pt},
				xlabel={SSIM},
				ylabel={PSNR},
				legend cell align=left,
				legend style={at={(0.15,0.9)},anchor=north},
				x post scale=2,
				xtick distance = 0.001,
				x tick label style={
					/pgf/number format/.cd,
					precision=4,
					/tikz/.cd,
				}
				]
				
				\addplot[red, only marks] 
				table[x=ssim_pet_mat, y=psnr_pet_mat, col sep=comma]{images/patients/lphc.txt};
				\addlegendentry{\ac{PPET}} ;
				
				\addplot[blue, only marks] 
				table[x=ssim_pet_vae, y=psnr_pet_vae, col sep=comma]{images/patients/lphc.txt};
				\addlegendentry{\ac{VPET}} ;
				
				\addplot[green, only marks]
				table[x=ssim_pet, y=psnr_pet, col sep=comma]
				{images/patients/munet_lc.txt};
				\addlegendentry{UNet-PET};

			\end{axis}

		\end{tikzpicture}
	}

	\subfloat[\ac{CT} \label{subfig:lh_ssim_vs_psnr_ct}]{
		\centering
		\begin{tikzpicture}[scale=0.55]
			\begin{axis}[
				mark options={mark size = 3pt},
				xlabel={SSIM},
				ylabel={PSNR},
				legend cell align=left,
				legend style={at={(0.15,0.9)},anchor=north},
				x post scale=2,
				xtick distance = 0.005,
				x tick label style={
					/pgf/number format/.cd,
					precision=4,
					/tikz/.cd,
				}
				]
				
				\addplot[red, only marks] 
				table[x=ssim_ct_mat, y=psnr_ct_mat, col sep=comma]{images/patients/lphc.txt};
				\addlegendentry{\ac{PCT}} ;
				
				\addplot[blue, only marks] 
				table[x=ssim_ct_vae, y=psnr_ct_vae, col sep=comma]{images/patients/lphc.txt};
				\addlegendentry{\ac{VCT}} ;
				
				\addplot[green, only marks]
				table[x=ssim_ct, y=psnr_ct, col sep=comma]
				{images/patients/munet_hc.txt};
				\addlegendentry{UNet-CT} ;

			\end{axis}

		\end{tikzpicture}
	}
	\caption{\Ac{LH}---\Ac{PSNR} vs \ac{SSIM} scatter plot of the 10 \ac{PET}/\ac{CT} images reconstructed using \ac{MVAE} and \ac{PLS}:  \subref{subfig:lh_ssim_vs_psnr_pet} reconstructed \ac{PET} and \subref{subfig:lh_ssim_vs_psnr_ct} reconstructed \ac{CT}.	}  \label{fig:lh_scatter}
\end{figure}

Table~\ref{table:lh_table} show the metrics averaged over the 10 patients.

\begin{table}[]
	\tiny
	\centering
	\begin{tabular}{|l|llll|llll|}
		\hline
		\multirow{2}{*}{} & \multicolumn{4}{c|}{LC-\acs{PET}}  & \multicolumn{4}{c|}{HC-\acs{CT}}   \\
		\hline
		& \multicolumn{1}{l}{\acs{MVAE}}   & \multicolumn{1}{l}{\acs{PLS}}  & \multicolumn{1}{l}{U-Net} & \acs{MLEM}   & \multicolumn{1}{l}{\acs{MVAE}}  & \multicolumn{1}{l}{\acs{PLS}}    & \multicolumn{1}{l}{U-Net} & \acs{WLS}    \\
		\hline
		\acs{PSNR}  & \multicolumn{1}{l}{\DTLfetch{table_lphc_mean}{row}{0}{psnr_pet_vae}}  & \multicolumn{1}{l}{\DTLfetch{table_lphc_mean}{row}{0}{psnr_pet_mat}}  & \multicolumn{1}{l}{\DTLfetch{table_lc_mean}{row}{0}{psnr_pet}}    & \DTLfetch{table_lphc_mean}{row}{0}{psnr_pet_solo}  & 
		\multicolumn{1}{l}{\DTLfetch{table_lphc_mean}{row}{0}{psnr_ct_vae}}  & \multicolumn{1}{l}{\DTLfetch{table_lphc_mean}{row}{0}{psnr_ct_mat}}  & \multicolumn{1}{l}{\DTLfetch{table_hc_mean}{row}{0}{psnr_ct}}    & 
		\DTLfetch{table_lphc_mean}{row}{0}{psnr_ct_solo}  \\
		
		\acs{SSIM} & \multicolumn{1}{l}{\DTLfetch{table_lphc_mean}{row}{0}{ssim_pet_vae}} & 
		\multicolumn{1}{l}{\DTLfetch{table_lphc_mean}{row}{0}{ssim_pet_mat}} & 
		\multicolumn{1}{l}{\DTLfetch{table_lc_mean}{row}{0}{ssim_pet}}    & 
		\DTLfetch{table_lphc_mean}{row}{0}{ssim_pet_solo} & 
		\multicolumn{1}{l}{\DTLfetch{table_lphc_mean}{row}{0}{ssim_ct_vae}} & 
		\multicolumn{1}{l}{\DTLfetch{table_lphc_mean}{row}{0}{ssim_ct_mat}} & 
		\multicolumn{1}{l}{\DTLfetch{table_hc_mean}{row}{0}{ssim_ct}}    & 
		\DTLfetch{table_lphc_mean}{row}{0}{ssim_ct_solo}    \\
		\hline
	\end{tabular}
	\caption{\Ac{LH}---Metric for each method averaged over the 10 patients.} \label{table:lh_table}
\end{table}

%% file: content/discussion.tex
\section{Discussion}\label{sec:discussion}

This work follows up on our previous studies presented in \cite{pinton2023synergistic,pinton2023joint}, where we initially trained our models on full images. However, we encountered potential overfitting issues due to the lack of data for training, resulting in overly optimistic results. To address this challenge, we adopted a patch decomposition approach. It has been previously reported that training on repetitive and consistent patches yields better results than training on the entire image \cite{gupta2020patchvae}. To mitigate artifacts in the reconstructed images, our regularization strategy necessitates numerous overlapping patches. However, this comes at the expense of increased computational cost, as each patch requires its own latent variable. Besides, it is essential to remain vigilant about the potential for hallucinations. Future work may involve integrating more robust constraints and validation techniques to ensure the reliability and accuracy of the reconstructed images.

We demonstrated that our models successfully learn from two images simultaneously, suggesting their applicability  for synergistic image reconstruction. Results obtained with \ac{MNIST}-trained models distinctly showcase how our generative model-based regularizer effectively utilize information from both images for denoising. 

Our \ac{HL} and \ac{LH} experiments showed that \ac{MVAE} outperforms the \ac{PLS} technique  \cite{ehrhardt2014joint}  on low-count \ac{PET} (although it is outperformed by \ac{PLS} with high-count data), thus demonstrating that \ac{MVAE} is suitable for low-dose imaging. However it  is outperformed by U-Net denoising methods that are specifically trained to process this noise level.

The main drawback of \acp{VAE} is their tendency to generate blurry images. While this phenomenon has limited impact on \ac{PET} due to its lower intrinsic resolution, it is more problematic for \ac{CT}. In a previous version of this work, we implemented a multibranch \ac{GAN} which produced sharper images than those produced with \ac{MVAE}, with the \ac{MNIST} dataset. However the optimization with respect to the latent variable could not be achieved by \ac{LBFGS} and therefore we deployed a computationally expensive  \ac{PSO} \cite{pso1995} (global optimizer). While this is reasonable with small images, is becomes unpractical if applied on a high number of patches.       

Although we have demonstrated that multibranch generative models offer a viable approach for learned synergistic reconstruction, it is important to explore alternative options beyond \acp{VAE}. \Acp{DM} have shown significant promise in generating high-quality images from training datasets \cite{dhariwal2021diffusion}. These models can be seamlessly integrated into a \ac{PML} reconstruction framework, as demonstrated by \ac{DPS} \cite{chung2023diffusion}. Recent advancements in spectral \ac{CT} reconstruction have further highlighted the potential of \acp{DM}. More specifically, \acp{DM} are capable of capturing multichannel information and employing multichannel \ac{DPS} leads to superior spectral \ac{CT} images compared to conventional techniques \cite{vazia2024diffusion,vazia2024spectral}. Therefore, the future of learned synergistic reconstruction may shift towards leveraging \acp{DM}.

%% file: content/conclusion.tex
\section{Conclusion}\label{sec:conclusion}

In conclusion, our study highlights the utility of generative models for learned synergistic reconstruction in medical imaging. By training on pairs of images, our approach harnesses the power of \acp{VAE} to improve denoising and reconstruction outcomes. While challenges such as patch decomposition and inherent model limitations persist, our results demonstrate promising advancements in leveraging generative models for enhancing image quality and information exchange between modalities. Moving forward, further exploration of alternative models, such as \acp{DM}, may offer additional avenues for enhancing imaging outcomes in medical diagnostics and research. Overall, our findings contribute to the growing body of literature on learned synergistic reconstruction methods and pave the way for future developments in medical multimodal imaging technology.

%% file: content/training_vae.tex
In this section we summarize the \ac{MVAE} training strategy taking inspiration from \citeauthor{duff2021regularising}~\cite{duff2021regularising} with a generalization the multichannel setting.

The $K$-channel patches from the training dataset are represented by a random array $\boldU = \{\boldu_k\} = \{\boldu_1,\dots,\boldu_K\}\in \calU^K$, such that for all $k$, $\boldu_k \in \calU$ represents a patch in channel $k$. We denote by $p^{\ast}\colon \calU^K \rightarrow \R^+$ the empirical \ac{PDF} of $\boldU$, which corresponds to randomly selecting a patient image from the training dataset and extracting $K$ patches (one patch per channel at the same location for all $k$). We assume the latent space $\calZ$ is endowed with a \ac{PDF} $p_0: \calZ \rightarrow \mathbb{R}^+$ (generally a standard normal \ac{PDF}). The training consists in learning a parameter $\boldtheta$ such that  $\boldG_{\boldtheta}^{\mathrm{mult}}(\boldz)$, $\boldz\sim p_0$, has a \ac{PDF} that generalizes $p^{\ast}$ such that is approximates the true \ac{PDF} of $\boldU$. Denoting by $p_{\boldtheta}\colon \calU^K \rightarrow \R^+$  the probability distribution of  $\boldG_{\boldtheta}^{\mathrm{mult}}(\boldz)$
the training is achieved by minimizing a ``distance'' $p^{\ast}$ and $p_{\boldtheta}$:
\begin{equation}
	\min_{\boldtheta} \, d \left( p^{\ast}   \,||\,  p_{\boldtheta}   \right)
\end{equation}
For \acp{VAE}, the distance considered is the \ac{KL} divergence, denoted $d_{\mathrm{KL}}$, which when applied to  $p^{\ast}$ and  $p_{\boldtheta}$ gives us
\begin{align}
	d_{\mathrm{KL}} ( p^{\ast}   \,||\,  p_{\boldtheta}   ) & = \mathbb{E}_{\boldU \sim p^{\ast}  } \left[ \log \left(  \frac{p^{\ast}(\boldU)}{       p_{\boldtheta}(\boldU)    } \right) \right]   \\
	& = {}  \mathbb{E}_{\boldU \sim p^{\ast}  }   \left[\log p^{\ast}(\boldU)\right] -  \mathbb{E}_{\boldU \sim p^{\ast}  }    [\log p_{\boldtheta}(\boldU)] \, .  \label{eq:kl}
\end{align} 
Minimizing $d_{\mathrm{KL}} ( p^{\ast}   \,||\,  p_{\boldtheta}   )$ is therefore equivalent to maximizing $\mathbb{E}_{\boldU \sim p^{\ast}  }    [\log p_{\boldtheta}(\boldU)]$.  

The \ac{PDF} $p_{\boldtheta}$ is untractable and therefore we employ the following parametric inference model:
\begin{align}
	\boldU \mid \boldz & \sim \mathscr{N}\left( \boldG_{\boldtheta}(\boldz) ,  \rho^2 \mathrm{id}_{\calU^K} \right)   \\
	\boldz \mid \boldU & \sim \mathscr{N}\left( \boldmu_{\boldpsi} (\boldU)   , \mathrm{diag}\left(\boldsigma^2_{\boldpsi} (\boldU) \right) \right)  
\end{align} 
where $\boldmu_{\boldpsi},\boldsigma_{\boldpsi}^2\colon \calU^K \to \calZ$ are respectively the multichannel \emph{encoder mean} and multichannel \emph{encoder variance} (parametrized by $\boldpsi$), each of which mapping a $K$-channel patch $\boldU = \{\boldu_1,\dots,\boldu_K\}$ to a single latent variable. In our work, these encoders are built with $K=2$ encoders whose output are merged (via a concatenation) into a single vector out of which the mean and variance vectors are generated (cf. left part of Figure~\ref{fig:mvae-archi}). Denoting by $p_\boldtheta(\cdot| \boldz)$ and $q_\boldpsi(\cdot| \boldU)$ the conditional \acp{PDF} of $\boldU | \boldz$ and $\boldz | \boldU$ respectively, the \ac{PDF} $p_{\boldtheta}$ can be derived using the Bayes' rule and marginalization as
\begin{align}
	p_{\boldtheta}(\boldU) & = \int_{\boldZ}  \frac{p_{\boldtheta}(\boldU \mid\boldz)p_0(\boldz)}{q_\boldpsi(\boldz\mid \boldU)}  q_\boldpsi(\boldz\mid \boldU) \, \rmd \boldz\\
						   & = \mathbb{E}_{\boldz|\boldU} \left[ \frac{p_{\boldtheta}(\boldU \mid\boldz)p_0(\boldz)}{q_\boldpsi(\boldz\mid \boldU)}\right]	\, .  \label{eq:marg}
\end{align}
The $\log$ function being concave, we can derive the following \ac{ELBO} for $\log p_{\boldtheta}(\boldU)$ using the Jensen inequality:
\begin{align}
	\log p_{\boldtheta}(\boldU) & \ge \mathbb{E}_{\boldz|\boldU} \left[ \log \left( \frac{p_{\boldtheta}(\boldU \mid\boldz)p_0(\boldz)}{q_\boldpsi(\boldz\mid \boldU)} \right) \right] \\
	              & = \mathbb{E}_{\boldz|\boldU} \left[ p_{\boldtheta}(\boldU \mid\boldz)  - d_{\mathrm{KL}} (q_\boldpsi(\cdot| \boldU) \,||\, p_0 )\right] \, .
\end{align}	
Finally, an approximate minimizer of \eqref{eq:kl} can be obtain by minimizing the expectation of the negative  \ac{ELBO}, i.e.,
\begin{dmath}\label{eq:vae}
	\min_{\boldtheta,\boldpsi} \,  \mathbb{E}_{\boldU \sim{}p^\ast} \bigg[   \mathbb{E}_{\boldz | \boldU} \left[ \frac{1}{2\rho^2 }\left\| \boldU -  \boldG^{\mathrm{mult}}_{\boldtheta}(\boldz) \right\|_2^2 \right]   + d_{\mathrm{KL}}\left\{ q_\boldpsi(\cdot| \boldU) \, || \,  p_0 \right\}  \bigg]  \, .
\end{dmath}



%% file: content/big_fig1.tex
\newlength{\tempdimc}
\setlength{\tempdimc}{0.206\linewidth}
\setlength{\dimvert}{\tempdimc}	

\begin{figure*}
	\centering
	\begin{tabular}{p{0.0\tempdimc}p{.8\tempdimc}p{0.8\tempdimc}p{0.8\tempdimc}p{0.8\tempdimc}p{0.8\tempdimc}p{0.8\tempdimc}p{0.8\tempdimc}p{0.8\tempdimc}p{0.8\tempdimc}}
		& \hfill \footnotesize Patient 2 &  \hfill\footnotesize Patient 3 &  \hfill\footnotesize Patient 4 &  \hfill\footnotesize Patient 5 &  \hfill \footnotesize Patient 6 &   \hfill\footnotesize Patient 7 &   \hfill\footnotesize Patient 8 &  \hfill \footnotesize Patient 9 & \hfill \footnotesize Patient 10  \\ 
		\rowname{\footnotesize \acs{VPET}} 
		& 
		\begin{overpic}[width=\tempdimc,height=\tempdimc]{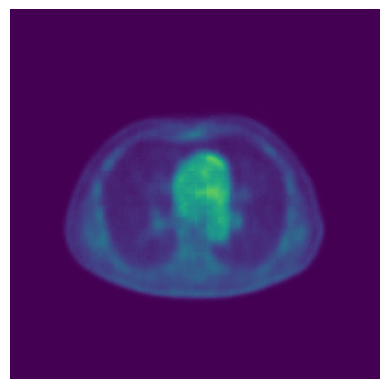}
			\put(-2,3){
				\textcolor{red}{\scriptsize PSNR={\DTLfetch{table1}{patient_no}{340}{psnr_pet_vae}} }
			}
			\put(-2,86){
				\textcolor{red}{\scriptsize SSIM={\DTLfetch{table1}{patient_no}{340}{ssim_pet_vae}} }
			}
		\end{overpic}
		&  
		\begin{overpic}[width=\tempdimc,height=\tempdimc]{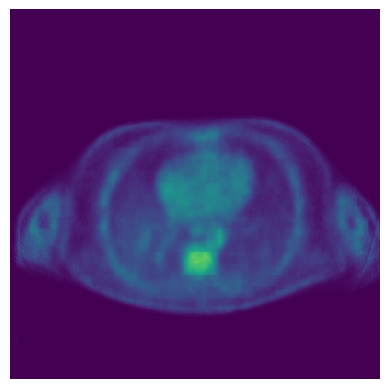}
			\put(-2,3){
				\textcolor{red}{\scriptsize PSNR=\DTLfetch{table1}{patient_no}{341}{psnr_pet_vae} }
			}
			\put(-2,86){
				\textcolor{red}{\scriptsize SSIM=\DTLfetch{table1}{patient_no}{341}{ssim_pet_vae} }
				
			}
		\end{overpic}
		&  \begin{overpic}[width=\tempdimc,height=\tempdimc]{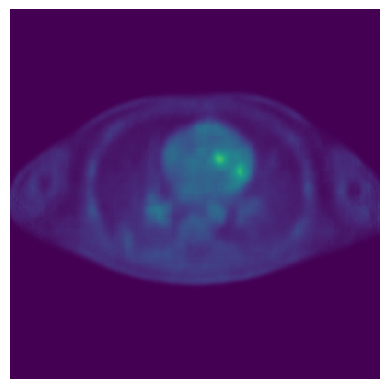}
			\put(-2,3){
				\textcolor{red}{\scriptsize PSNR={\DTLfetch{table1}{patient_no}{345}{psnr_pet_vae}}}
			}
			\put(-2,86){
				\textcolor{red}{\scriptsize SSIM={\DTLfetch{table1}{patient_no}{345}{ssim_pet_vae}}}
			}
		\end{overpic}
		& \begin{overpic}[width=\tempdimc,height=\tempdimc]{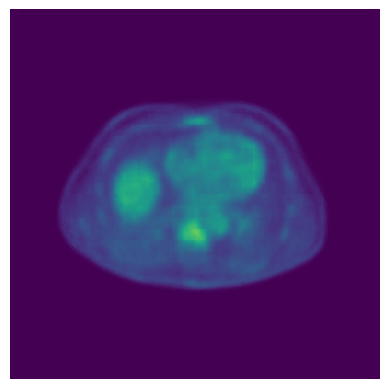}
			\put(-2,3){
				\textcolor{red}{\scriptsize PSNR={\DTLfetch{table1}{patient_no}{348}{psnr_pet_vae}}}
			}
			\put(-2,86){
				\textcolor{red}{\scriptsize SSIM={\DTLfetch{table1}{patient_no}{348}{ssim_pet_vae}}}
			}
		\end{overpic}
		&  \begin{overpic}[width=\tempdimc,height=\tempdimc]{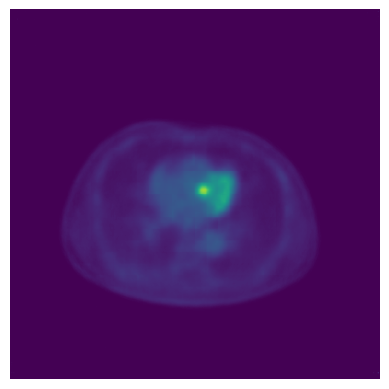}
			\put(-2,3){
				\textcolor{red}{\scriptsize PSNR={\DTLfetch{table1}{patient_no}{351}{psnr_pet_vae}}}
			}
			\put(-2,86){
				\textcolor{red}{\scriptsize SSIM={\DTLfetch{table1}{patient_no}{351}{ssim_pet_vae}}}
			}
		\end{overpic}
		& \begin{overpic}[width=\tempdimc,height=\tempdimc]{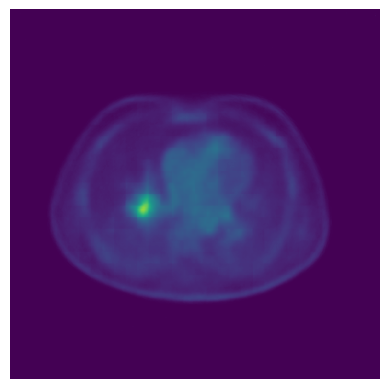}
			\put(-2,3){
				\textcolor{red}{\scriptsize PSNR={\DTLfetch{table1}{patient_no}{370}{psnr_pet_vae}}}
			}
			\put(-2,86){
				\textcolor{red}{\scriptsize SSIM={\DTLfetch{table1}{patient_no}{370}{ssim_pet_vae}}}
			}
		\end{overpic}
		& \begin{overpic}[width=\tempdimc,height=\tempdimc]{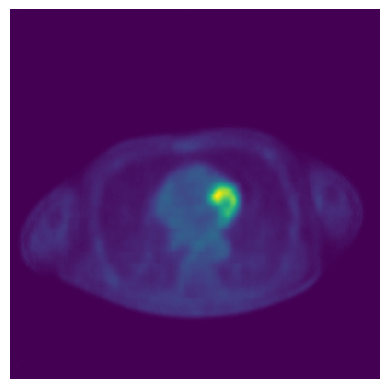}
			\put(-2,3){
				\textcolor{red}{\scriptsize PSNR={\DTLfetch{table1}{patient_no}{371}{psnr_pet_vae}}}
			}
			\put(-2,86){
				\textcolor{red}{\scriptsize SSIM={\DTLfetch{table1}{patient_no}{371}{ssim_pet_vae}}}
			}
		\end{overpic}
		& \begin{overpic}[width=\tempdimc,height=\tempdimc]{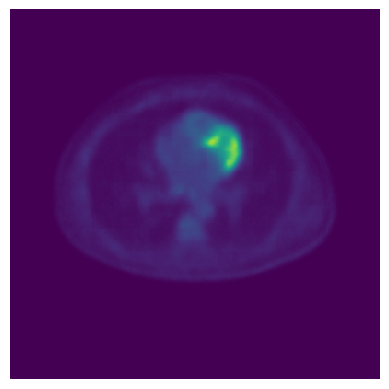}
			\put(-2,3){
				\textcolor{red}{\scriptsize PSNR={\DTLfetch{table1}{patient_no}{374}{psnr_pet_vae}}}
			}
			\put(-2,86){
				\textcolor{red}{\scriptsize SSIM={\DTLfetch{table1}{patient_no}{374}{ssim_pet_vae}}}
			}
		\end{overpic}
		& \begin{overpic}[width=\tempdimc,height=\tempdimc]{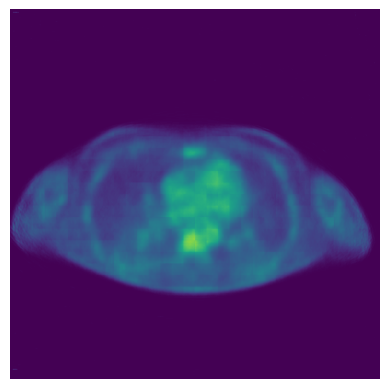}
			\put(-2,3){
				\textcolor{red}{\scriptsize PSNR={\DTLfetch{table1}{patient_no}{375}{psnr_pet_vae}}}
			}
			\put(-2,86){
				\textcolor{red}{\scriptsize SSIM={\DTLfetch{table1}{patient_no}{375}{ssim_pet_vae}}}
			}
		\end{overpic}
		\\
		\rowname{\footnotesize \acs{PPET}}  &  \begin{overpic}[width=\tempdimc,height=\tempdimc]{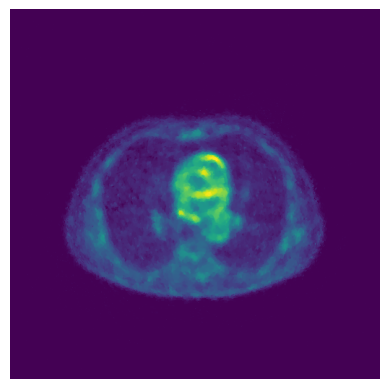}
			\put(-2,3){
				\textcolor{red}{\scriptsize PSNR=\DTLfetch{table1}{patient_no}{340}{psnr_pet_mat} }
			}
			\put(-2,86){
				\textcolor{red}{\scriptsize SSIM=\DTLfetch{table1}{patient_no}{340}{ssim_pet_mat} }
			}
		\end{overpic}
		&  
		\begin{overpic}[width=\tempdimc,height=\tempdimc]{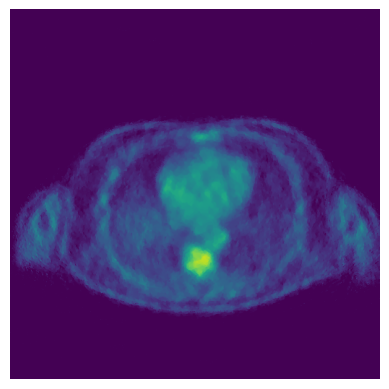}
			\put(-2,3){
				\textcolor{red}{\scriptsize PSNR=\DTLfetch{table1}{patient_no}{341}{psnr_pet_mat} }
			}
			\put(-2,86){
				\textcolor{red}{\scriptsize SSIM=\DTLfetch{table1}{patient_no}{341}{ssim_pet_mat} }
			}
		\end{overpic}
		&  \begin{overpic}[width=\tempdimc,height=\tempdimc]{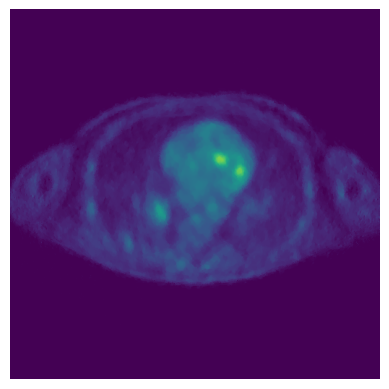}
			\put(-2,3){
				\textcolor{red}{\scriptsize PSNR=\DTLfetch{table1}{patient_no}{345}{psnr_pet_mat}}
			}
			\put(-2,86){
				\textcolor{red}{\scriptsize SSIM=\DTLfetch{table1}{patient_no}{345}{ssim_pet_mat} }
			}
		\end{overpic}
		& \begin{overpic}[width=\tempdimc,height=\tempdimc]{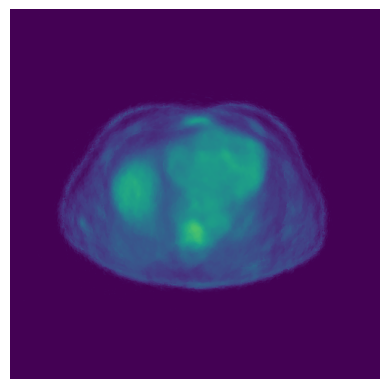}
			\put(-2,3){
				\textcolor{red}{\scriptsize PSNR=\DTLfetch{table1}{patient_no}{348}{psnr_pet_mat} }
			}
			\put(-2,86){
				\textcolor{red}{\scriptsize SSIM=\DTLfetch{table1}{patient_no}{348}{ssim_pet_mat} }
			}
		\end{overpic}
		&  \begin{overpic}[width=\tempdimc,height=\tempdimc]{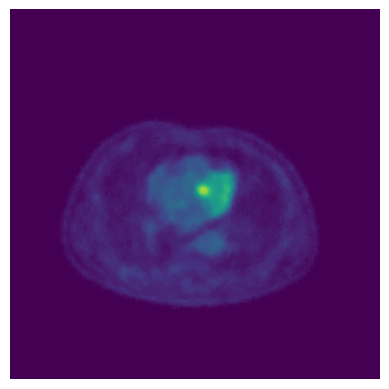}
			\put(-2,3){
				\textcolor{red}{\scriptsize PSNR=\DTLfetch{table1}{patient_no}{351}{psnr_pet_mat} }
			}
			\put(-2,86){
				\textcolor{red}{\scriptsize SSIM=\DTLfetch{table1}{patient_no}{351}{ssim_pet_mat} }
			}
		\end{overpic}
		& \begin{overpic}[width=\tempdimc,height=\tempdimc]{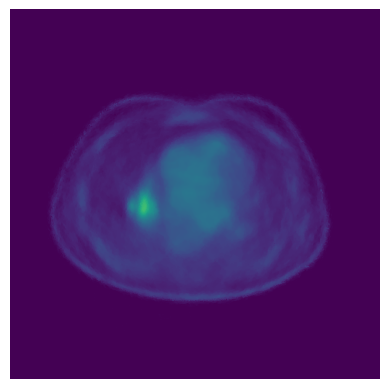}
			\put(-2,3){
				\textcolor{red}{\scriptsize PSNR=\DTLfetch{table1}{patient_no}{370}{psnr_pet_mat} }
			}
			\put(-2,86){
				\textcolor{red}{\scriptsize SSIM=\DTLfetch{table1}{patient_no}{370}{ssim_pet_mat} }
			}
		\end{overpic}
		& \begin{overpic}[width=\tempdimc,height=\tempdimc]{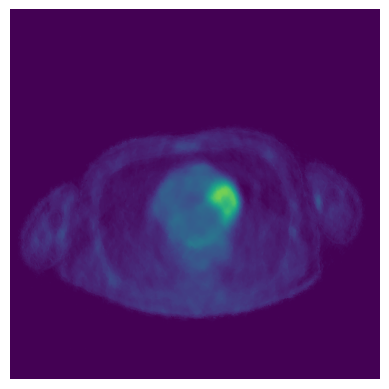}
			\put(-2,3){
				\textcolor{red}{\scriptsize PSNR=\DTLfetch{table1}{patient_no}{371}{psnr_pet_mat} }
			}
			\put(-2,86){
				\textcolor{red}{\scriptsize SSIM=\DTLfetch{table1}{patient_no}{371}{ssim_pet_mat} }
			}
		\end{overpic}
		& \begin{overpic}[width=\tempdimc,height=\tempdimc]{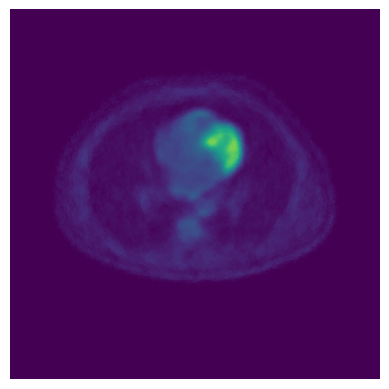}
			\put(-2,3){
				\textcolor{red}{\scriptsize PSNR=\DTLfetch{table1}{patient_no}{374}{psnr_pet_mat} }
			}
			\put(-2,86){
				\textcolor{red}{\scriptsize SSIM=\DTLfetch{table1}{patient_no}{374}{ssim_pet_mat} }
			}
		\end{overpic}
		& \begin{overpic}[width=\tempdimc,height=\tempdimc]{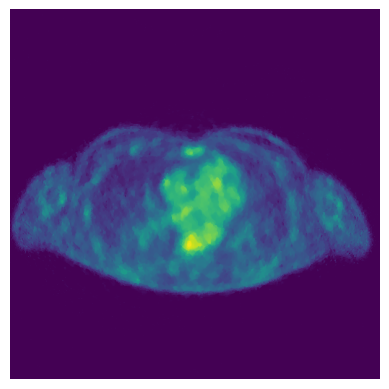}
			\put(-2,3){
				\textcolor{red}{\scriptsize PSNR=\DTLfetch{table1}{patient_no}{375}{psnr_pet_mat} }
			}
			\put(-2,86){
				\textcolor{red}{\scriptsize SSIM=\DTLfetch{table1}{patient_no}{375}{ssim_pet_mat} }
			}
		\end{overpic} \\
		\rowname{\footnotesize \acs{PETunet}}  
		&  \begin{overpic}[width=\tempdimc,height=\tempdimc]{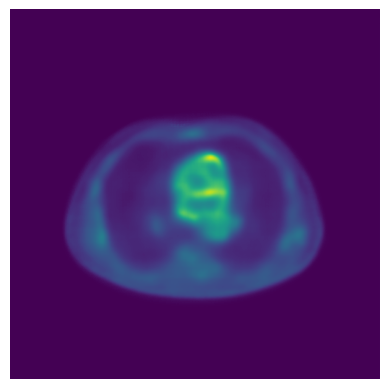}
			\put(-2,3){
				\textcolor{red}{\scriptsize 
					PSNR=\DTLfetch{table2}{patient_no}{340}{psnr_pet}
				}
			}
			\put(-2,86){
				\textcolor{red}{\scriptsize 
					SSIM=\DTLfetch{table2}{patient_no}{340}{ssim_pet}
				}
			}
		\end{overpic}
		&  
		\begin{overpic}[width=\tempdimc,height=\tempdimc]{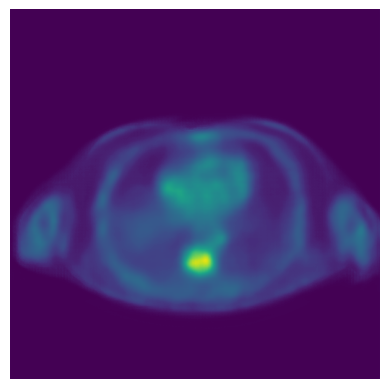}
			\put(-2,3){
				\textcolor{red}{\scriptsize PSNR=\DTLfetch{table2}{patient_no}{341}{psnr_pet} }
			}
			\put(-2,86){
				\textcolor{red}{\scriptsize SSIM=\DTLfetch{table2}{patient_no}{341}{ssim_pet} }
			}
		\end{overpic}
		&  \begin{overpic}[width=\tempdimc,height=\tempdimc]{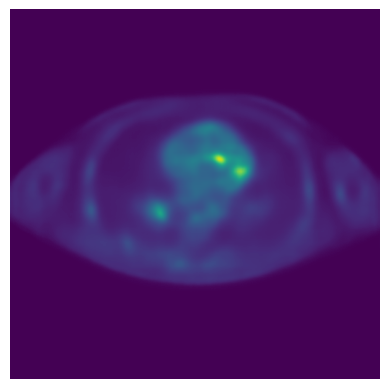}
			\put(-2,3){
				\textcolor{red}{\scriptsize PSNR=\DTLfetch{table2}{patient_no}{345}{psnr_pet}}
			}
			\put(-2,86){
				\textcolor{red}{\scriptsize SSIM=\DTLfetch{table2}{patient_no}{345}{ssim_pet}}
			}
		\end{overpic}
		& \begin{overpic}[width=\tempdimc,height=\tempdimc]{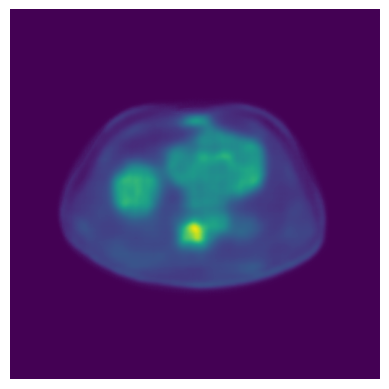}
			\put(-2,3){
				\textcolor{red}{\scriptsize PSNR=\DTLfetch{table2}{patient_no}{348}{psnr_pet} }
			}
			\put(-2,86){
				\textcolor{red}{\scriptsize SSIM=\DTLfetch{table2}{patient_no}{348}{ssim_pet}}
			}
		\end{overpic}
		&  \begin{overpic}[width=\tempdimc,height=\tempdimc]{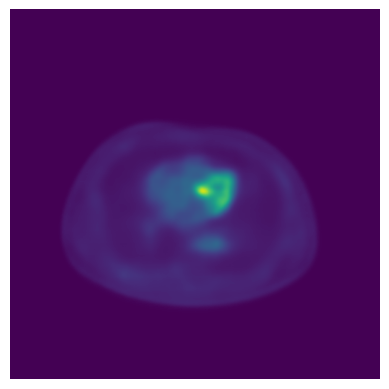}
			\put(-2,3){
				\textcolor{red}{\scriptsize PSNR=\DTLfetch{table2}{patient_no}{351}{psnr_pet} }
			}
			\put(-2,86){
				\textcolor{red}{\scriptsize SSIM=\DTLfetch{table2}{patient_no}{351}{ssim_pet} }
			}
		\end{overpic}
		& \begin{overpic}[width=\tempdimc,height=\tempdimc]{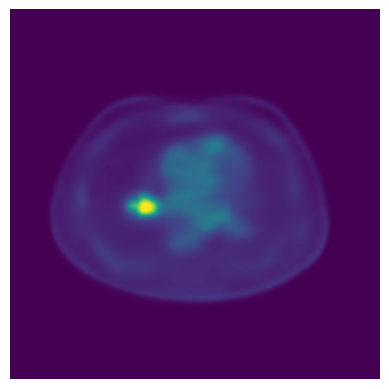}
			\put(-2,3){
				\textcolor{red}{\scriptsize PSNR=\DTLfetch{table2}{patient_no}{370}{psnr_pet} }
			}
			\put(-2,86){
				\textcolor{red}{\scriptsize SSIM=\DTLfetch{table2}{patient_no}{370}{ssim_pet} }
			}
		\end{overpic}
		& \begin{overpic}[width=\tempdimc,height=\tempdimc]{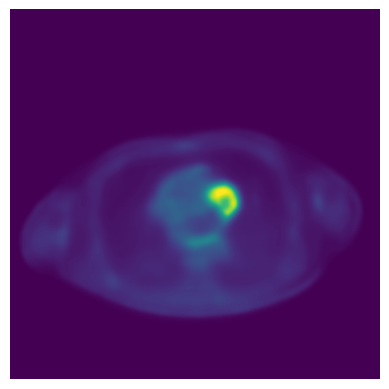}
			\put(-2,3){
				\textcolor{red}{\scriptsize PSNR=\DTLfetch{table2}{patient_no}{371}{psnr_pet} }
			}
			\put(-2,86){
				\textcolor{red}{\scriptsize SSIM=\DTLfetch{table2}{patient_no}{371}{ssim_pet} }
			}
		\end{overpic}
		& \begin{overpic}[width=\tempdimc,height=\tempdimc]{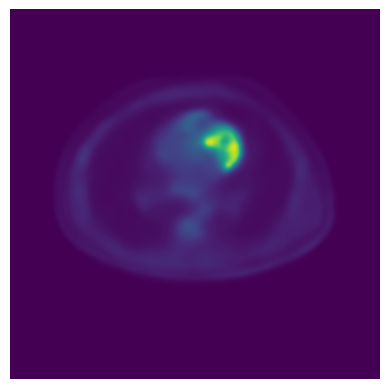}
			\put(-2,3){
				\textcolor{red}{\scriptsize PSNR=\DTLfetch{table2}{patient_no}{374}{psnr_pet} }
			}
			\put(-2,86){
				\textcolor{red}{\scriptsize SSIM=\DTLfetch{table2}{patient_no}{374}{ssim_pet} }
			}
		\end{overpic}
		& \begin{overpic}[width=\tempdimc,height=\tempdimc]{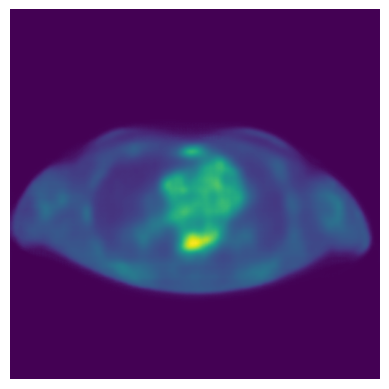}
			\put(-2,3){
				\textcolor{red}{\scriptsize PSNR=\DTLfetch{table2}{patient_no}{375}{psnr_pet} }
			}
			\put(-2,86){
				\textcolor{red}{\scriptsize SSIM=\DTLfetch{table2}{patient_no}{375}{ssim_pet} }
			}
		\end{overpic} \\
		\rowname{\footnotesize \acs{EMPET}}  & 
		\begin{overpic}[width=\tempdimc,height=\tempdimc]{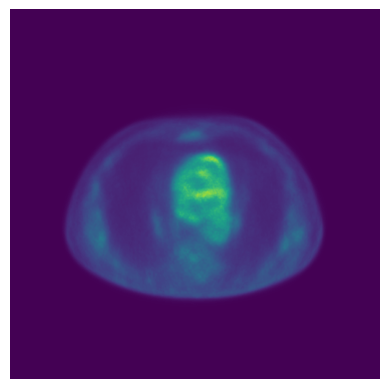}
			\put(-2,3){
				\textcolor{red}{\scriptsize PSNR=\DTLfetch{table1}{patient_no}{340}{psnr_pet_solo} }
			}
			\put(-2,86){
				\textcolor{red}{\scriptsize SSIM=\DTLfetch{table1}{patient_no}{340}{ssim_pet_solo} }
			}
		\end{overpic}
		&  
		\begin{overpic}[width=\tempdimc,height=\tempdimc]{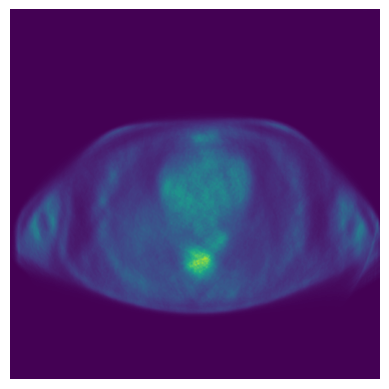}
			\put(-2,3){
				\textcolor{red}{\scriptsize PSNR=\DTLfetch{table1}{patient_no}{341}{psnr_pet_solo} }
			}
			\put(-2,86){
				\textcolor{red}{\scriptsize SSIM=\DTLfetch{table1}{patient_no}{341}{ssim_pet_solo} }
			}
		\end{overpic}
		&  \begin{overpic}[width=\tempdimc,height=\tempdimc]{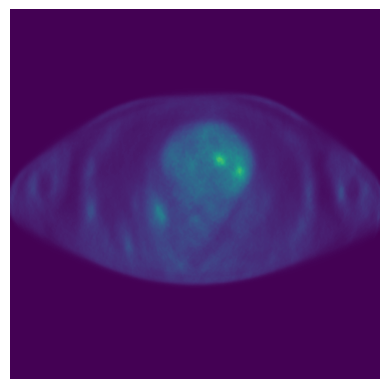}
			\put(-2,3){
				\textcolor{red}{\scriptsize PSNR=\DTLfetch{table1}{patient_no}{345}{psnr_pet_solo} }
			}
			\put(-2,86){
				\textcolor{red}{\scriptsize SSIM=\DTLfetch{table1}{patient_no}{345}{ssim_pet_solo} }
			}
		\end{overpic}
		& \begin{overpic}[width=\tempdimc,height=\tempdimc]{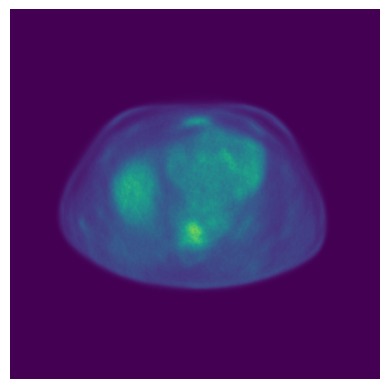}
			\put(-2,3){
				\textcolor{red}{\scriptsize PSNR=\DTLfetch{table1}{patient_no}{348}{psnr_pet_solo} }
			}
			\put(-2,86){
				\textcolor{red}{\scriptsize SSIM=\DTLfetch{table1}{patient_no}{348}{ssim_pet_solo} }
			}
		\end{overpic}
		&  \begin{overpic}[width=\tempdimc,height=\tempdimc]{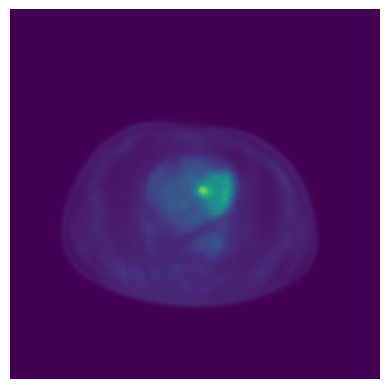}
			\put(-2,3){
				\textcolor{red}{\scriptsize PSNR=\DTLfetch{table1}{patient_no}{351}{psnr_pet_solo} }
			}
			\put(-2,86){
				\textcolor{red}{\scriptsize SSIM=\DTLfetch{table1}{patient_no}{351}{ssim_pet_solo} }
			}
		\end{overpic}
		& \begin{overpic}[width=\tempdimc,height=\tempdimc]{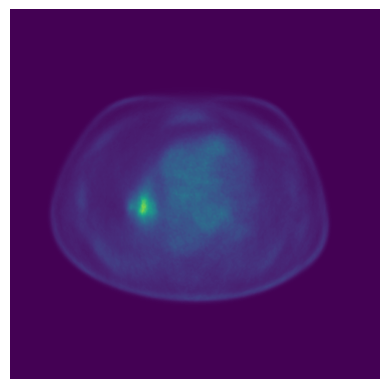}
			\put(-2,3){
				\textcolor{red}{\scriptsize PSNR=\DTLfetch{table1}{patient_no}{370}{psnr_pet_solo} }
			}
			\put(-2,86){
				\textcolor{red}{\scriptsize SSIM=\DTLfetch{table1}{patient_no}{370}{ssim_pet_solo} }
			}
		\end{overpic}
		& \begin{overpic}[width=\tempdimc,height=\tempdimc]{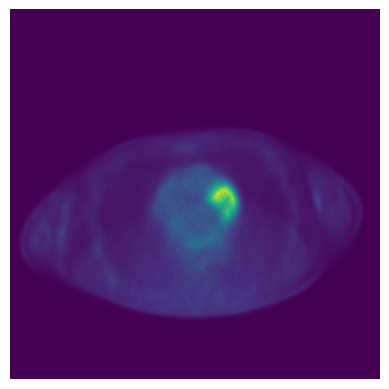}
			\put(-2,3){
				\textcolor{red}{\scriptsize PSNR=\DTLfetch{table1}{patient_no}{371}{psnr_pet_solo} }
			}
			\put(-2,86){
				\textcolor{red}{\scriptsize SSIM=\DTLfetch{table1}{patient_no}{371}{ssim_pet_solo} }
			}
		\end{overpic}
		& \begin{overpic}[width=\tempdimc,height=\tempdimc]{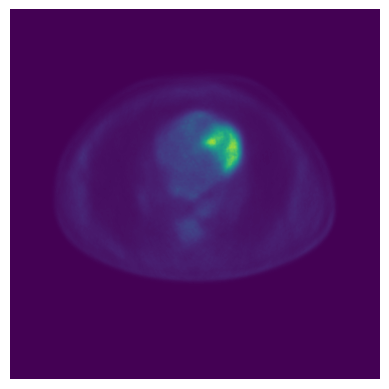}
			\put(-2,3){
				\textcolor{red}{\scriptsize PSNR=\DTLfetch{table1}{patient_no}{374}{psnr_pet_solo} }
			}
			\put(-2,86){
				\textcolor{red}{\scriptsize SSIM=\DTLfetch{table1}{patient_no}{374}{ssim_pet_solo} }
			}
		\end{overpic}
		& \begin{overpic}[width=\tempdimc,height=\tempdimc]{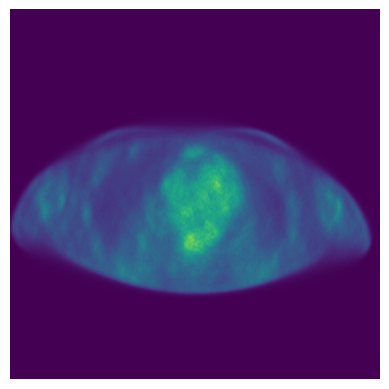}
			\put(-2,3){
				\textcolor{red}{\scriptsize PSNR=\DTLfetch{table1}{patient_no}{375}{psnr_pet_solo} }
			}
			\put(-2,86){
				\textcolor{red}{\scriptsize SSIM=\DTLfetch{table1}{patient_no}{375}{ssim_pet_solo} }
			}
		\end{overpic} \\
		\rowname{\footnotesize \acs{VCT}} & 
		\begin{overpic}[width=\tempdimc,height=\tempdimc]{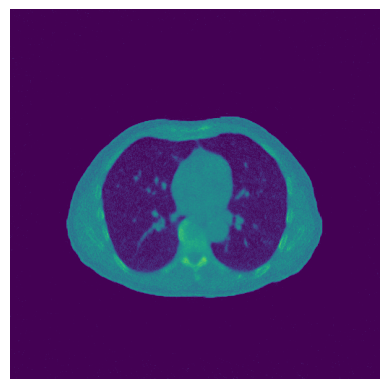}
			\put(-2,3){
				\textcolor{red}{\scriptsize PSNR=\DTLfetch{table1}{patient_no}{340}{psnr_ct_vae} }
			}
			\put(-2,86){
				\textcolor{red}{\scriptsize SSIM=\DTLfetch{table1}{patient_no}{340}{ssim_ct_vae} }
			}
		\end{overpic}
		&  
		\begin{overpic}[width=\tempdimc,height=\tempdimc]{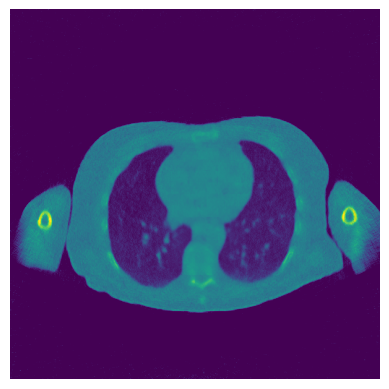}
			\put(-2,3){
				\textcolor{red}{\scriptsize PSNR=\DTLfetch{table1}{patient_no}{341}{psnr_ct_vae} }
			}
			\put(-2,86){
				\textcolor{red}{\scriptsize SSIM=\DTLfetch{table1}{patient_no}{341}{ssim_ct_vae} }
			}
		\end{overpic}
		&  \begin{overpic}[width=\tempdimc,height=\tempdimc]{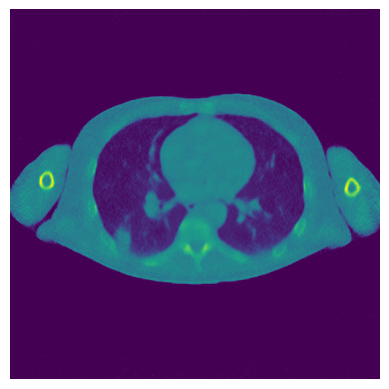}
			\put(-2,3){
				\textcolor{red}{\scriptsize PSNR=\DTLfetch{table1}{patient_no}{345}{psnr_ct_vae} }
			}
			\put(-2,86){
				\textcolor{red}{\scriptsize SSIM=\DTLfetch{table1}{patient_no}{345}{ssim_ct_vae} }
			}
		\end{overpic}
		& \begin{overpic}[width=\tempdimc,height=\tempdimc]{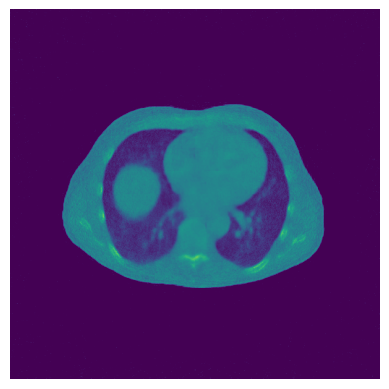}
			\put(-2,3){
				\textcolor{red}{\scriptsize PSNR=\DTLfetch{table1}{patient_no}{348}{psnr_ct_vae} }
			}
			\put(-2,86){
				\textcolor{red}{\scriptsize SSIM=\DTLfetch{table1}{patient_no}{348}{ssim_ct_vae} }
			}
		\end{overpic}
		&  \begin{overpic}[width=\tempdimc,height=\tempdimc]{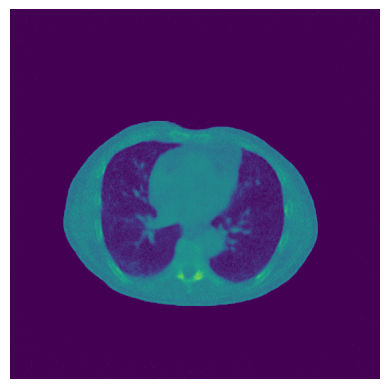}
			\put(-2,3){
				\textcolor{red}{\scriptsize PSNR=\DTLfetch{table1}{patient_no}{351}{psnr_ct_vae} }
			}
			\put(-2,86){
				\textcolor{red}{\scriptsize SSIM=\DTLfetch{table1}{patient_no}{351}{ssim_ct_vae} }
			}
		\end{overpic}
		& \begin{overpic}[width=\tempdimc,height=\tempdimc]{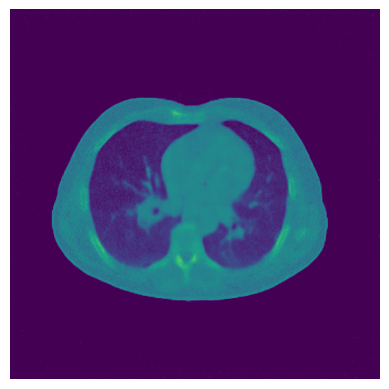}
			\put(-2,3){
				\textcolor{red}{\scriptsize PSNR=\DTLfetch{table1}{patient_no}{370}{psnr_ct_vae} }
			}
			\put(-2,86){
				\textcolor{red}{\scriptsize SSIM=\DTLfetch{table1}{patient_no}{370}{ssim_ct_vae} }
			}
		\end{overpic}
		& \begin{overpic}[width=\tempdimc,height=\tempdimc]{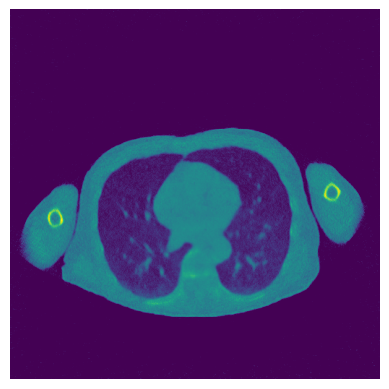}
			\put(-2,3){
				\textcolor{red}{\scriptsize PSNR=\DTLfetch{table1}{patient_no}{371}{psnr_ct_vae} }
			}
			\put(-2,86){
				\textcolor{red}{\scriptsize SSIM=\DTLfetch{table1}{patient_no}{371}{ssim_ct_vae} }
			}
		\end{overpic}
		& \begin{overpic}[width=\tempdimc,height=\tempdimc]{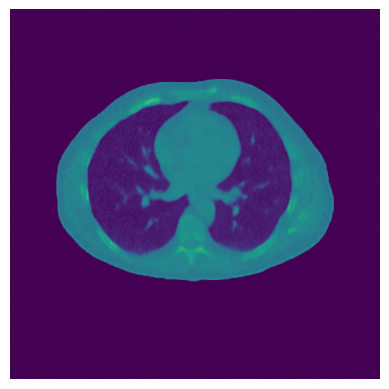}
			\put(-2,3){
				\textcolor{red}{\scriptsize PSNR=\DTLfetch{table1}{patient_no}{374}{psnr_ct_vae} }
			}
			\put(-2,86){
				\textcolor{red}{\scriptsize SSIM=\DTLfetch{table1}{patient_no}{374}{ssim_ct_vae} }
			}
		\end{overpic}
		& \begin{overpic}[width=\tempdimc,height=\tempdimc]{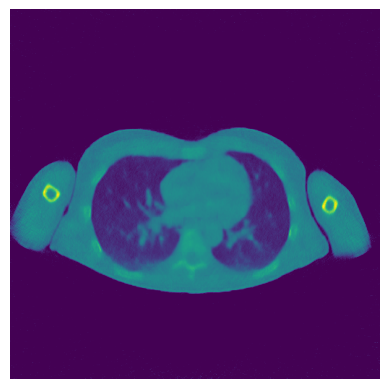}
			\put(-2,3){
				\textcolor{red}{\scriptsize PSNR=\DTLfetch{table1}{patient_no}{375}{psnr_ct_vae} }
			}
			\put(-2,86){
				\textcolor{red}{\scriptsize SSIM=\DTLfetch{table1}{patient_no}{375}{ssim_ct_vae} }
			}
		\end{overpic}
		\\
		\rowname{\footnotesize \acs{PCT}}  &  
		\begin{overpic}[width=\tempdimc,height=\tempdimc]{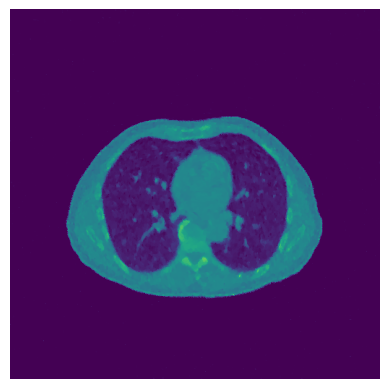}
			\put(-2,3){
				\textcolor{red}{\scriptsize PSNR=\DTLfetch{table1}{patient_no}{340}{psnr_ct_mat} }
			}
			\put(-2,86){
				\textcolor{red}{\scriptsize SSIM=\DTLfetch{table1}{patient_no}{340}{ssim_ct_mat} }
			}
		\end{overpic}
		&  
		\begin{overpic}[width=\tempdimc,height=\tempdimc]{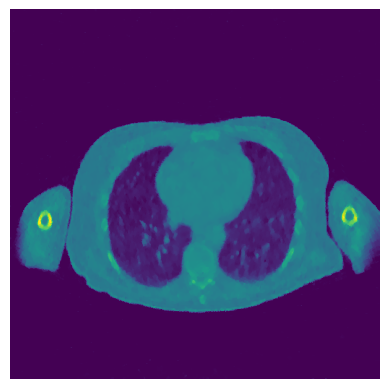}
			\put(-2,3){
				\textcolor{red}{\scriptsize PSNR=\DTLfetch{table1}{patient_no}{341}{psnr_ct_mat} }
			}
			\put(-2,86){
				\textcolor{red}{\scriptsize SSIM=\DTLfetch{table1}{patient_no}{341}{ssim_ct_mat} }
			}
		\end{overpic}
		&  \begin{overpic}[width=\tempdimc,height=\tempdimc]{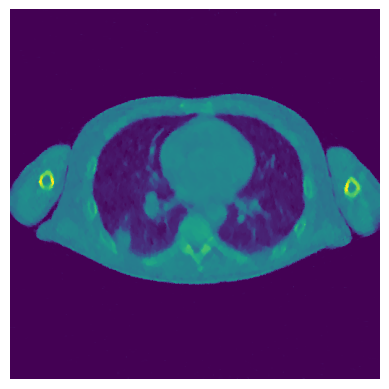}
			\put(-2,3){
				\textcolor{red}{\scriptsize PSNR=\DTLfetch{table1}{patient_no}{345}{psnr_ct_mat} }
			}
			\put(-2,86){
				\textcolor{red}{\scriptsize SSIM=\DTLfetch{table1}{patient_no}{345}{ssim_ct_mat} }
			}
		\end{overpic}
		& \begin{overpic}[width=\tempdimc,height=\tempdimc]{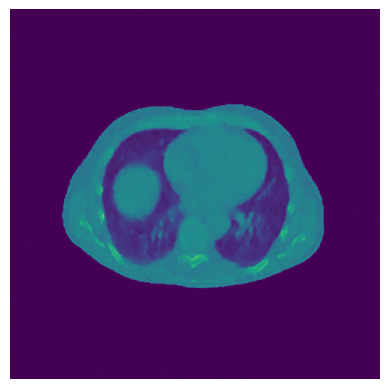}
			\put(-2,3){
				\textcolor{red}{\scriptsize PSNR=\DTLfetch{table1}{patient_no}{348}{psnr_ct_mat} }
			}
			\put(-2,86){
				\textcolor{red}{\scriptsize SSIM=\DTLfetch{table1}{patient_no}{348}{ssim_ct_mat} }
			}
		\end{overpic}
		&  \begin{overpic}[width=\tempdimc,height=\tempdimc]{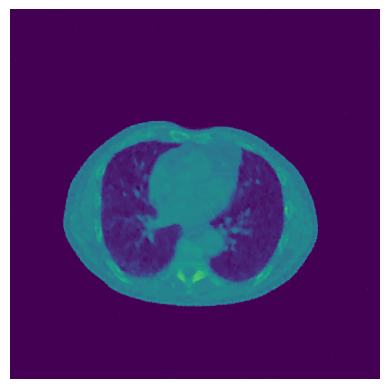}
			\put(-2,3){
				\textcolor{red}{\scriptsize PSNR=\DTLfetch{table1}{patient_no}{351}{psnr_ct_mat} }
			}
			\put(-2,86){
				\textcolor{red}{\scriptsize SSIM=\DTLfetch{table1}{patient_no}{351}{ssim_ct_mat} }
			}
		\end{overpic}
		& \begin{overpic}[width=\tempdimc,height=\tempdimc]{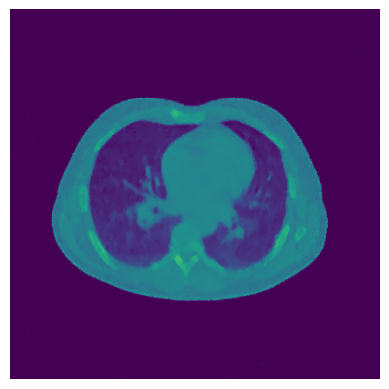}
			\put(-2,3){
				\textcolor{red}{\scriptsize PSNR=\DTLfetch{table1}{patient_no}{370}{psnr_ct_mat} }
			}
			\put(-2,86){
				\textcolor{red}{\scriptsize SSIM=\DTLfetch{table1}{patient_no}{370}{ssim_ct_mat} }
			}
		\end{overpic}
		& \begin{overpic}[width=\tempdimc,height=\tempdimc]{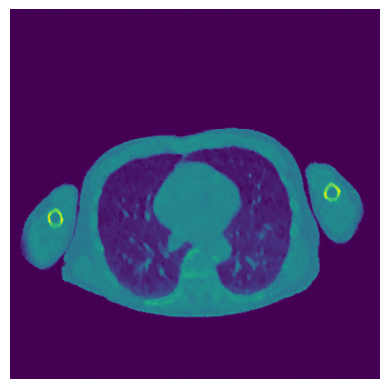}
			\put(-2,3){
				\textcolor{red}{\scriptsize PSNR=\DTLfetch{table1}{patient_no}{371}{psnr_ct_mat} }
			}
			\put(-2,86){
				\textcolor{red}{\scriptsize SSIM=\DTLfetch{table1}{patient_no}{371}{ssim_ct_mat} }
			}
		\end{overpic}
		& \begin{overpic}[width=\tempdimc,height=\tempdimc]{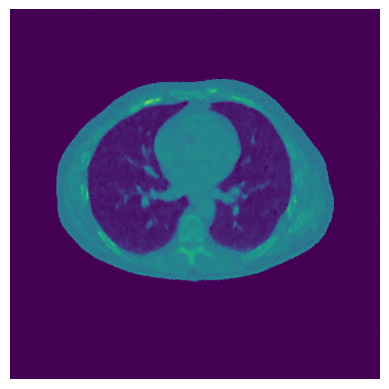}
			\put(-2,3){
				\textcolor{red}{\scriptsize PSNR=\DTLfetch{table1}{patient_no}{374}{psnr_ct_mat} }
			}
			\put(-2,86){
				\textcolor{red}{\scriptsize SSIM=\DTLfetch{table1}{patient_no}{374}{ssim_ct_mat} }
			}
		\end{overpic}
		& \begin{overpic}[width=\tempdimc,height=\tempdimc]{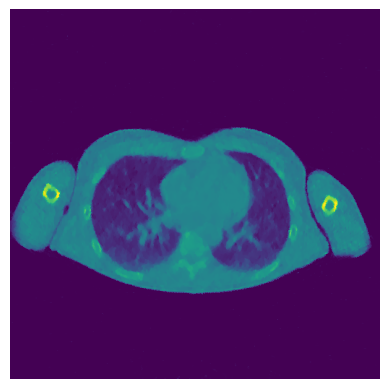}
			\put(-2,3){
				\textcolor{red}{\scriptsize PSNR=\DTLfetch{table1}{patient_no}{375}{psnr_ct_mat} }
			}
			\put(-2,86){
				\textcolor{red}{\scriptsize SSIM=\DTLfetch{table1}{patient_no}{375}{ssim_ct_mat} }
			}
		\end{overpic} \\
		\rowname{\footnotesize \acs{CTunet}}  
		&  \begin{overpic}[width=\tempdimc,height=\tempdimc]{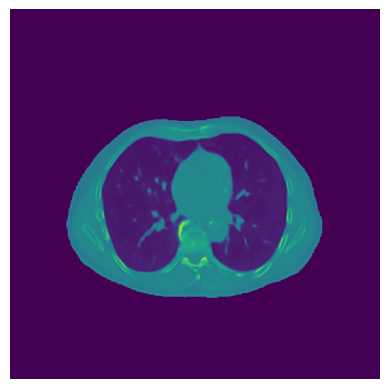}
			\put(-2,3){
				\textcolor{red}{\scriptsize PSNR=\DTLfetch{table3}{patient_no}{340}{psnr_ct} }
			}
			\put(-2,86){
				\textcolor{red}{\scriptsize SSIM=\DTLfetch{table3}{patient_no}{340}{ssim_ct} }
			}
		\end{overpic}
		&  
		\begin{overpic}[width=\tempdimc,height=\tempdimc]{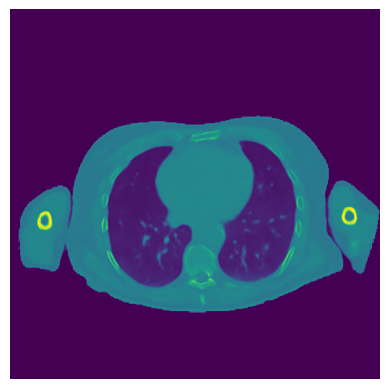}
			\put(-2,3){
				\textcolor{red}{\scriptsize PSNR=\DTLfetch{table3}{patient_no}{341}{psnr_ct} }
			}
			\put(-2,86){
				\textcolor{red}{\scriptsize SSIM=\DTLfetch{table3}{patient_no}{341}{ssim_ct} }
			}
		\end{overpic}
		&  \begin{overpic}[width=\tempdimc,height=\tempdimc]{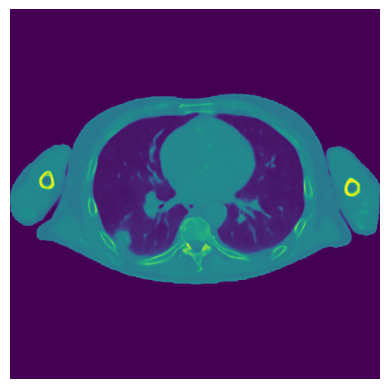}
			\put(-2,3){
				\textcolor{red}{\scriptsize PSNR=\DTLfetch{table3}{patient_no}{345}{psnr_ct} }
			}
			\put(-2,86){
				\textcolor{red}{\scriptsize SSIM=\DTLfetch{table3}{patient_no}{345}{ssim_ct} }
			}
		\end{overpic}
		& \begin{overpic}[width=\tempdimc,height=\tempdimc]{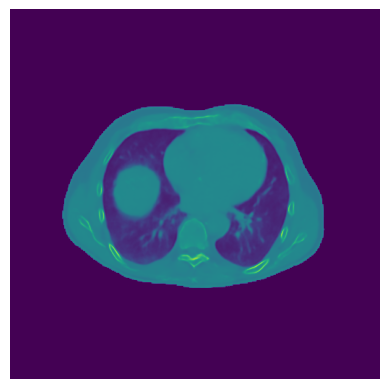}
			\put(-2,3){
				\textcolor{red}{\scriptsize PSNR=\DTLfetch{table3}{patient_no}{348}{psnr_ct} }
			}
			\put(-2,86){
				\textcolor{red}{\scriptsize SSIM=\DTLfetch{table3}{patient_no}{348}{ssim_ct} }
			}
		\end{overpic}
		&  \begin{overpic}[width=\tempdimc,height=\tempdimc]{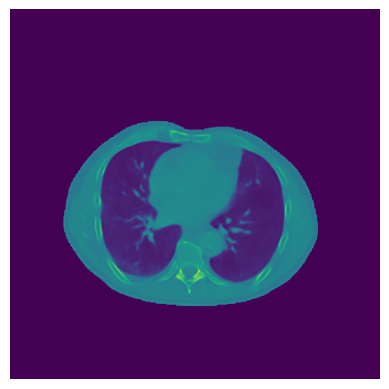}
			\put(-2,3){
				\textcolor{red}{\scriptsize PSNR=\DTLfetch{table3}{patient_no}{351}{psnr_ct} }
			}
			\put(-2,86){
				\textcolor{red}{\scriptsize SSIM=\DTLfetch{table3}{patient_no}{351}{ssim_ct} }
			}
		\end{overpic}
		& \begin{overpic}[width=\tempdimc,height=\tempdimc]{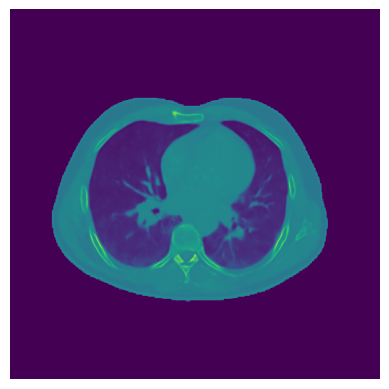}
			\put(-2,3){
				\textcolor{red}{\scriptsize PSNR=\DTLfetch{table3}{patient_no}{370}{psnr_ct} }
			}
			\put(-2,86){
				\textcolor{red}{\scriptsize SSIM=\DTLfetch{table3}{patient_no}{370}{ssim_ct} }
			}
		\end{overpic}
		& \begin{overpic}[width=\tempdimc,height=\tempdimc]{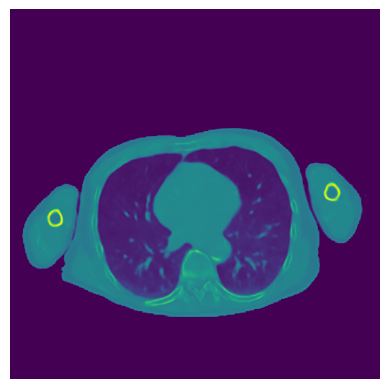}
			\put(-2,3){
				\textcolor{red}{\scriptsize PSNR=\DTLfetch{table3}{patient_no}{371}{psnr_ct} }
			}
			\put(-2,86){
				\textcolor{red}{\scriptsize SSIM=\DTLfetch{table3}{patient_no}{371}{ssim_ct} }
			}
		\end{overpic}
		& \begin{overpic}[width=\tempdimc,height=\tempdimc]{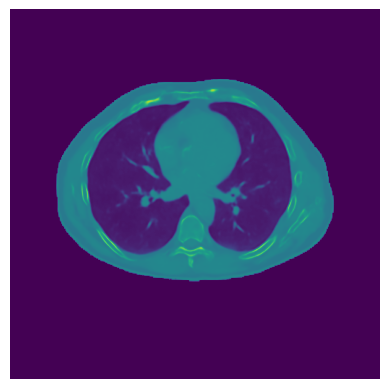}
			\put(-2,3){
				\textcolor{red}{\scriptsize PSNR=\DTLfetch{table3}{patient_no}{374}{psnr_ct} }
			}
			\put(-2,86){
				\textcolor{red}{\scriptsize SSIM=\DTLfetch{table3}{patient_no}{374}{ssim_ct} }
			}
		\end{overpic}
		& \begin{overpic}[width=\tempdimc,height=\tempdimc]{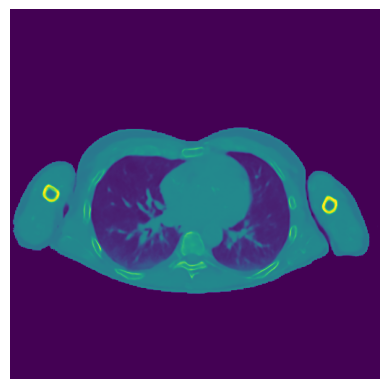}
			\put(-2,3){
				\textcolor{red}{\scriptsize PSNR=\DTLfetch{table3}{patient_no}{375}{psnr_ct} }
			}
			\put(-2,86){
				\textcolor{red}{\scriptsize SSIM=\DTLfetch{table3}{patient_no}{375}{ssim_ct} }
			}
		\end{overpic} \\
		\rowname{\footnotesize \acs{WCT}}  & 
		\begin{overpic}[width=\tempdimc,height=\tempdimc]{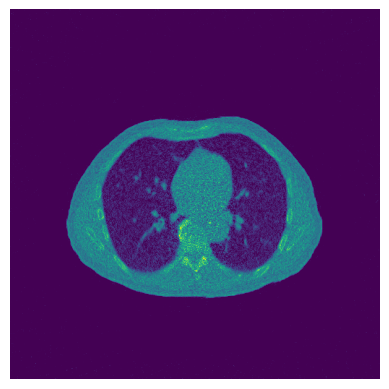}
			\put(-2,3){
				\textcolor{red}{\scriptsize PSNR=\DTLfetch{table1}{patient_no}{340}{psnr_ct_solo} }
			}
			\put(-2,86){
				\textcolor{red}{\scriptsize SSIM=\DTLfetch{table1}{patient_no}{340}{ssim_ct_solo} }
			}
		\end{overpic}
		&  
		\begin{overpic}[width=\tempdimc,height=\tempdimc]{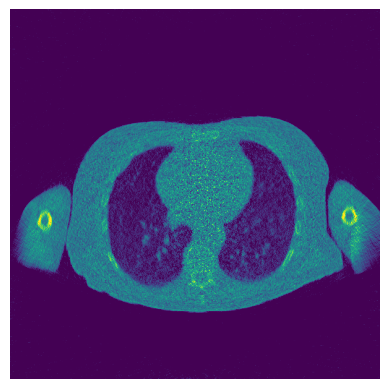}
			\put(-2,3){
				\textcolor{red}{\scriptsize PSNR=\DTLfetch{table1}{patient_no}{341}{psnr_ct_solo}}
			}
			\put(-2,86){
				\textcolor{red}{\scriptsize SSIM=\DTLfetch{table1}{patient_no}{341}{ssim_ct_solo} }
			}
		\end{overpic}
		&  \begin{overpic}[width=\tempdimc,height=\tempdimc]{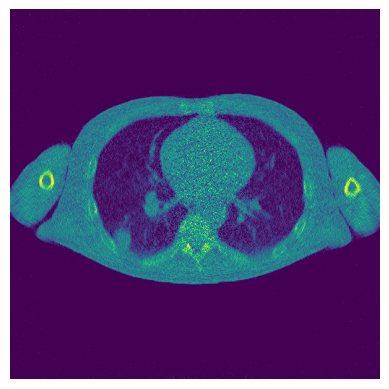}
			\put(-2,3){
				\textcolor{red}{\scriptsize PSNR=\DTLfetch{table1}{patient_no}{345}{psnr_ct_solo} }
			}
			\put(-2,86){
				\textcolor{red}{\scriptsize SSIM=\DTLfetch{table1}{patient_no}{345}{ssim_ct_solo} }
			}
		\end{overpic}
		& \begin{overpic}[width=\tempdimc,height=\tempdimc]{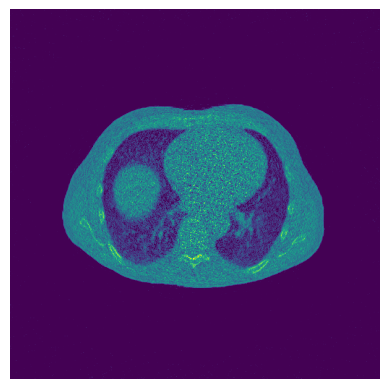}
			\put(-2,3){
				\textcolor{red}{\scriptsize PSNR=\DTLfetch{table1}{patient_no}{348}{psnr_ct_solo} }
			}
			\put(-2,86){
				\textcolor{red}{\scriptsize SSIM=\DTLfetch{table1}{patient_no}{348}{ssim_ct_solo} }
			}
		\end{overpic}
		&  \begin{overpic}[width=\tempdimc,height=\tempdimc]{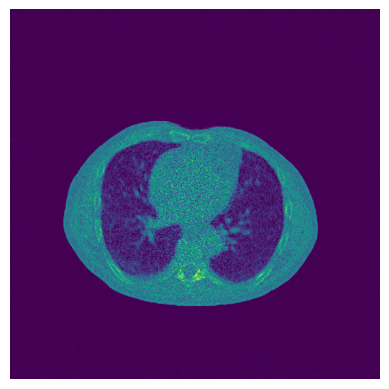}
			\put(-2,3){
				\textcolor{red}{\scriptsize PSNR=\DTLfetch{table1}{patient_no}{351}{psnr_ct_solo} }
			}
			\put(-2,86){
				\textcolor{red}{\scriptsize SSIM=\DTLfetch{table1}{patient_no}{351}{ssim_ct_solo} }
			}
		\end{overpic}
		& \begin{overpic}[width=\tempdimc,height=\tempdimc]{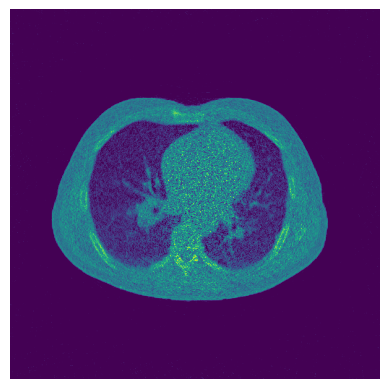}
			\put(-2,3){
				\textcolor{red}{\scriptsize PSNR=\DTLfetch{table1}{patient_no}{370}{psnr_ct_solo} }
			}
			\put(-2,86){
				\textcolor{red}{\scriptsize SSIM=\DTLfetch{table1}{patient_no}{370}{ssim_ct_solo} }
			}
		\end{overpic}
		& \begin{overpic}[width=\tempdimc,height=\tempdimc]{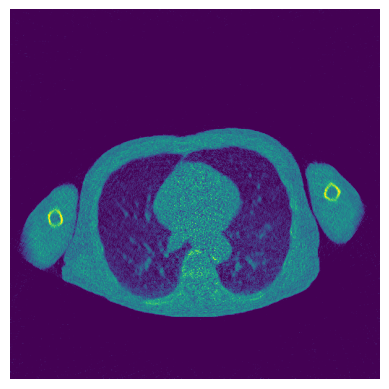}
			\put(-2,3){
				\textcolor{red}{\scriptsize PSNR=\DTLfetch{table1}{patient_no}{371}{psnr_ct_solo} }
			}
			\put(-2,86){
				\textcolor{red}{\scriptsize SSIM=\DTLfetch{table1}{patient_no}{371}{ssim_ct_solo} }
			}
		\end{overpic}
		& \begin{overpic}[width=\tempdimc,height=\tempdimc]{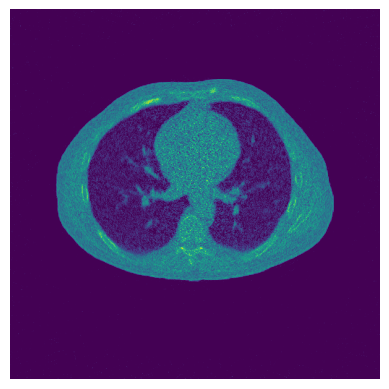}
			\put(-2,3){
				\textcolor{red}{\scriptsize PSNR=\DTLfetch{table1}{patient_no}{374}{psnr_ct_solo} }
			}
			\put(-2,86){
				\textcolor{red}{\scriptsize SSIM=\DTLfetch{table1}{patient_no}{374}{ssim_ct_solo} }
			}
		\end{overpic}
		& \begin{overpic}[width=\tempdimc,height=\tempdimc]{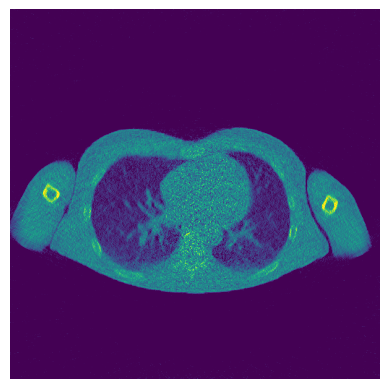}
			\put(-2,3){
				\textcolor{red}{\scriptsize PSNR=\DTLfetch{table1}{patient_no}{375}{psnr_ct_solo} }
			}
			\put(-2,86){
				\textcolor{red}{\scriptsize SSIM=\DTLfetch{table1}{patient_no}{375}{ssim_ct_solo} }
			}
		\end{overpic}

	\end{tabular}
	\caption{\Ac{HL}---Reconstructed images of the nine other patients.}\label{fig:bigfig1}
\end{figure*}

%% file: content/big_fig2.tex
\begin{figure*}
	\centering
	\begin{tabular}{p{0.0\tempdimc}p{.8\tempdimc}p{0.8\tempdimc}p{0.8\tempdimc}p{0.8\tempdimc}p{0.8\tempdimc}p{0.8\tempdimc}p{0.8\tempdimc}p{0.8\tempdimc}p{0.8\tempdimc}}
		& \hfill \footnotesize Patient 2 &  \hfill\footnotesize Patient 3 &  \hfill\footnotesize Patient 4 &  \hfill\footnotesize Patient 5 &  \hfill \footnotesize Patient 6 &   \hfill\footnotesize Patient 7 &   \hfill\footnotesize Patient 8 &  \hfill \footnotesize Patient 9 & \hfill \footnotesize Patient 10  \\ 
		\rowname{\footnotesize \acs{VPET}} 
		& 
		\begin{overpic}[width=\tempdimc,height=\tempdimc]{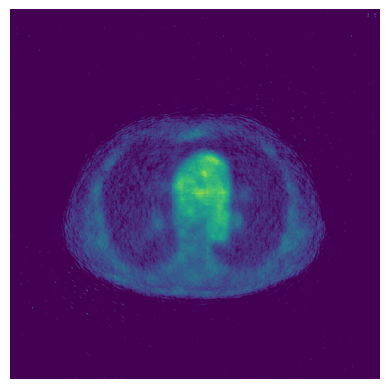}
			\put(-2,3){
				\textcolor{red}{\scriptsize PSNR={\DTLfetch{table4}{patient_no}{340}{psnr_pet_vae}} }
			}
			\put(-2,86){
				\textcolor{red}{\scriptsize SSIM={\DTLfetch{table4}{patient_no}{340}{ssim_pet_vae}} }
			}
		\end{overpic}
		&  
		\begin{overpic}[width=\tempdimc,height=\tempdimc]{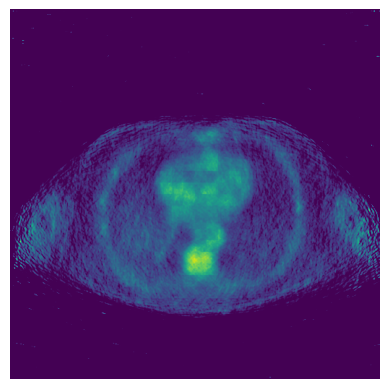}
			\put(-2,3){
				\textcolor{red}{\scriptsize PSNR={\DTLfetch{table4}{patient_no}{341}{psnr_pet_vae}} }
			}
			\put(-2,86){
				\textcolor{red}{\scriptsize SSIM={\DTLfetch{table4}{patient_no}{341}{ssim_pet_vae}} }
			}
		\end{overpic}
		&  \begin{overpic}[width=\tempdimc,height=\tempdimc]{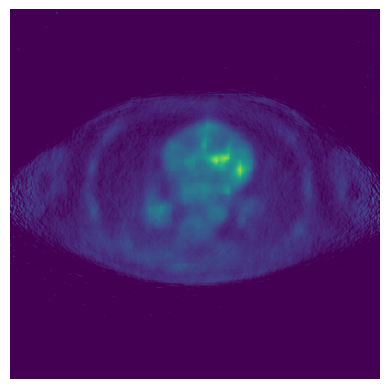}
			\put(-2,3){
				\textcolor{red}{\scriptsize PSNR={\DTLfetch{table4}{patient_no}{345}{psnr_pet_vae}} }
			}
			\put(-2,86){
				\textcolor{red}{\scriptsize SSIM={\DTLfetch{table4}{patient_no}{345}{ssim_pet_vae}} }
			}
		\end{overpic}
		& \begin{overpic}[width=\tempdimc,height=\tempdimc]{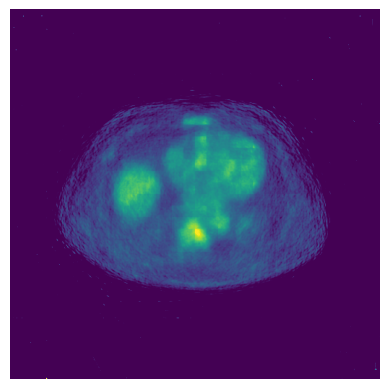}
			\put(-2,3){
				\textcolor{red}{\scriptsize PSNR={\DTLfetch{table4}{patient_no}{348}{psnr_pet_vae}} }
			}
			\put(-2,86){
				\textcolor{red}{\scriptsize SSIM={\DTLfetch{table4}{patient_no}{348}{ssim_pet_vae}} }
			}
		\end{overpic}
		&  \begin{overpic}[width=\tempdimc,height=\tempdimc]{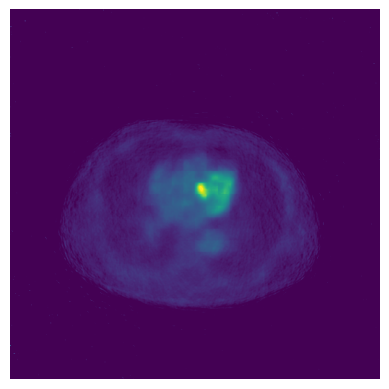}
			\put(-2,3){
				\textcolor{red}{\scriptsize PSNR={\DTLfetch{table4}{patient_no}{351}{psnr_pet_vae}} }
			}
			\put(-2,86){
				\textcolor{red}{\scriptsize SSIM={\DTLfetch{table4}{patient_no}{351}{ssim_pet_vae}} }
			}
		\end{overpic}
		& \begin{overpic}[width=\tempdimc,height=\tempdimc]{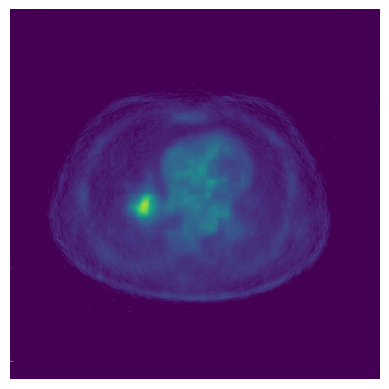}
			\put(-2,3){
				\textcolor{red}{\scriptsize PSNR={\DTLfetch{table4}{patient_no}{370}{psnr_pet_vae}} }
			}
			\put(-2,86){
				\textcolor{red}{\scriptsize SSIM={\DTLfetch{table4}{patient_no}{370}{ssim_pet_vae}} }
			}
		\end{overpic}
		& \begin{overpic}[width=\tempdimc,height=\tempdimc]{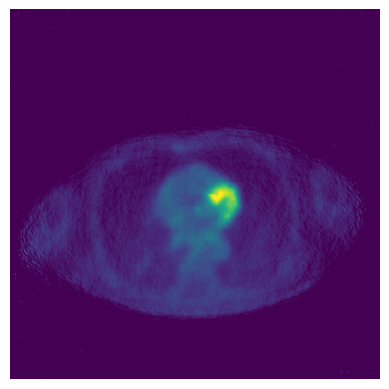}
			\put(-2,3){
				\textcolor{red}{\scriptsize PSNR={\DTLfetch{table4}{patient_no}{371}{psnr_pet_vae}} }
			}
			\put(-2,86){
				\textcolor{red}{\scriptsize SSIM={\DTLfetch{table4}{patient_no}{371}{ssim_pet_vae}} }
			}
		\end{overpic}
		& \begin{overpic}[width=\tempdimc,height=\tempdimc]{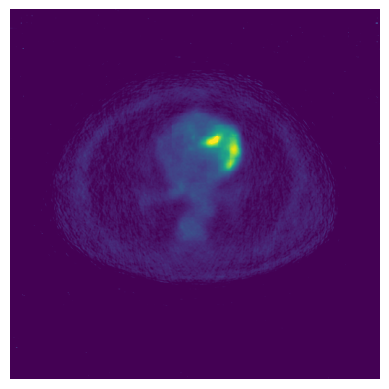}
			\put(-2,3){
				\textcolor{red}{\scriptsize PSNR={\DTLfetch{table4}{patient_no}{374}{psnr_pet_vae}} }
			}
			\put(-2,86){
				\textcolor{red}{\scriptsize SSIM={\DTLfetch{table4}{patient_no}{374}{ssim_pet_vae}} }
			}
		\end{overpic}
		& \begin{overpic}[width=\tempdimc,height=\tempdimc]{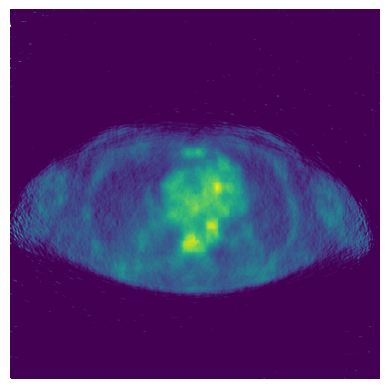}
			\put(-2,3){
				\textcolor{red}{\scriptsize PSNR={\DTLfetch{table4}{patient_no}{375}{psnr_pet_vae}} }
			}
			\put(-2,86){
				\textcolor{red}{\scriptsize SSIM={\DTLfetch{table4}{patient_no}{375}{ssim_pet_vae}} }
			}
		\end{overpic}
		\\
		\rowname{\footnotesize \acs{PPET}}  &  \begin{overpic}[width=\tempdimc,height=\tempdimc]{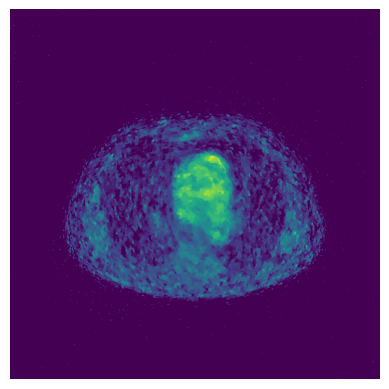}
			\put(-2,3){
				\textcolor{red}{\scriptsize PSNR={\DTLfetch{table4}{patient_no}{340}{psnr_pet_mat}} }
			}
			\put(-2,86){
				\textcolor{red}{\scriptsize SSIM={\DTLfetch{table4}{patient_no}{340}{ssim_pet_mat}} }
			}
		\end{overpic}
		&  
		\begin{overpic}[width=\tempdimc,height=\tempdimc]{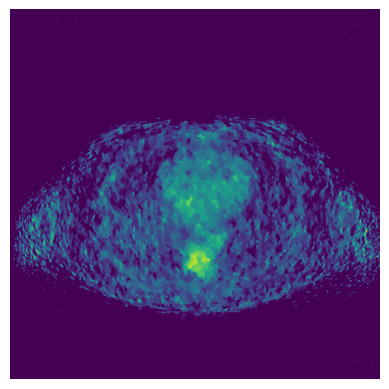}
			\put(-2,3){
				\textcolor{red}{\scriptsize PSNR={\DTLfetch{table4}{patient_no}{341}{psnr_pet_mat}} }
			}
			\put(-2,86){
				\textcolor{red}{\scriptsize SSIM={\DTLfetch{table4}{patient_no}{341}{ssim_pet_mat}} }
			}
		\end{overpic}
		&  \begin{overpic}[width=\tempdimc,height=\tempdimc]{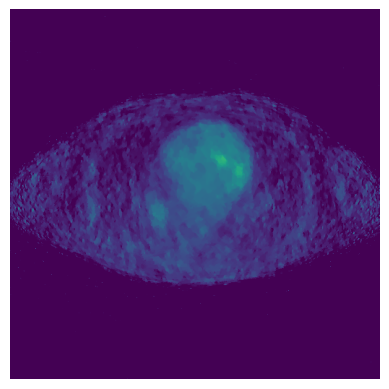}
			\put(-2,3){
				\textcolor{red}{\scriptsize PSNR={\DTLfetch{table4}{patient_no}{345}{psnr_pet_mat}} }
			}
			\put(-2,86){
				\textcolor{red}{\scriptsize SSIM={\DTLfetch{table4}{patient_no}{345}{ssim_pet_mat}} }
			}
		\end{overpic}
		& \begin{overpic}[width=\tempdimc,height=\tempdimc]{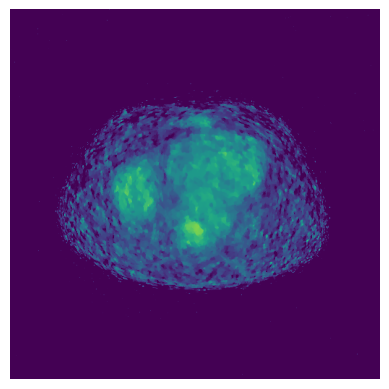}
			\put(-2,3){
				\textcolor{red}{\scriptsize PSNR={\DTLfetch{table4}{patient_no}{348}{psnr_pet_mat}} }
			}
			\put(-2,86){
				\textcolor{red}{\scriptsize SSIM={\DTLfetch{table4}{patient_no}{348}{ssim_pet_mat}} }
			}
		\end{overpic}
		&  \begin{overpic}[width=\tempdimc,height=\tempdimc]{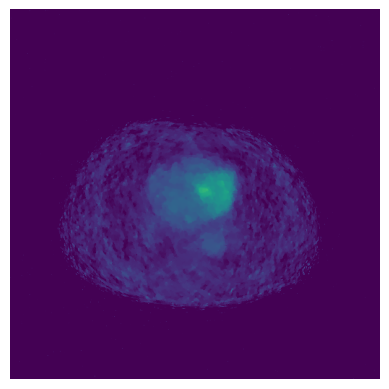}
			\put(-2,3){
				\textcolor{red}{\scriptsize PSNR={\DTLfetch{table4}{patient_no}{351}{psnr_pet_mat}} }
			}
			\put(-2,86){
				\textcolor{red}{\scriptsize SSIM={\DTLfetch{table4}{patient_no}{351}{ssim_pet_mat}} }
			}
		\end{overpic}
		& \begin{overpic}[width=\tempdimc,height=\tempdimc]{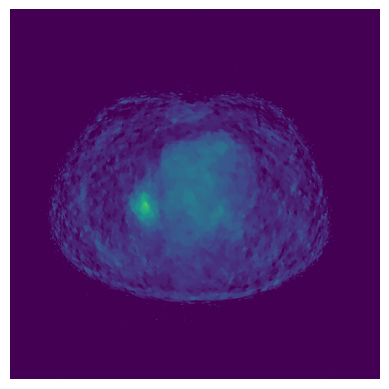}
			\put(-2,3){
				\textcolor{red}{\scriptsize PSNR={\DTLfetch{table4}{patient_no}{370}{psnr_pet_mat}} }
			}
			\put(-2,86){
				\textcolor{red}{\scriptsize SSIM={\DTLfetch{table4}{patient_no}{370}{ssim_pet_mat}} }
			}
		\end{overpic}
		& \begin{overpic}[width=\tempdimc,height=\tempdimc]{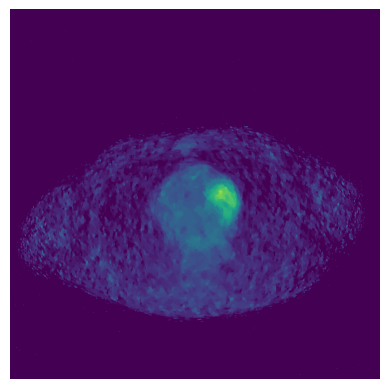}
			\put(-2,3){
				\textcolor{red}{\scriptsize PSNR={\DTLfetch{table4}{patient_no}{371}{psnr_pet_mat}} }
			}
			\put(-2,86){
				\textcolor{red}{\scriptsize SSIM={\DTLfetch{table4}{patient_no}{371}{ssim_pet_mat}} }
			}
		\end{overpic}
		& \begin{overpic}[width=\tempdimc,height=\tempdimc]{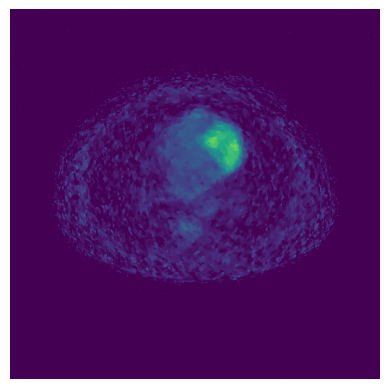}
			\put(-2,3){
				\textcolor{red}{\scriptsize PSNR={\DTLfetch{table4}{patient_no}{374}{psnr_pet_mat}} }
			}
			\put(-2,86){
				\textcolor{red}{\scriptsize SSIM={\DTLfetch{table4}{patient_no}{374}{ssim_pet_mat}} }
			}
		\end{overpic}
		& \begin{overpic}[width=\tempdimc,height=\tempdimc]{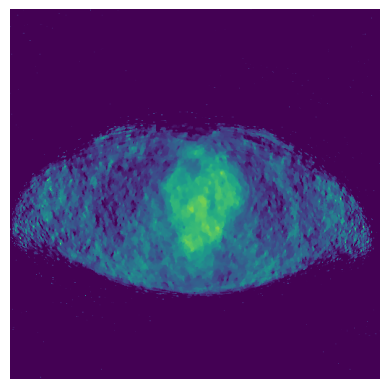}
			\put(-2,3){
				\textcolor{red}{\scriptsize PSNR={\DTLfetch{table4}{patient_no}{375}{psnr_pet_mat}} }
			}
			\put(-2,86){
				\textcolor{red}{\scriptsize SSIM={\DTLfetch{table4}{patient_no}{375}{ssim_pet_mat}} }
			}
		\end{overpic} \\
		\rowname{\footnotesize \acs{PETunet}}  
		&  \begin{overpic}[width=\tempdimc,height=\tempdimc]{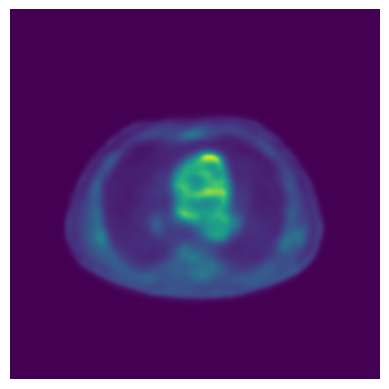}
			\put(-2,3){
				\textcolor{red}{\scriptsize PSNR=\DTLfetch{table3}{patient_no}{340}{psnr_pet} }
			}
			\put(-2,86){
				\textcolor{red}{\scriptsize SSIM=\DTLfetch{table3}{patient_no}{340}{ssim_pet}
				}
			}
		\end{overpic}
		&  
		\begin{overpic}[width=\tempdimc,height=\tempdimc]{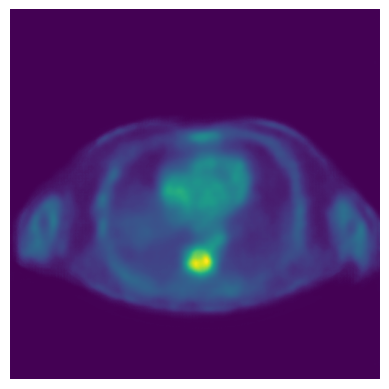}
			\put(-2,3){
				\textcolor{red}{\scriptsize PSNR=\DTLfetch{table3}{patient_no}{341}{psnr_pet} }
			}
			\put(-2,86){
				\textcolor{red}{\scriptsize SSIM=\DTLfetch{table3}{patient_no}{341}{ssim_pet} }
			}
		\end{overpic}
		&  \begin{overpic}[width=\tempdimc,height=\tempdimc]{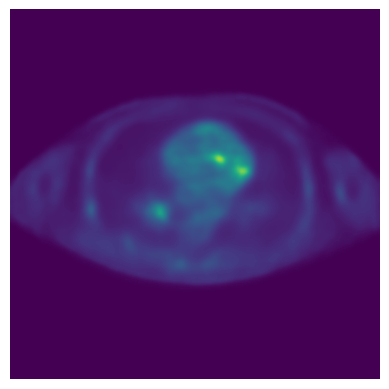}
			\put(-2,3){
				\textcolor{red}{\scriptsize PSNR=\DTLfetch{table3}{patient_no}{345}{psnr_pet} }
			}
			\put(-2,86){
				\textcolor{red}{\scriptsize SSIM=\DTLfetch{table3}{patient_no}{345}{ssim_pet} }
			}
		\end{overpic}
		& \begin{overpic}[width=\tempdimc,height=\tempdimc]{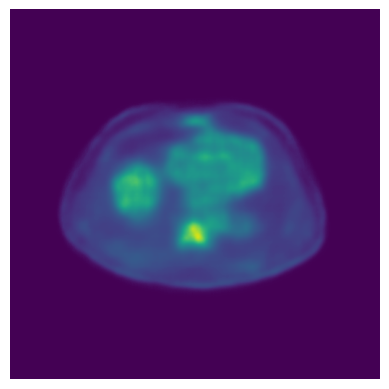}
			\put(-2,3){
				\textcolor{red}{\scriptsize PSNR=\DTLfetch{table3}{patient_no}{348}{psnr_pet} }
			}
			\put(-2,86){
				\textcolor{red}{\scriptsize SSIM=\DTLfetch{table3}{patient_no}{348}{ssim_pet} }
			}
		\end{overpic}
		&  \begin{overpic}[width=\tempdimc,height=\tempdimc]{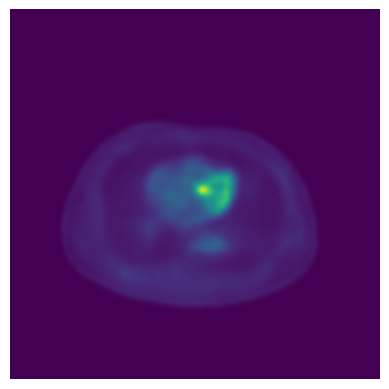}
			\put(-2,3){
				\textcolor{red}{\scriptsize PSNR=\DTLfetch{table3}{patient_no}{351}{psnr_pet} }
			}
			\put(-2,86){
				\textcolor{red}{\scriptsize SSIM=\DTLfetch{table3}{patient_no}{351}{ssim_pet} }
			}
		\end{overpic}
		& \begin{overpic}[width=\tempdimc,height=\tempdimc]{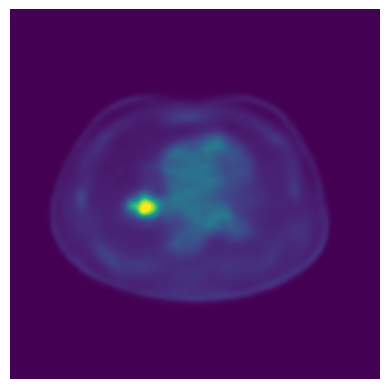}
			\put(-2,3){
				\textcolor{red}{\scriptsize PSNR=\DTLfetch{table3}{patient_no}{370}{psnr_pet} }
			}
			\put(-2,86){
				\textcolor{red}{\scriptsize SSIM=\DTLfetch{table3}{patient_no}{370}{ssim_pet} }
			}
		\end{overpic}
		& \begin{overpic}[width=\tempdimc,height=\tempdimc]{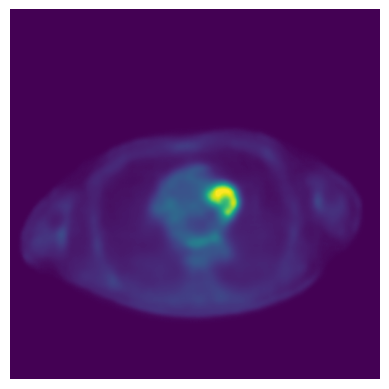}
			\put(-2,3){
				\textcolor{red}{\scriptsize PSNR=\DTLfetch{table3}{patient_no}{371}{psnr_pet} }
			}
			\put(-2,86){
				\textcolor{red}{\scriptsize SSIM=\DTLfetch{table3}{patient_no}{371}{ssim_pet} }
			}
		\end{overpic}
		& \begin{overpic}[width=\tempdimc,height=\tempdimc]{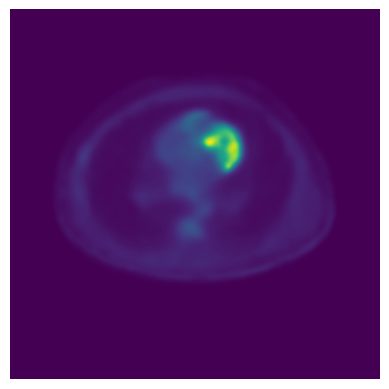}
			\put(-2,3){
				\textcolor{red}{\scriptsize PSNR=\DTLfetch{table3}{patient_no}{374}{psnr_pet} }
			}
			\put(-2,86){
				\textcolor{red}{\scriptsize SSIM=\DTLfetch{table3}{patient_no}{374}{ssim_pet} }
			}
		\end{overpic}
		& \begin{overpic}[width=\tempdimc,height=\tempdimc]{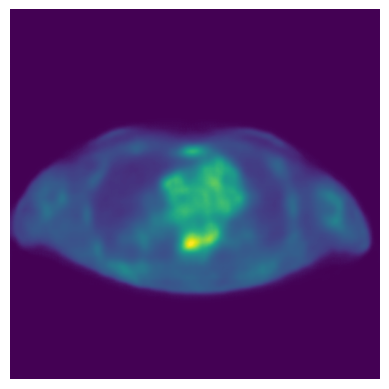}
			\put(-2,3){
				\textcolor{red}{\scriptsize PSNR=\DTLfetch{table3}{patient_no}{375}{psnr_pet} }
			}
			\put(-2,86){
				\textcolor{red}{\scriptsize SSIM=\DTLfetch{table3}{patient_no}{375}{ssim_pet} }
			}
		\end{overpic} \\
		\rowname{\footnotesize \acs{EMPET}}  & 
		\begin{overpic}[width=\tempdimc,height=\tempdimc]{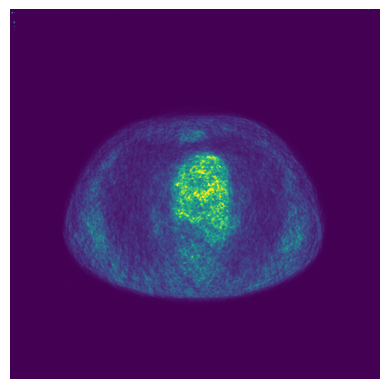}
			\put(-2,3){
				\textcolor{red}{\scriptsize PSNR={\DTLfetch{table4}{patient_no}{340}{psnr_pet_solo}} }
			}
			\put(-2,86){
				\textcolor{red}{\scriptsize SSIM={\DTLfetch{table4}{patient_no}{340}{ssim_pet_solo}} }
			}
		\end{overpic}
		&  
		\begin{overpic}[width=\tempdimc,height=\tempdimc]{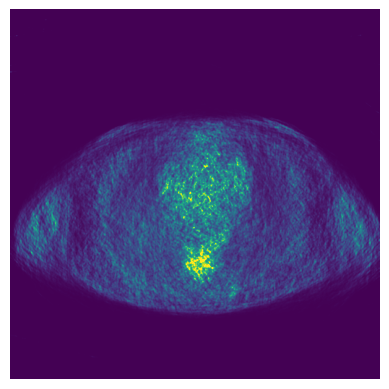}
			\put(-2,3){
				\textcolor{red}{\scriptsize PSNR={\DTLfetch{table4}{patient_no}{341}{psnr_pet_solo}} }
			}
			\put(-2,86){
				\textcolor{red}{\scriptsize SSIM={\DTLfetch{table4}{patient_no}{341}{ssim_pet_solo}} }
			}
		\end{overpic}
		&  \begin{overpic}[width=\tempdimc,height=\tempdimc]{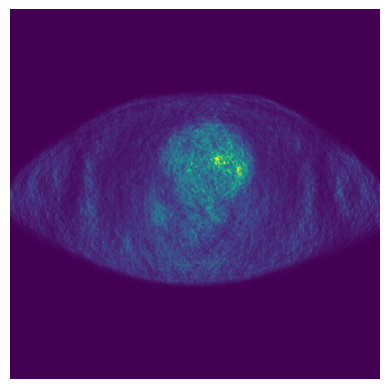}
			\put(-2,3){
				\textcolor{red}{\scriptsize PSNR={\DTLfetch{table4}{patient_no}{345}{psnr_pet_solo}} }
			}
			\put(-2,86){
				\textcolor{red}{\scriptsize SSIM={\DTLfetch{table4}{patient_no}{345}{ssim_pet_solo}} }
			}
		\end{overpic}
		& \begin{overpic}[width=\tempdimc,height=\tempdimc]{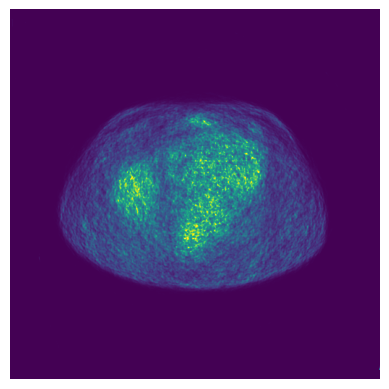}
			\put(-2,3){
				\textcolor{red}{\scriptsize PSNR={\DTLfetch{table4}{patient_no}{348}{psnr_pet_solo}} }
			}
			\put(-2,86){
				\textcolor{red}{\scriptsize SSIM={\DTLfetch{table4}{patient_no}{348}{ssim_pet_solo}} }
			}
		\end{overpic}
		&  \begin{overpic}[width=\tempdimc,height=\tempdimc]{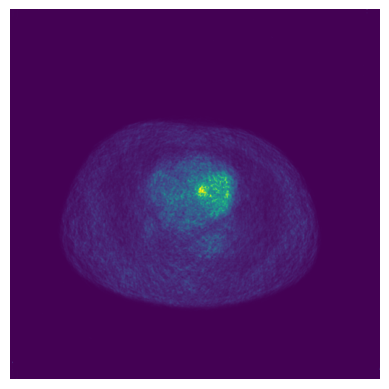}
			\put(-2,3){
				\textcolor{red}{\scriptsize PSNR={\DTLfetch{table4}{patient_no}{351}{psnr_pet_solo}} }
			}
			\put(-2,86){
				\textcolor{red}{\scriptsize SSIM={\DTLfetch{table4}{patient_no}{351}{ssim_pet_solo}} }
			}
		\end{overpic}
		& \begin{overpic}[width=\tempdimc,height=\tempdimc]{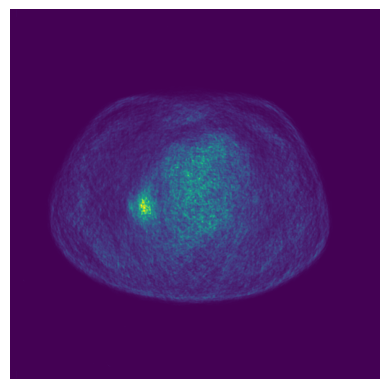}
			\put(-2,3){
				\textcolor{red}{\scriptsize PSNR={\DTLfetch{table4}{patient_no}{370}{psnr_pet_solo}} }
			}
			\put(-2,86){
				\textcolor{red}{\scriptsize SSIM={\DTLfetch{table4}{patient_no}{370}{ssim_pet_solo}} }
			}
		\end{overpic}
		& \begin{overpic}[width=\tempdimc,height=\tempdimc]{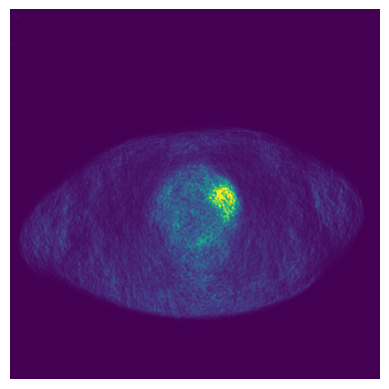}
			\put(-2,3){
				\textcolor{red}{\scriptsize PSNR={\DTLfetch{table4}{patient_no}{371}{psnr_pet_solo}} }
			}
			\put(-2,86){
				\textcolor{red}{\scriptsize SSIM={\DTLfetch{table4}{patient_no}{371}{ssim_pet_solo}} }
			}
		\end{overpic}
		& \begin{overpic}[width=\tempdimc,height=\tempdimc]{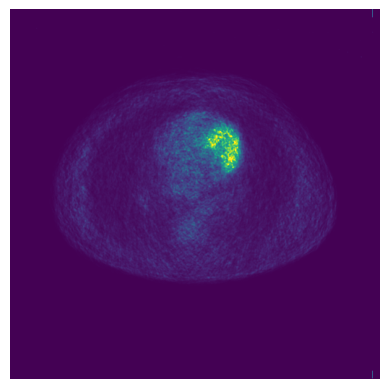}
			\put(-2,3){
				\textcolor{red}{\scriptsize PSNR={\DTLfetch{table4}{patient_no}{374}{psnr_pet_solo}} }
			}
			\put(-2,86){
				\textcolor{red}{\scriptsize SSIM={\DTLfetch{table4}{patient_no}{374}{ssim_pet_solo}} }
			}
		\end{overpic}
		& \begin{overpic}[width=\tempdimc,height=\tempdimc]{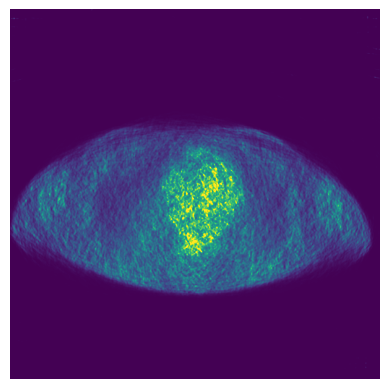}
			\put(-2,3){
				\textcolor{red}{\scriptsize PSNR={\DTLfetch{table4}{patient_no}{375}{psnr_pet_solo}} }
			}
			\put(-2,86){
				\textcolor{red}{\scriptsize SSIM={\DTLfetch{table4}{patient_no}{375}{ssim_pet_solo}} }
			}
		\end{overpic} \\
		\rowname{\footnotesize \acs{VCT}} & 
		\begin{overpic}[width=\tempdimc,height=\tempdimc]{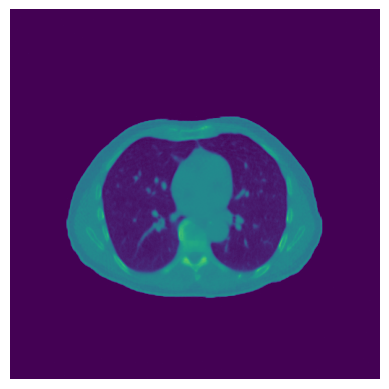}
			\put(-2,3){
				\textcolor{red}{\scriptsize PSNR={\DTLfetch{table4}{patient_no}{340}{psnr_ct_vae}} }
			}
			\put(-2,86){
				\textcolor{red}{\scriptsize SSIM={\DTLfetch{table4}{patient_no}{340}{ssim_ct_vae}} }
			}
		\end{overpic}
		&  
		\begin{overpic}[width=\tempdimc,height=\tempdimc]{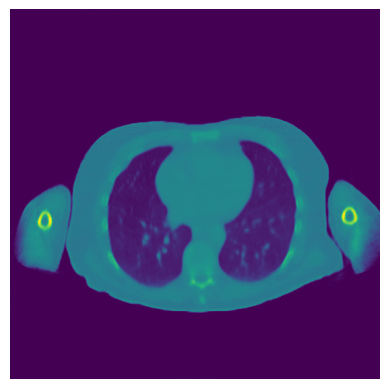}
			\put(-2,3){
				\textcolor{red}{\scriptsize PSNR={\DTLfetch{table4}{patient_no}{341}{psnr_ct_vae}} }
			}
			\put(-2,86){
				\textcolor{red}{\scriptsize SSIM={\DTLfetch{table4}{patient_no}{341}{ssim_ct_vae}} }
			}
		\end{overpic}
		&  \begin{overpic}[width=\tempdimc,height=\tempdimc]{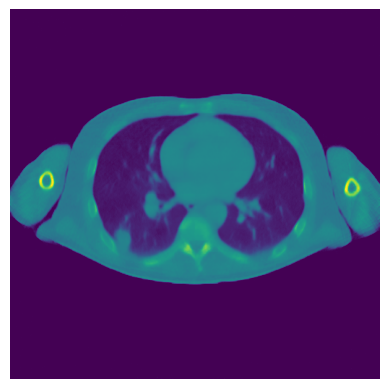}
			\put(-2,3){
				\textcolor{red}{\scriptsize PSNR={\DTLfetch{table4}{patient_no}{345}{psnr_ct_vae}} }
			}
			\put(-2,86){
				\textcolor{red}{\scriptsize SSIM={\DTLfetch{table4}{patient_no}{345}{ssim_ct_vae}} }
			}
		\end{overpic}
		& \begin{overpic}[width=\tempdimc,height=\tempdimc]{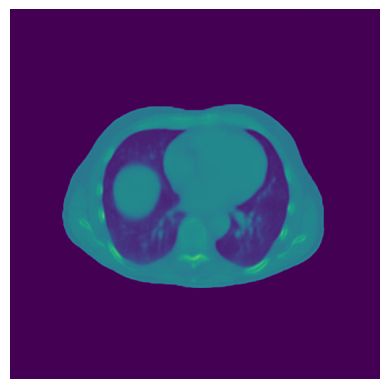}
			\put(-2,3){
				\textcolor{red}{\scriptsize PSNR={\DTLfetch{table4}{patient_no}{348}{psnr_ct_vae}} }
			}
			\put(-2,86){
				\textcolor{red}{\scriptsize SSIM={\DTLfetch{table4}{patient_no}{348}{ssim_ct_vae}} }
			}
		\end{overpic}
		&  \begin{overpic}[width=\tempdimc,height=\tempdimc]{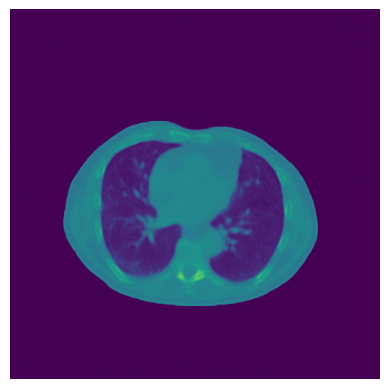}
			\put(-2,3){
				\textcolor{red}{\scriptsize PSNR={\DTLfetch{table4}{patient_no}{351}{psnr_ct_vae}} }
			}
			\put(-2,86){
				\textcolor{red}{\scriptsize SSIM={\DTLfetch{table4}{patient_no}{351}{ssim_ct_vae}} }
			}
		\end{overpic}
		& \begin{overpic}[width=\tempdimc,height=\tempdimc]{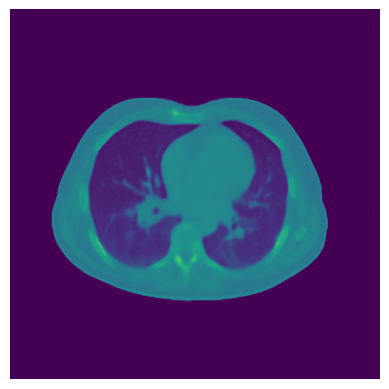}
			\put(-2,3){
				\textcolor{red}{\scriptsize PSNR={\DTLfetch{table4}{patient_no}{370}{psnr_ct_vae}} }
			}
			\put(-2,86){
				\textcolor{red}{\scriptsize SSIM={\DTLfetch{table4}{patient_no}{370}{ssim_ct_vae}} }
			}
		\end{overpic}
		& \begin{overpic}[width=\tempdimc,height=\tempdimc]{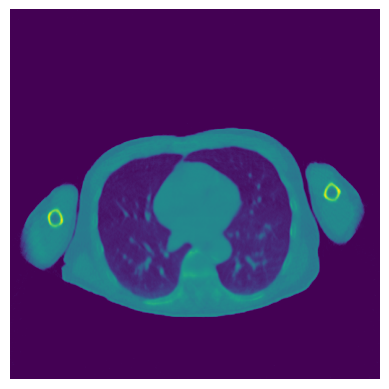}
			\put(-2,3){
				\textcolor{red}{\scriptsize PSNR={\DTLfetch{table4}{patient_no}{371}{psnr_ct_vae}} }
			}
			\put(-2,86){
				\textcolor{red}{\scriptsize SSIM={\DTLfetch{table4}{patient_no}{371}{ssim_ct_vae}} }
			}
		\end{overpic}
		& \begin{overpic}[width=\tempdimc,height=\tempdimc]{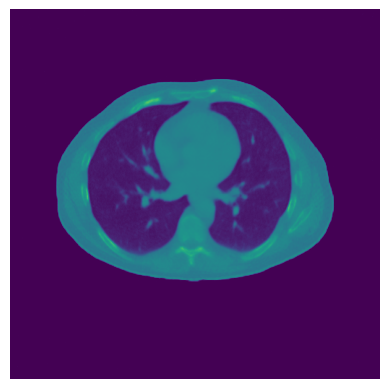}
			\put(-2,3){
				\textcolor{red}{\scriptsize PSNR={\DTLfetch{table4}{patient_no}{374}{psnr_ct_vae}} }
			}
			\put(-2,86){
				\textcolor{red}{\scriptsize SSIM={\DTLfetch{table4}{patient_no}{374}{ssim_ct_vae}} }
			}
		\end{overpic}
		& \begin{overpic}[width=\tempdimc,height=\tempdimc]{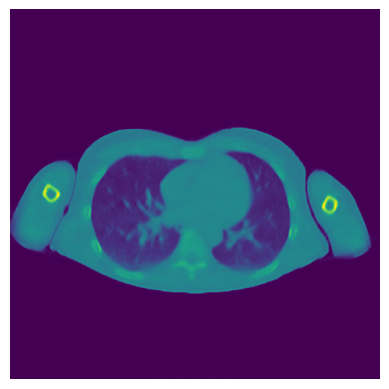}
			\put(-2,3){
				\textcolor{red}{\scriptsize PSNR={\DTLfetch{table4}{patient_no}{375}{psnr_ct_vae}} }
			}
			\put(-2,86){
				\textcolor{red}{\scriptsize SSIM={\DTLfetch{table4}{patient_no}{375}{ssim_ct_vae}} }
			}
		\end{overpic}
		\\
		\rowname{\footnotesize \acs{PCT}}  &  \begin{overpic}[width=\tempdimc,height=\tempdimc]{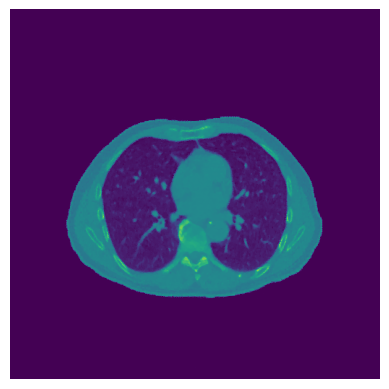}
			\put(-2,3){
				\textcolor{red}{\scriptsize PSNR={\DTLfetch{table4}{patient_no}{340}{psnr_ct_mat}} }
			}
			\put(-2,86){
				\textcolor{red}{\scriptsize SSIM={\DTLfetch{table4}{patient_no}{340}{ssim_ct_mat}} }
			}
		\end{overpic}
		&  
		\begin{overpic}[width=\tempdimc,height=\tempdimc]{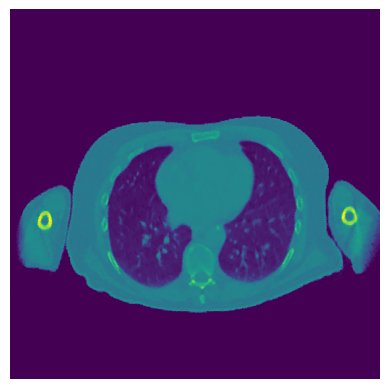}
			\put(-2,3){
				\textcolor{red}{\scriptsize PSNR={\DTLfetch{table4}{patient_no}{341}{psnr_ct_mat}} }
			}
			\put(-2,86){
				\textcolor{red}{\scriptsize SSIM={\DTLfetch{table4}{patient_no}{341}{ssim_ct_mat}} }
			}
		\end{overpic}
		&  \begin{overpic}[width=\tempdimc,height=\tempdimc]{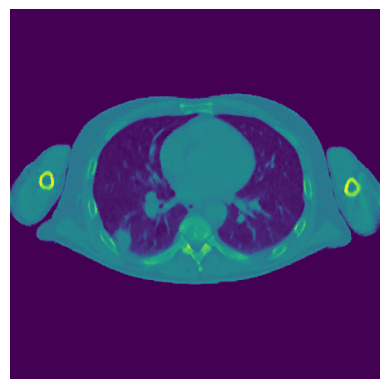}
			\put(-2,3){
				\textcolor{red}{\scriptsize PSNR={\DTLfetch{table4}{patient_no}{345}{psnr_ct_mat}} }
			}
			\put(-2,86){
				\textcolor{red}{\scriptsize SSIM={\DTLfetch{table4}{patient_no}{345}{ssim_ct_mat}} }
			}
		\end{overpic}
		& \begin{overpic}[width=\tempdimc,height=\tempdimc]{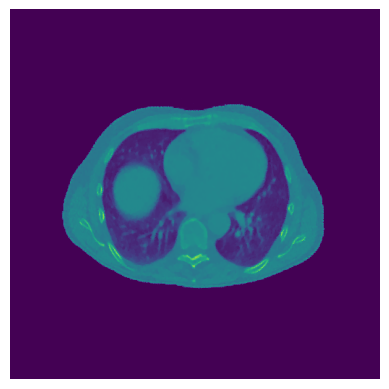}
			\put(-2,3){
				\textcolor{red}{\scriptsize PSNR={\DTLfetch{table4}{patient_no}{348}{psnr_ct_mat}} }
			}
			\put(-2,86){
				\textcolor{red}{\scriptsize SSIM={\DTLfetch{table4}{patient_no}{348}{ssim_ct_mat}} }
			}
		\end{overpic}
		&  \begin{overpic}[width=\tempdimc,height=\tempdimc]{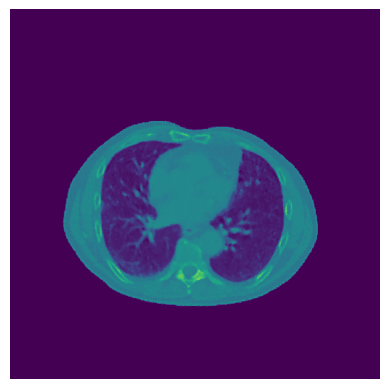}
			\put(-2,3){
				\textcolor{red}{\scriptsize PSNR={\DTLfetch{table4}{patient_no}{351}{psnr_ct_mat}} }
			}
			\put(-2,86){
				\textcolor{red}{\scriptsize SSIM={\DTLfetch{table4}{patient_no}{351}{ssim_ct_mat}} }
			}
		\end{overpic}
		& \begin{overpic}[width=\tempdimc,height=\tempdimc]{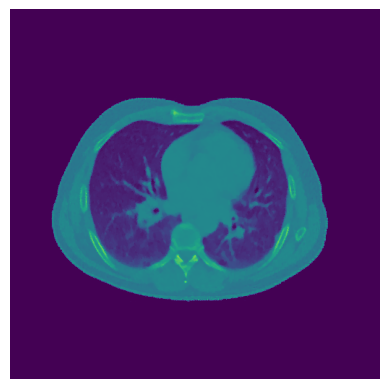}
			\put(-2,3){
				\textcolor{red}{\scriptsize PSNR={\DTLfetch{table4}{patient_no}{370}{psnr_ct_mat}} }
			}
			\put(-2,86){
				\textcolor{red}{\scriptsize SSIM={\DTLfetch{table4}{patient_no}{370}{ssim_ct_mat}} }
			}
		\end{overpic}
		& \begin{overpic}[width=\tempdimc,height=\tempdimc]{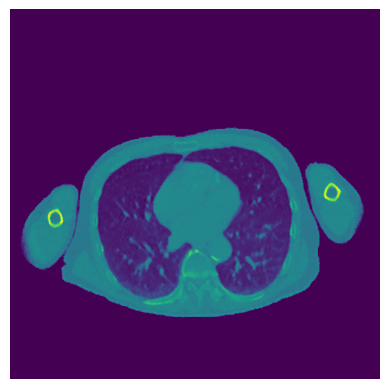}
			\put(-2,3){
				\textcolor{red}{\scriptsize PSNR={\DTLfetch{table4}{patient_no}{371}{psnr_ct_mat}} }
			}
			\put(-2,86){
				\textcolor{red}{\scriptsize SSIM={\DTLfetch{table4}{patient_no}{371}{ssim_ct_mat}} }
			}
		\end{overpic}
		& \begin{overpic}[width=\tempdimc,height=\tempdimc]{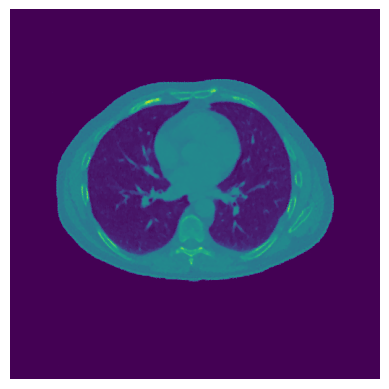}
			\put(-2,3){
				\textcolor{red}{\scriptsize PSNR={\DTLfetch{table4}{patient_no}{374}{psnr_ct_mat}} }
			}
			\put(-2,86){
				\textcolor{red}{\scriptsize SSIM={\DTLfetch{table4}{patient_no}{374}{ssim_ct_mat}} }
			}
		\end{overpic}
		& \begin{overpic}[width=\tempdimc,height=\tempdimc]{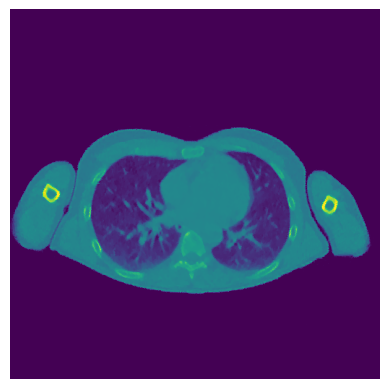}
			\put(-2,3){
				\textcolor{red}{\scriptsize PSNR={\DTLfetch{table4}{patient_no}{375}{psnr_ct_mat}} }
			}
			\put(-2,86){
				\textcolor{red}{\scriptsize SSIM={\DTLfetch{table4}{patient_no}{375}{ssim_ct_mat}} }
			}
		\end{overpic} \\
		\rowname{\footnotesize \acs{CTunet}}  
		&  \begin{overpic}[width=\tempdimc,height=\tempdimc]{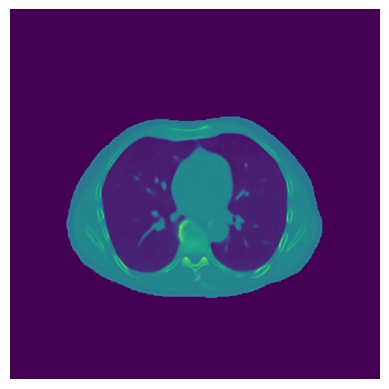}
			\put(-2,3){
				\textcolor{red}{\scriptsize PSNR=\DTLfetch{table2}{patient_no}{340}{psnr_ct} }
			}
			\put(-2,86){
				\textcolor{red}{\scriptsize SSIM=\DTLfetch{table2}{patient_no}{340}{ssim_ct} }
			}
		\end{overpic}
		&  
		\begin{overpic}[width=\tempdimc,height=\tempdimc]{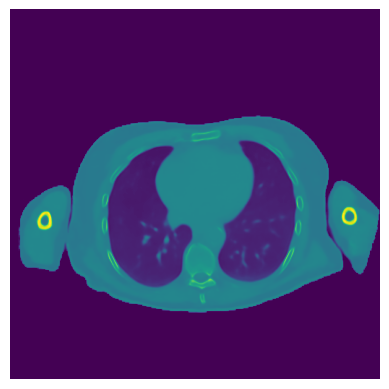}
			\put(-2,3){
				\textcolor{red}{\scriptsize PSNR=\DTLfetch{table2}{patient_no}{341}{psnr_ct} }
			}
			\put(-2,86){
				\textcolor{red}{\scriptsize SSIM=\DTLfetch{table2}{patient_no}{341}{ssim_ct} }
			}
		\end{overpic}
		&  \begin{overpic}[width=\tempdimc,height=\tempdimc]{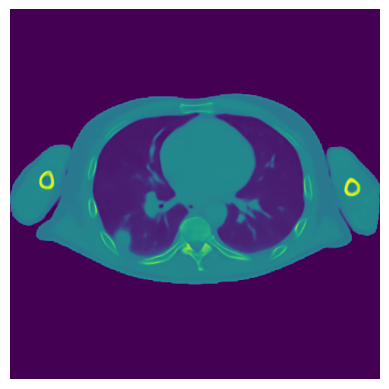}
			\put(-2,3){
				\textcolor{red}{\scriptsize PSNR=\DTLfetch{table2}{patient_no}{345}{psnr_ct} }
			}
			\put(-2,86){
				\textcolor{red}{\scriptsize SSIM=\DTLfetch{table2}{patient_no}{345}{ssim_ct} }
			}
		\end{overpic}
		& \begin{overpic}[width=\tempdimc,height=\tempdimc]{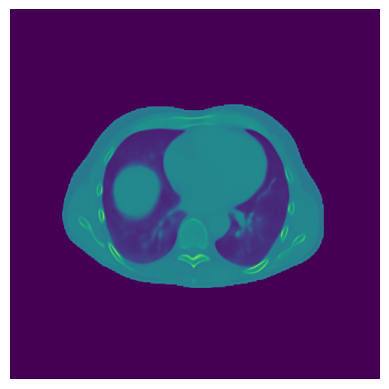}
			\put(-2,3){
				\textcolor{red}{\scriptsize PSNR=\DTLfetch{table2}{patient_no}{348}{psnr_ct} }
			}
			\put(-2,86){
				\textcolor{red}{\scriptsize SSIM=\DTLfetch{table2}{patient_no}{348}{ssim_ct} }
			}
		\end{overpic}
		&  \begin{overpic}[width=\tempdimc,height=\tempdimc]{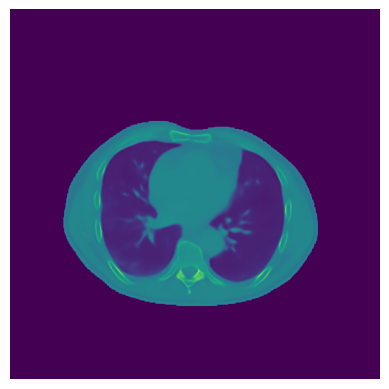}
			\put(-2,3){
				\textcolor{red}{\scriptsize PSNR=\DTLfetch{table2}{patient_no}{351}{psnr_ct} }
			}
			\put(-2,86){
				\textcolor{red}{\scriptsize SSIM=\DTLfetch{table2}{patient_no}{351}{ssim_ct} }
			}
		\end{overpic}
		& \begin{overpic}[width=\tempdimc,height=\tempdimc]{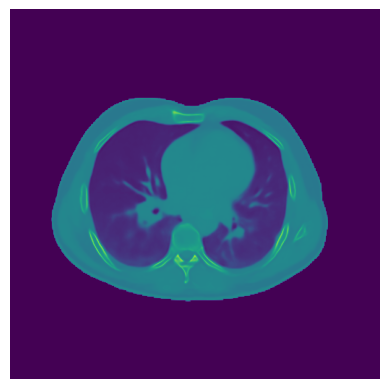}
			\put(-2,3){
				\textcolor{red}{\scriptsize PSNR=\DTLfetch{table2}{patient_no}{370}{psnr_ct} }
			}
			\put(-2,86){
				\textcolor{red}{\scriptsize SSIM=\DTLfetch{table2}{patient_no}{370}{ssim_ct} }
			}
		\end{overpic}
		& \begin{overpic}[width=\tempdimc,height=\tempdimc]{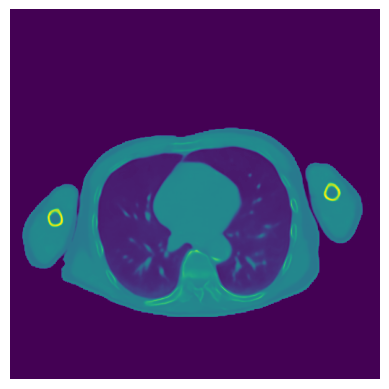}
			\put(-2,3){
				\textcolor{red}{\scriptsize PSNR=\DTLfetch{table2}{patient_no}{371}{psnr_ct} }
			}
			\put(-2,86){
				\textcolor{red}{\scriptsize SSIM=\DTLfetch{table2}{patient_no}{371}{ssim_ct} }
			}
		\end{overpic}
		& \begin{overpic}[width=\tempdimc,height=\tempdimc]{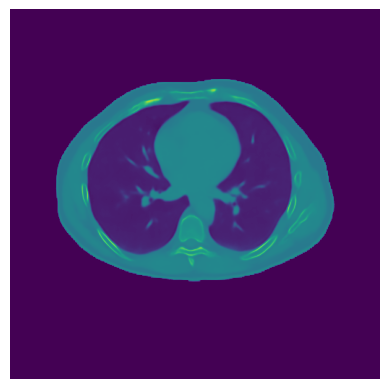}
			\put(-2,3){
				\textcolor{red}{\scriptsize PSNR=\DTLfetch{table2}{patient_no}{374}{psnr_ct} }
			}
			\put(-2,86){
				\textcolor{red}{\scriptsize SSIM=\DTLfetch{table2}{patient_no}{374}{ssim_ct} }
			}
		\end{overpic}
		& \begin{overpic}[width=\tempdimc,height=\tempdimc]{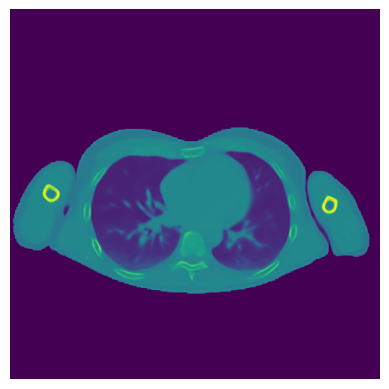}
			\put(-2,3){
				\textcolor{red}{\scriptsize PSNR=\DTLfetch{table2}{patient_no}{375}{psnr_ct} }
			}
			\put(-2,86){
				\textcolor{red}{\scriptsize SSIM=\DTLfetch{table2}{patient_no}{375}{ssim_ct} }
			}
		\end{overpic} \\
		\rowname{\footnotesize \acs{WCT}}  & 
		\begin{overpic}[width=\tempdimc,height=\tempdimc]{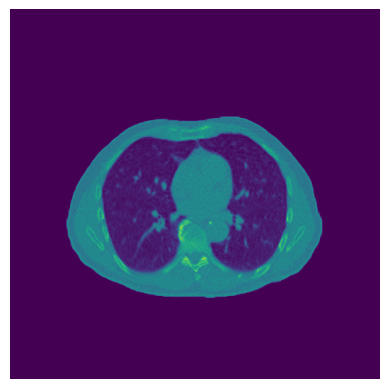}
			\put(-2,3){
				\textcolor{red}{\scriptsize PSNR={\DTLfetch{table4}{patient_no}{340}{psnr_ct_solo}} }
			}
			\put(-2,86){
				\textcolor{red}{\scriptsize SSIM={\DTLfetch{table4}{patient_no}{340}{ssim_ct_solo}} }
			}
		\end{overpic}
		&  
		\begin{overpic}[width=\tempdimc,height=\tempdimc]{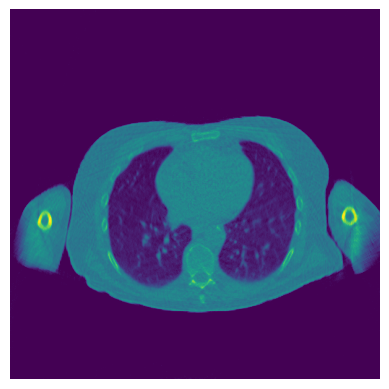}
			\put(-2,3){
				\textcolor{red}{\scriptsize PSNR={\DTLfetch{table4}{patient_no}{341}{psnr_ct_solo}} }
			}
			\put(-2,86){
				\textcolor{red}{\scriptsize SSIM={\DTLfetch{table4}{patient_no}{341}{ssim_ct_solo}} }
			}
		\end{overpic}
		&  \begin{overpic}[width=\tempdimc,height=\tempdimc]{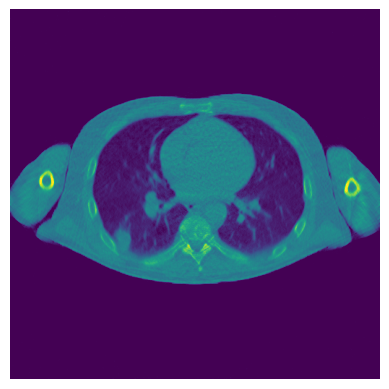}
			\put(-2,3){
				\textcolor{red}{\scriptsize PSNR={\DTLfetch{table4}{patient_no}{345}{psnr_ct_solo}} }
			}
			\put(-2,86){
				\textcolor{red}{\scriptsize SSIM={\DTLfetch{table4}{patient_no}{345}{ssim_ct_solo}} }
			}
		\end{overpic}
		& \begin{overpic}[width=\tempdimc,height=\tempdimc]{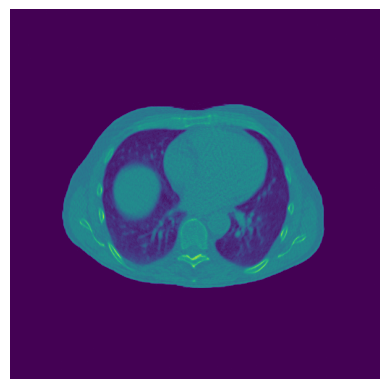}
			\put(-2,3){
				\textcolor{red}{\scriptsize PSNR={\DTLfetch{table4}{patient_no}{348}{psnr_ct_solo}} }
			}
			\put(-2,86){
				\textcolor{red}{\scriptsize SSIM={\DTLfetch{table4}{patient_no}{348}{ssim_ct_solo}} }
			}
		\end{overpic}
		&  \begin{overpic}[width=\tempdimc,height=\tempdimc]{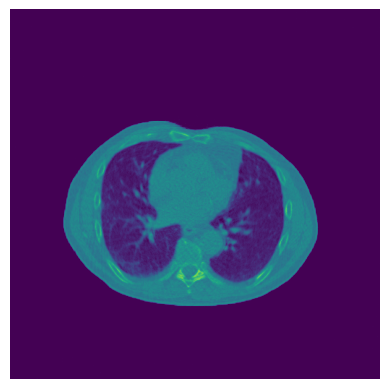}
			\put(-2,3){
				\textcolor{red}{\scriptsize PSNR={\DTLfetch{table4}{patient_no}{351}{psnr_ct_solo}} }
			}
			\put(-2,86){
				\textcolor{red}{\scriptsize SSIM={\DTLfetch{table4}{patient_no}{351}{ssim_ct_solo}} }
			}
		\end{overpic}
		& \begin{overpic}[width=\tempdimc,height=\tempdimc]{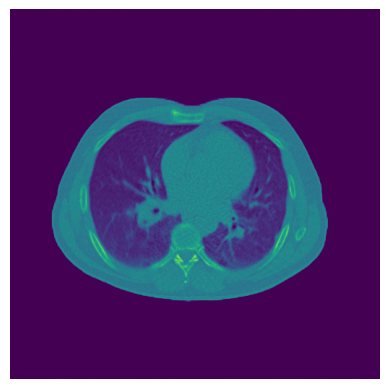}
			\put(-2,3){
				\textcolor{red}{\scriptsize PSNR={\DTLfetch{table4}{patient_no}{370}{psnr_ct_solo}} }
			}
			\put(-2,86){
				\textcolor{red}{\scriptsize SSIM={\DTLfetch{table4}{patient_no}{370}{ssim_ct_solo}} }
			}
		\end{overpic}
		& \begin{overpic}[width=\tempdimc,height=\tempdimc]{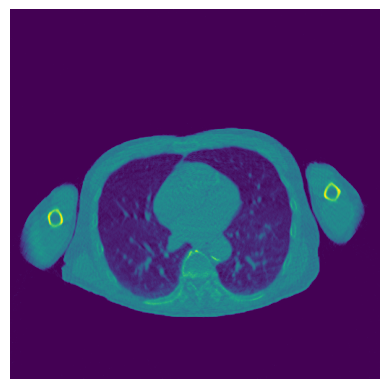}
			\put(-2,3){
				\textcolor{red}{\scriptsize PSNR={\DTLfetch{table4}{patient_no}{371}{psnr_ct_solo}} }
			}
			\put(-2,86){
				\textcolor{red}{\scriptsize SSIM={\DTLfetch{table4}{patient_no}{371}{ssim_ct_solo}} }
			}
		\end{overpic}
		& \begin{overpic}[width=\tempdimc,height=\tempdimc]{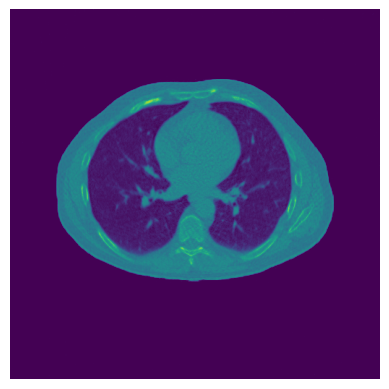}
			\put(-2,3){
				\textcolor{red}{\scriptsize PSNR={\DTLfetch{table4}{patient_no}{374}{psnr_ct_solo}} }
			}
			\put(-2,86){
				\textcolor{red}{\scriptsize SSIM={\DTLfetch{table4}{patient_no}{374}{ssim_ct_solo}} }
			}
		\end{overpic}
		& \begin{overpic}[width=\tempdimc,height=\tempdimc]{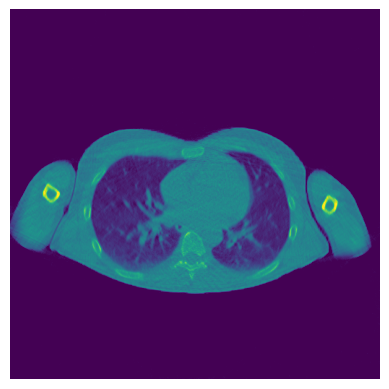}
			\put(-2,3){
				\textcolor{red}{\scriptsize PSNR={\DTLfetch{table4}{patient_no}{375}{psnr_ct_solo}} }
			}
			\put(-2,86){
				\textcolor{red}{\scriptsize SSIM={\DTLfetch{table4}{patient_no}{375}{ssim_ct_solo}} }
			}
		\end{overpic}
		
	\end{tabular}
	\caption{\Ac{LH}---Reconstructed images of the nine other patients.}\label{fig:bigfig2}
\end{figure*}